\documentclass[smallextended]{svjour3}       

\usepackage{natbib}
\usepackage{epubtk}    
\usepackage{booktabs}  
\usepackage{graphicx}  
\usepackage{journaux}
\usepackage{amsmath,amssymb}
\usepackage{color}
\usepackage[latin9]{inputenc}
\newcommand{\degr}{\ensuremath{^{\circ}}}

\newcommand{\modif}[1]{\textcolor{black}{#1}}
\newcommand{\modifeng}[1]{\textcolor{black}{#1}}

\begin{document}

\title{Asteroseismology of solar-type stars}

\author{Rafael A. Garc\'\i a \and J\'er\^ome Ballot}
\institute{Rafael A. Garc\'\i a
		\at IRFU, CEA, Universit\'e Paris-Saclay, F-91191 Gif-sur-Yvette, France
		\at 
		 AIM, CEA, CNRS, Universit\'e Paris-Saclay, Universit\'e Paris Diderot, Sorbonne Paris Cit\'e, F-91191 Gif-sur-Yvette, France\\\email{rafael.garcia@cea.fr} \\ http://irfu.cea.fr/Sap/ \\ ORCID:0000-0002-8854-3776
		\and
        J\'er\^ome Ballot
        \at IRAP, CNRS,  Universit\'e de Toulouse, UPS-OMP, CNES 14 avenue Edouard Belin, 31400 Toulouse, France \\\email{jerome.ballot@irap.omp.eu\\ ORCID:0000-0002-9649-1013}
        }

\date{}
\maketitle
%
\begin{abstract}
\modifeng{Until the last few decades}, investigations of stellar interiors \modifeng{had} been restricted to theoretical studies only constrained by observations of their global properties and external characteristics. However, in the last \modifeng{thirty years} the field has been revolutionized by the ability to perform seismic investigations of stellar interiors. \modifeng{This revolution begun} with the Sun, where helioseismology has been yielding information competing with what can be inferred about the Earth's interior from geoseismology.  The last \modifeng{two} decades have witnessed the advent of asteroseismology of solar-like stars, thanks to a dramatic development of new observing facilities providing the first reliable results on the interiors of distant stars. The coming years will see a huge development in this field.
In this review we focus on solar-type stars, i.e., cool main-sequence stars where oscillations are stochastically excited by surface convection. After a short introduction and a historical overview of the discipline, we review the observational techniques generally used, and we describe the theory behind stellar oscillations in cool main-sequence stars. We continue \modifeng{with} a complete description of the normal mode analyses \modifeng{through} which it is possible to extract the physical information about the structure and dynamics of the stars. We \modifeng{then summarize} the lessons that we have learned and discuss unsolved issues and questions that are still unanswered.

\end{abstract}

\epubtkKeywords{stars: oscillations -- stars: solar-like }

\tableofcontents  \markboth{Index}{Index}

\newpage


\section{Introduction}
\label{sec:intro}

Helio- and asteroseismology \modifeng{allow us to} study the internal structure and dynamics of the Sun and other stars by means of their resonant oscillations \citep[e.g.][and references therein]{Gou1985,STCDap1993,2002RvMP...74.1073C,2010aste.book.....A,2016LRSP...13....2B}. These vibrations manifest themselves by motions of the stellar photosphere and by temperature and density changes implying modulations of the positions of the Fraunhofer lines and of the stellar luminosity respectively. 

\modifeng{Repeated sequences of stochastic excitation and damping by turbulent motions in the external convective layers lead to a suite of resonant modes in the Sun}
\citep{1977ApJ...212..243G,1988ApJ...326..462G,1992MNRAS.255..639B,1994ApJ...424..466G,2001A&A...370..136S,2008A&A...478..163B}. The stars where their modes are excited in this way are usually called ``solar-like \modifeng{pulsators}" or simply ``solar-like" stars even though their structure and dynamics could be different compared to the actual Sun, \modif{ covering main-sequence (MS), sub-giant and red-giant stars} \citep[e.g.][]{2009Natur.459..398D,2010ApJ...713L.176B,GarStello2015,2017A&ARv..25....1H}. The oscillation periods of solar-like stars range from minutes to years \citep[e.g.][]{2010A&A...517A..22M,2013A&A...559A.137M,2013ApJ...765L..41S,2014ApJ...788L..10S,2014ApJS..210....1C}.

\begin{figure}[!htb]
\center
\includegraphics[width = 0.90\textwidth]{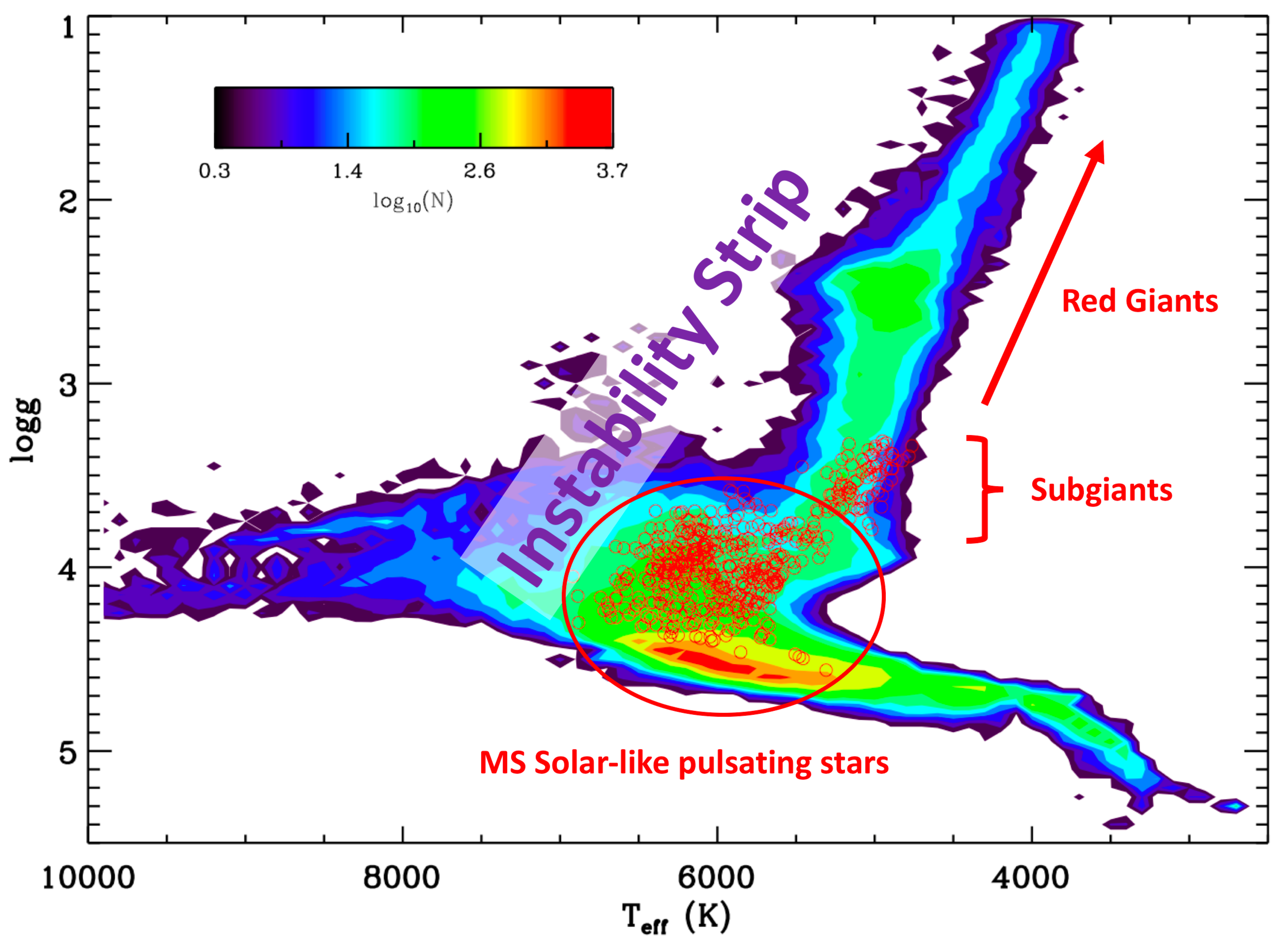}
\caption[Kiel diagram showing observed \emph{Kepler} MS Solar-like stars]{\label{HRdiag_Fig}  Kiel diagram, i.e. the logarithm of the surface gravity, $\log g$, as a function of the effective temperature, $T_{\rm{eff}}$. The diagram is color coded by the logarithm of the number of stars, N, observed by \emph{Kepler} \citep{2017ApJS..229...30M}. Open red circles locate MS and sub-giant stars where pulsations have been measured by \emph{Kepler} \citep{2014ApJS..210....1C}. \modif{In this review, we focus on cool MS solar-like stars that will be referred as solar-type stars.} }     
\end{figure}

\modif{ In this review, we focus on ``Solar-\modif{type}" stars defined as cool main-sequence dwarfs} located below the red edge of the classical instability strip (spectral types from \modifeng{late F, G and K dwarfs}, see the zone encircled by the red circle in Figure~\ref{HRdiag_Fig}) with a near-surface convective zone that excites the oscillation modes. \modif{However, to put these stars in context, we will sometimes discuss sub-giants and early red giants. In such a way, the continuity towards more evolved stars will be ensured}.

There are other mechanisms exciting stellar oscillations in more massive and luminous main-sequence stars: a) the heat-engine mechanism (also known as $\kappa$ or opacity-driven mechanism), related to the changes in the opacity profile due to temperature variations, and responsible for the pulsation in the instability strip \modif{and white dwarfs} \citep[e.g.][]{1926Obs....49...88E}; b) the $\epsilon$ mechanism, where the nuclear reaction rate changes as a consequence of the contraction and expansion of the star \citep[e.g.][]{2006CoAst.147...93L}.; c) tidal effects, where non-radial oscillations can be forced in stars belonging to multiple systems \citep[e.g.][]{2011ApJS..197....4W}. All of these pulsating stars are usually referred to by the generic term ``classical \modifeng{pulsators}" (e.g. $\delta$ Scuti, $\gamma$ Doradus, RR Lyrae, Cepheids, etc) and their study is out of the scope of this review \citep[for more information on these variable stars see, for example,][]{2010aste.book.....A}.
  
\section{Asteroseismology of solar-type stars in a helioseismic context}
\label{sec:astero_context}

\begin{figure}[!htb]
\center
\includegraphics[width = 0.75\textwidth]{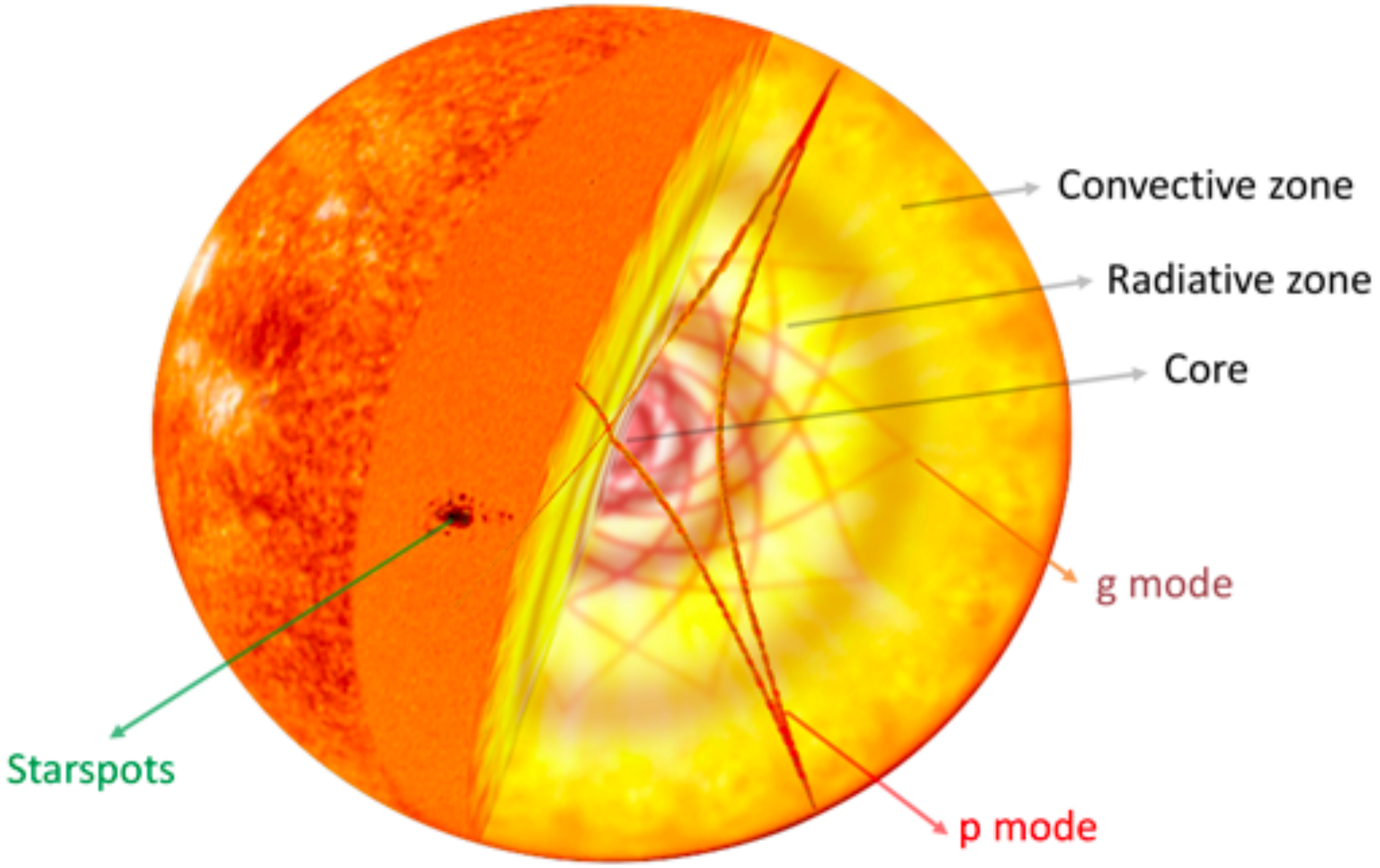}
\caption[Artist's view of the structure of a solar-like star]{\label{fig_star_struct} Artistic view of the structure of a typical 1 M$_\odot$ cool main-sequence solar-like pulsating star where two modes are propagating, a low-degree p mode and a g mode. The structure of the star is divided in two main regions: the inner radiative zone (including the core where the nuclear reactions take place) and the external convective zone. In the Sun, the convective envelope extends inward from the surface 30\% of its radius. }
\end{figure}

The best known star representative of  solar-like stars is the Sun. \modifeng{Over} the last 30 years, helioseismology has proven its ability to study the structure and dynamics of the solar interior in a stratified \modifeng{(layer-by-layer)} and latitudinally varying way (see Figure~\ref{fig_star_struct}). Starting from the photosphere, the internal structure is divided into a convective envelope followed by a radiative zone that includes the inner core where the nuclear reactions to transform Hydrogen in Helium take place. In more massive stars ($\gtrsim$ 1.2 M$_\odot$) a convective core develops with a total size that is not well constrained yet \citep[e.g.][]{1965ApJ...142.1468S,1991A&A...252..179Z,2016A&A...589A..93D}.

 Seismic \modifeng{analysis} tools were first applied to our closest star, the Sun, in order to infer its radial and latitudinal internal structure and dynamics. Therefore, the sound-speed profile \citep[e.g.][]{BasJCD1997,STCBas1997}, the density profile \citep[e.g.][]{2009ApJ...699.1403B}, the internal rotation in the convective zone \citep[e.g.][]{ThoToo1996} and in the radiative region \citep[e.g.][]{ElsHow1995,BasJCD1997,1999MNRAS.308..405C,CouGar2003,2004SoPh..220..269G,2008ApJ...679.1636E,2008SoPh..251..119G} or the conditions and properties of the solar core  \citep[e.g.][]{STCCou2001,STCGar2004,2007Sci...316.1591G,2008AN....329..476G,2008SoPh..251..135G,2009ApJ...699.1403B,2010A&ARv..18..197A} have been studied and well determined. Moreover, \modifeng{the characterization of the p-mode properties has led to the determination of} other quantities such as the position of the base of the convection zone \citep[e.g.][]{JCDGou1985,BalSTC2004} or the Helium abundance \citep[e.g.][]{1983prhe.work..117G,1991Natur.349...49V} with high precision. With all of these observational constraints, the standard solar and stellar evolution models have been significnatly improved, reducing the uncertainties in the calculation of the stellar ages when individual p-mode frequencies are considered (see for more detail the reviews by \citet{2014EAS....65...99L} and \citet{2014EAS....65..177L}). However, new asteroseismic observations of many other stars \citep[e.g.][]{2011Sci...332..213C,2011ApJ...743..143H,2017ApJ...835..172L} covering a larger fraction of the H-R diagram, allow us to test stellar evolution under many different conditions \cite[e.g.][]{2010Ap&SS.328...51C} while putting the Sun in its evolutionary context. 
 
In asteroseismology, due to the absence of spatial resolution in the observations, only low-degree modes (those with a small number of nodal lines at the surface of the star) are measured. Therefore compared to the Sun, less detailed information is available on stellar interiors. On the other hand, some pulsating solar-like stars offer the possibility to observe mixed modes, i.e., modes with mixed character resulting from the coupling between p and g modes \citep{2008ApJ...687.1180A,2010ApJ...713..935B,2010ApJ...713L.169C,2010A&A...515A..87D,2011Sci...332..205B}. The study of these modes allows us to better constrain the structure and dynamics of the deep radiative interiors \cite[e.g.][]{2010A&A...514A..31D,2010ApJ...723.1583M,2011Natur.471..608B}. Unfortunately, neither mixed-modes nor pure g modes have been identified individually in main-sequence solar-like stars so far because they become evanescent in the convective region and their surface amplitudes are small compared to the granulation signal. Thus, all  of the information that we are obtaining for these stars comes from the characterization of p modes. However, it is important to note that for the special case of the Sun, the global signatures of the dipolar g modes have been measured \citep{2007Sci...316.1591G} with GOLF/SoHO, as well as some individual low-frequency g modes through the study of the perturbations induced by the g modes on the acoustic modes \citep{2017A&A...604A..40F}. \modif{Both results are still controversial as shown for example by \cite{2018SoPh..293...95S}, who demonstrated that the latter detection of individual modes is highly dependent on the selection of the parameters used in the analysis}.

\begin{figure}[!htb]
\center
\includegraphics[width = 0.8\textwidth]{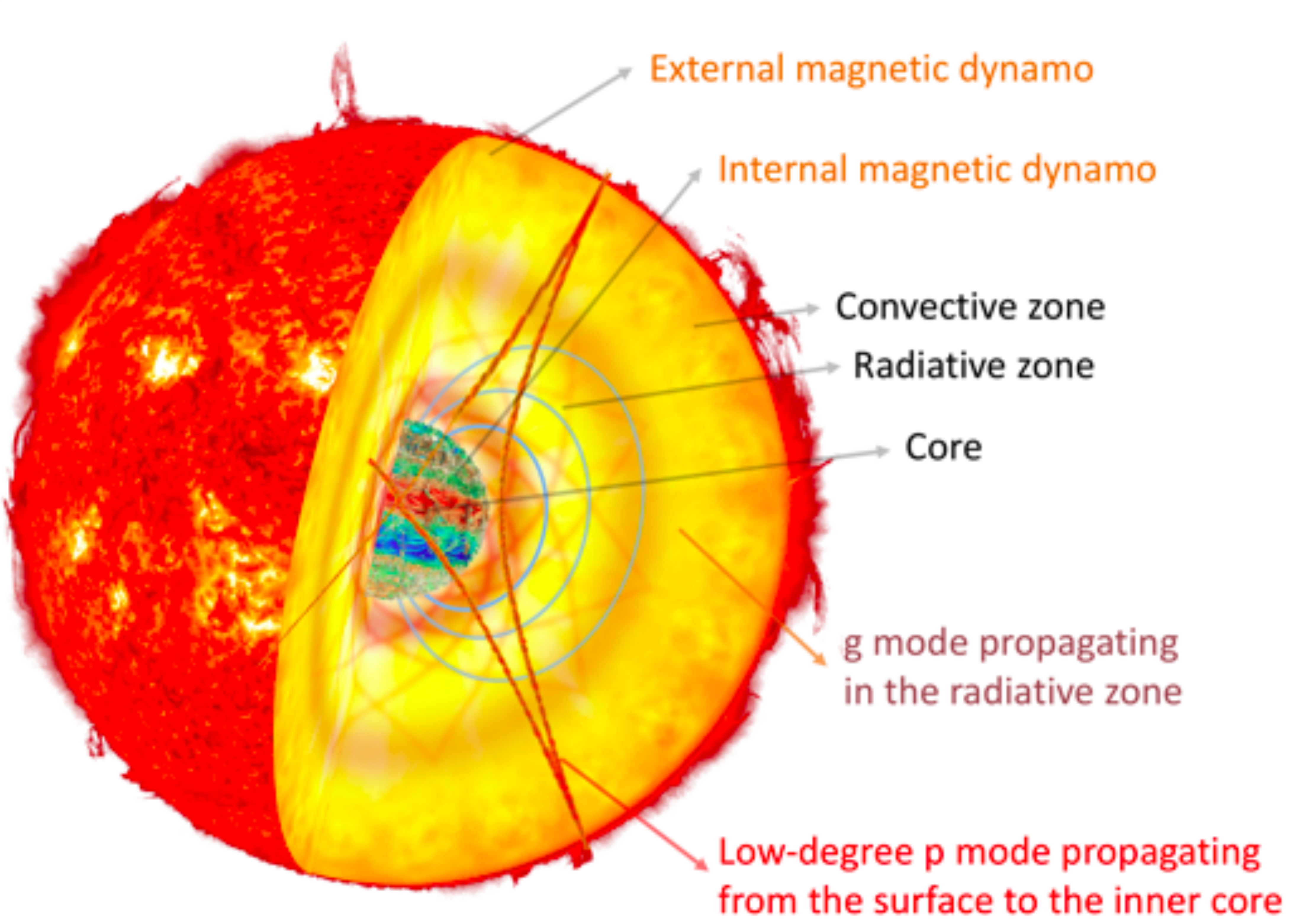}
\caption[Artist's view of the structure of a solar-like star above 1.2 M$_\odot$ with an hypothetical core dynamo]{\label{artist_core_mag} Artist's view of the structure of an F-type main-sequence solar-like star with a convective core (Mass~$\gtrsim 1.2$~M$_\odot$) and two possible dynamos, one at the external convective zone and another one in the convective core. }
\end{figure}

Stars are also known to be rotating magnetic objects. Such dynamical processes affect the internal structure and evolution of stars \citep[e.g.][]{2004ApJ...614.1073B,2008sf2a.conf..341Z,2010MNRAS.402..271D,2010A&A...519A.116E,2013ApJ...776...67V,2013EPJWC..4301005E}. Thus, it is necessary to go beyond the classical modeling of stellar interiors and evolution by taking into account transport and mixing mechanisms both on dynamical and secular time-scales \citep[e.g.][]{2004A&A...425..229M,2005A&A...440..653M,2010ApJ...715.1539T,2013ApJ...775L...1T,2017A&A...599A..18E}.  
\modifeng{A new generation of stellar models is fundamental to understand current and future observations of stellar surfaces and interiors. Indeed, more and more constraints on the stellar rotation profiles are being obtained \cite[e.g.][]{2008A&A...484..517M,2010ApJ...715.1539T,2012ApJ...756...19D,2014A&A...564A..27D,2015MNRAS.452.2654B,2016ApJ...817...65D,2017A&A...603A...6N}, while anomalies at stellar surfaces are found \citep[e.g.][]{2007ApJ...668..594M,2007A&A...469.1145Z}. Furthermore, hints on magnetic fields either at the surface \citep[e.g.][]{2010A&A...509A..32J} or in the inner cores of the stars \citep{2016Natur.529..364S} have been suggested and could be due to on-going dynamos developing in the convective cores of stars above 1.2 to 1.4 M$_\odot$ (see Fig.~\ref{artist_core_mag}).}


\section{Asteroseismic observations of solar-like stars}
\label{sec:obs}

The requirements needed to perform asteroseismic studies of distant stars are shared with helioseismology and any other seismic studies. Stable and uninterrupted observations are ideal because most of the analyses are performed in the frequency domain, \modifeng{requiring long observations} to increase the frequency resolution. 

When preparing the observations it is also mandatory to choose a sampling rate rapid enough \modifeng{that} the Nyquist frequency is well above the acoustic cut-off frequency of the oscillation modes. Conversely, when the stellar oscillations are just above the Nyquist frequency, aliased peaks are reflected from the Nyquist frequency leaking into lower frequencies. In this case, it is still possible to do asteroseismology for ``super-Nyquist" oscillations as first shown by \citet{2013MNRAS.430.2986M} and then applied to solar-like stars by \citet{2014MNRAS.445..946C}. Low-mass cool main-sequence and sub-giant stars have a frequency of maximum power above $\sim$ 500-8000 $\mu$Hz, i.e., in the range of $\sim$2 to $\sim$30 minutes. Therefore, a sampling rate faster than $\sim$1 minute is recommended to avoid dealing with super-Nyquist asteroseismology.  

Continuity is needed to reduce the effect of gaps in the data. In particular, regular gaps -- seen as a Dirac Comb function -- should be avoided. Regular gaps are typical in single ground-based observations due to the day/night cycle or from a satellite due to regular operations such as angular momentum dumps of the reaction wheels used to stabilize the spacecraft. When regular gaps are present in the time series, the power of every stellar oscillation peak leaks into surrounding side-lobes due to the convolution by the Fourier transform of the signal with the window function. Examples of the impact of the NASA \emph{Kepler} window function on stellar oscillations can be found in \citet{2014A&A...568A..10G}. If the gaps are sparse, the level of noise in the spectrum increases as a function of the duty cycle \citep[see examples in][]{2015A&A...574A..18P} and the signal-to-noise ratio decreases. 

Finally, stable instruments are necessary to minimize any possible instrumental modulations that could generate peaks in the same frequency domain as the expected stellar oscillations. In the case of multi-site observations, it is recommended to have instruments as similar as possible. \modifeng{However, global asteroseismic observing campaigns have shown} that it is possible to use very different instruments, \modifeng{normalize the data, and recover} the stellar pulsations \citep[e.g.][]{2010ApJ...713..935B}.

\begin{figure}[!htb]
 \centering
 \includegraphics[height=.35\textheight, width=0.7\textwidth]{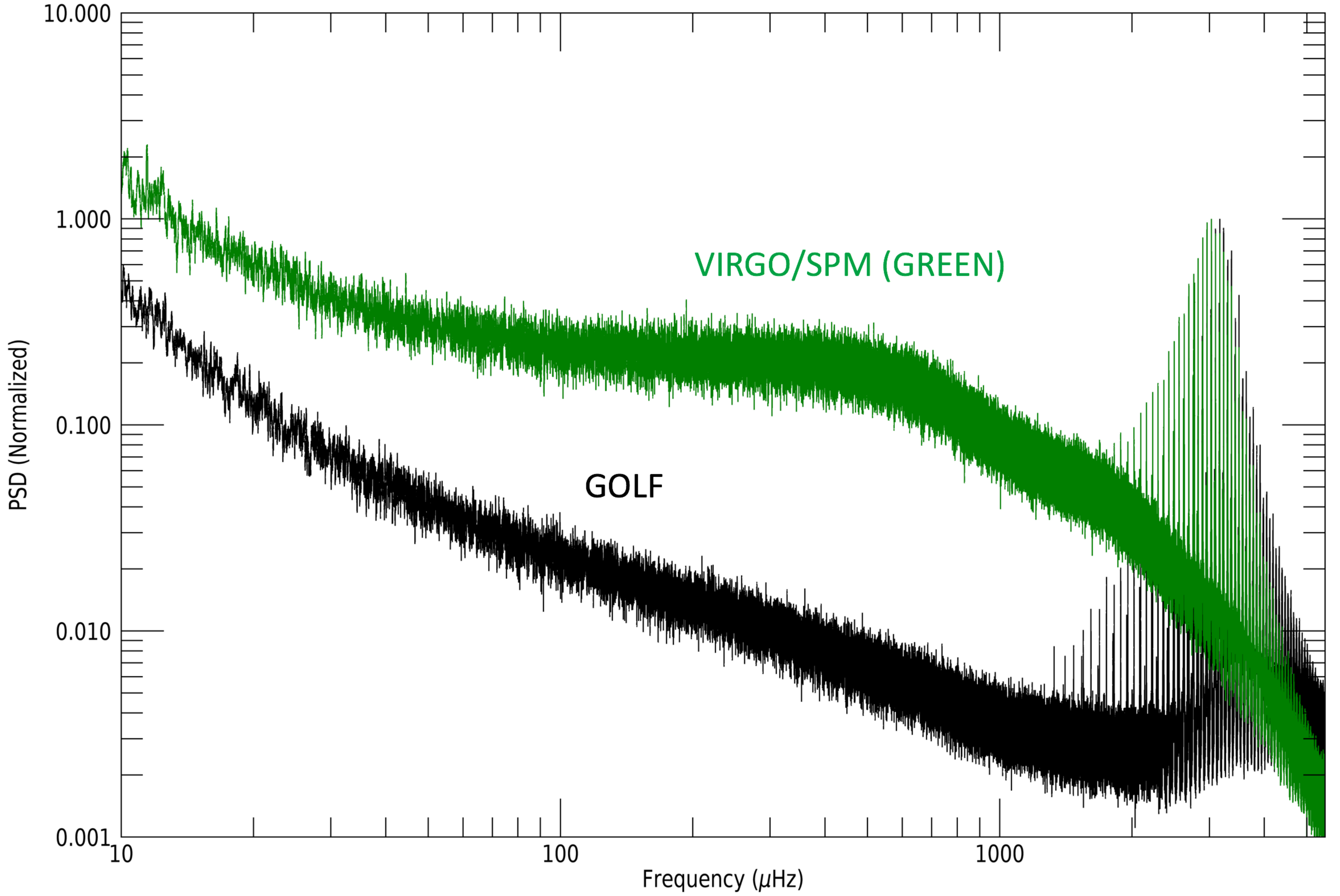}
  \caption[Comparison of the PSD obtained from velocity and intensity measurements from GOLF and VIRGO/SPM]{\label{Comp_Golf_Virgo} Comparison between the power spectrum density extracted from 21 years of Doppler velocity (by GOLF, in black) and from photometric measurements (by VIRGO/SPM green channel, in green).}
\end{figure}

Oscillations on the Sun and stars can be measured using two fundamental techniques: by measuring the Doppler velocity of the surface of the stars or by measuring the luminosity changes induced by the changes in temperature due to the pulsations. By using the instruments on board the ESA/NASA \emph{Solar and Heliospheric Observatory} \citep[SoHO,][]{DomFle1995}, it is possible to compare the solar power spectrum measured in Sun-as-a-star Doppler velocity by the \emph{Global Oscillations at Low Frequencies} \citep[GOLF,][]{GabGre1995,GabCha1997} instrument and using integrated photometry by the Sun photometers of the \emph{Variability of solar IRradiance and Gravity Oscillations instrument} \citep[VIRGO/SPMs][]{1995SoPh..162..101F}. The power spectral density obtained from the two instruments is shown in Fig.~\ref{Comp_Golf_Virgo}. To facilitate the comparison at all frequency ranges, the power spectral density has been normalized in such a way that the maximum of the p-mode bump has the same amplitude in both observables. The convective background is higher in photometry than in Doppler velocity (more than an order of magnitude at 0.5 mHz). Therefore, the signal-to-noise ratio of the acoustic modes is smaller in photometry (a factor $\sim$30) while it reaches a factor of $\sim$300 in Doppler velocity. Using Doppler velocity, it is then possible to characterize a larger number of modes at low frequency.

Observational asteroseismology of main-sequence cool dwarfs developed during the 1990s and the first years of the twenty-first century. The first solar-like star for which pulsations were observed was $\alpha$ Cen A. It was first observed in photometry by \citet{2000AAS...197.4604S,2001ESASP.464..391S} using the \emph{Wide-Field Infrared Explorer} satellite \citep[WIRE,][]{2000ApJ...532L.133B,2006MNRAS.371..935F}. Later, it was re-observed using Doppler velocity techniques from the ground \citep{2001A&A...374L...5B,2001ESASP.464..431M,2002A&A...390..205B}. 

Two other more evolved stars (sub-giants) were studied at that time too: $\alpha$ CMi (Procyon) and $\eta$ Boo. Procyon was observed by several teams from the ground using Doppler-velocity measurements \citep[][]{1991ApJ...368..599B,1998A&A...340..457M,1999A&A...351..993M,BouMae2004}. Procyon was also studied in photometry from space \citep{2004Natur.430...51M} by the Canadian satellite Microvariability and Oscillations of STars \citep[MOST,][]{1998JRASC..92Q.314M,2007CoAst.151....5G}, providing some controversial results \citep[see for example the discussions in][]{2005A&A...432L..43B,2005A&A...444L...5R,2007CoAst.150..106B}. \modifeng{Today, however, mode detection for} Procyon is well established thanks to a multi-site ground-based campaign \citep{2008ApJ...687.1180A,2010ApJ...713..935B}. Several individual modes were identified and the internal structure of the star was extracted using these asteroseismic constraints. $\eta$ Boo was first asteroseismically observed in the equivalent width of the Balmer lines by \citet[][]{1995AJ....109.1313K} (re-observed in radial velocity by \cite{ 2003AJ....126.1483K}) and its oscillations were independently confirmed by the MOST space photometric observations \citep{2005ApJ...635..547G}.

Although the highest signal-to-noise asteroseismic results are obtained by observing stars in Doppler velocity, most of the pulsating main-sequence, solar-like stars have been observed using photometric techniques. Indeed, photometric instruments have better performance outside the Earth's atmosphere. They require fewer photons per star allowing a high sampling rate while keeping the telescope size small. Moreover, many objects can be studied at a time. To give an example, the NASA \emph{Kepler} mission \citep{2010Sci...327..977B} allowed observing 512 stars at any time with a short cadence of 60 s.

\modif{Modern space-based asteroseismology of solar-type stars and sub-giants started with the \emph{Convection Rotation and Planetary Transits space mission}  \citep[CoRoT][]{2006cosp...36.3749B}, which observed around a dozen such targets. Originally, the objective was to study hot F stars because they were expected to have higher amplitude modes when compared to G and K dwarfs. Unfortunately, the widths of the modes in these stars were also very large. Although this was expected, it was found that the modes overlapped each other and it was extremely difficult to properly identify the modes and to extract precise p-mode parameters \citep[see the discussions in][]{2008A&A...488..705A,2009A&A...507L..13B}. Hence, the observing strategy evolved and cooler stars were prioritized. The same strategy was then followed later with \emph{Kepler}.} 

So far, cool main-sequence dwarfs and early sub-giants have been observed during five space missions: WIRE, MOST, CoRoT, \emph{Kepler}, including its second's life as K2 \citep{2014PASP..126..398H}, \modif{and the \emph{Transiting Exoplanet Survey Satellite} \citep[TESS, ][]{2014SPIE.9143E..20R}. TESS is primarily a mission for sub-giant stars as demonstrated by the theoretical studies already done \citep[][]{2017EPJWC.16001006C,2019arXiv190110148S} and corroborated by the first marginal detection of pulsations in the solar-type star $\pi$~Mensae \citep{2018A&A...619L..10G} and the clear detection of pulsations in the late sub-giant TOI-197 \citep{2019arXiv190101643H}. The small fraction of solar-type stars observed by TESS will be extremely useful as these targets will be very bright and ground-based complementary studies will contribute to better characterizing them. An additional space-based observatory, the ESA M3 Planetary Transits and Oscillations of stars (PLATO) mission \citep{2014ExA....38..249R} is expected to be launched around 2026}. From the ground, the SONG network \citep{2011JPhCS.271a2083G} is already running with two sites, \modif{Spain and China, with a site in Australia expected to be operative in 2020}. SONG will also be able to study solar-type stars although it will be best suited for sub-giants and red giants  \modif{\citep[e.g.][]{2017ApJ...836..142G,2019arXiv190106187A}.}

\subsection{Structure of the power spectrum density of a solar-type star}
\label{ssec:spectrum}

\begin{figure}[!htb]
\begin{center}
\includegraphics[width=.95\textwidth]{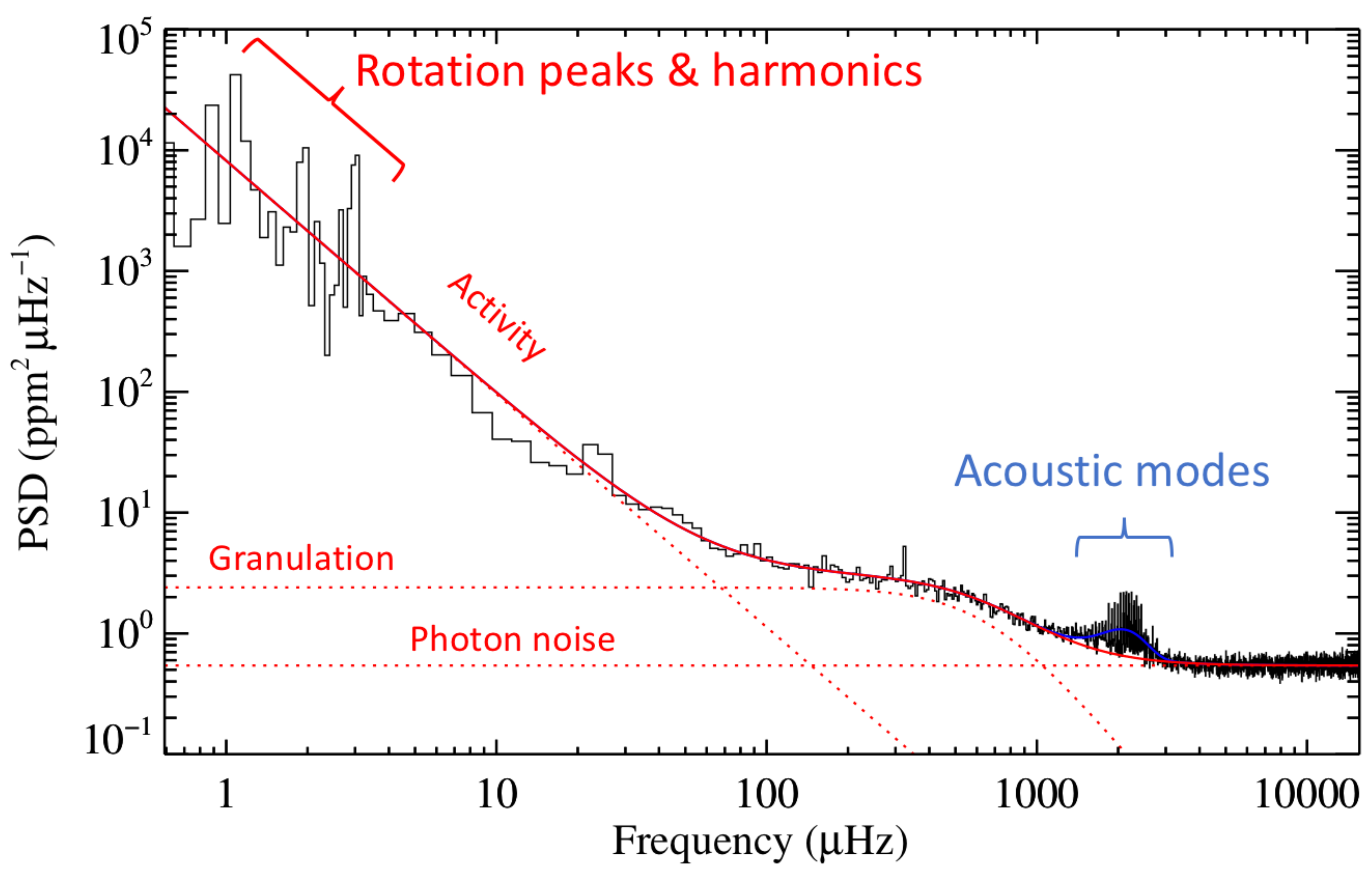}
\end{center}
\caption{\label{fig:PSD} Power spectrum density (PSD) of the CoRoT target HD52265 \citep{2011A&A...530A..97B}. Physical phenomena associated with each region of the PSD are indicated: photon noise, oscillations, convection (granulation), activity related slope, and rotation through the spot modulation of the emitted stellar flux. The continuous red line represents the fitted background components. The blue continuous line is the gaussian fit over the p-mode hump.}
\end{figure}

Asteroseismic analyses are mostly performed in the Fourier domain by computing the amplitude or power spectrum density. An example of the power spectrum density of HD~52265, a G0V star observed over 117 days by CoRoT, is shown in Fig.~\ref{fig:PSD}. Depending upon the frequency range to be explored, the spectrum is dominated by features related to different physical phenomena. 

Starting from the low-frequencies ($<$10 $\mu$Hz), the spectrum is dominated by a series of high peaks and their harmonics. These peaks correspond to the  surface differential rotation of the star because of the modulation induced in the mean stellar luminosity by dark spots crossing the visible face of the stellar disk \citep[e.g.][]{lrsp-2005-8}. The average surface rotation of this star is  12.3 $\pm$ 0.15 days \citep{2011A&A...530A..97B}. At higher frequencies, between 50 and 1000 $\mu$Hz, the spectrum is dominated by a continuum \citep{1985ESASP.235..199H}, which is the result of the turbulent movements at the surface of the star due to convection, such as granulation or supergranulation. At even higher frequencies, the p-mode envelope is visible. For this star, the acoustic modes are centered around 2000 $\mu$Hz, i.e., around 8 minutes. For reference, the oscillations of the Sun are centered around 3000 $\mu$Hz (5 minutes). Finally, close to the Nyquist frequency, the spectrum is flat and it is dominated by the photon noise of the instrument. This noise level depends on the properties of the instrument and it would eventually be above the p-mode hump in stars for which the modes have low amplitudes or when the star is distant and faint.

\begin{figure}[!htb]
\begin{center}
\includegraphics[width=.95\textwidth]{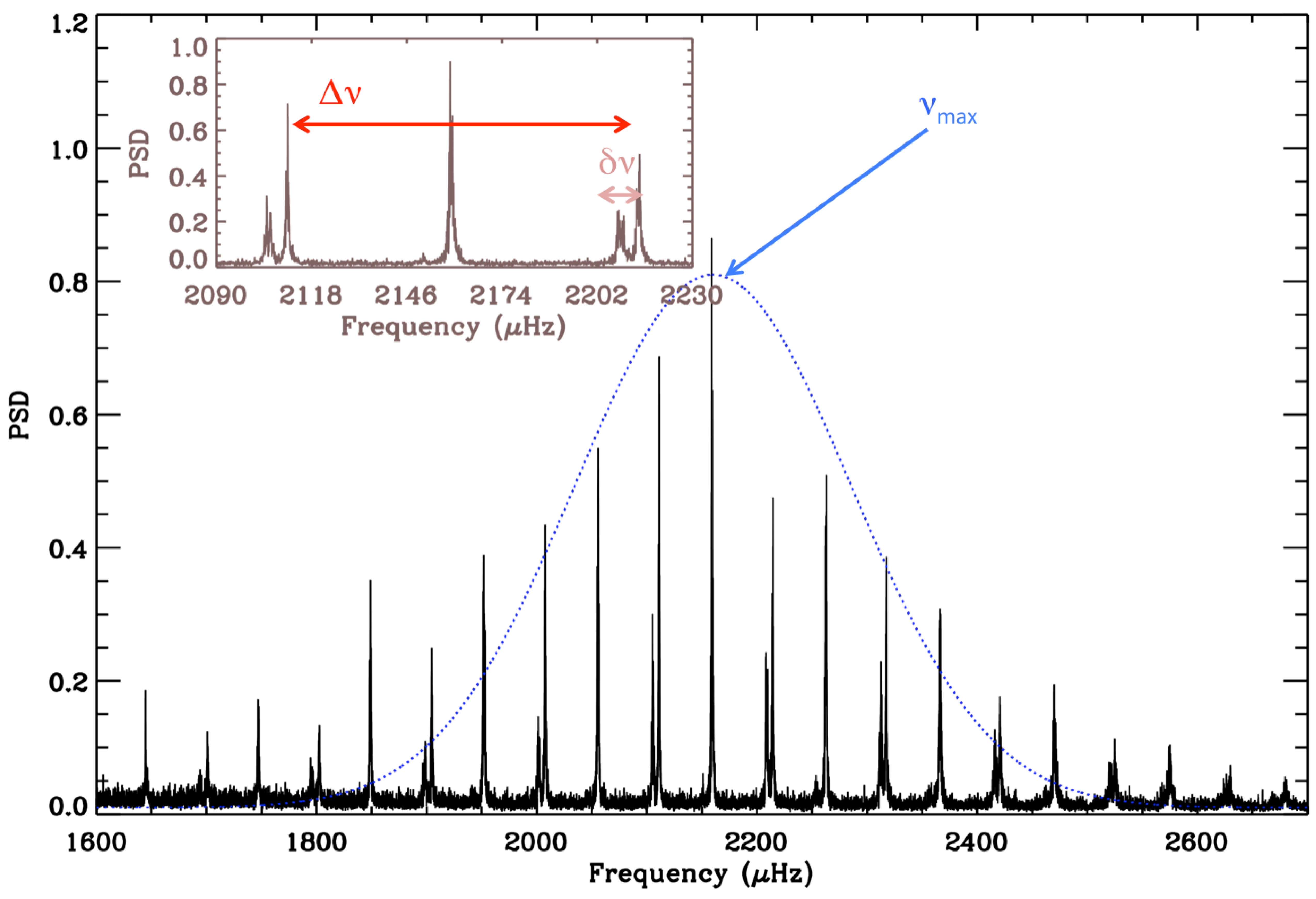}
\end{center}
\caption{\label{fig:Separation} Power spectrum density (in arbitrary units) of the \emph{Kepler} target 16~Cyg A. The blue dotted line represents the Gaussian fit to obtain the frequency at maximum power: $\nu_{\mathrm{max}}$. The inset is an enlargement showing the large frequency separation, $\Delta\nu$, between two consecutive modes of angular degree $\ell$=0 and the small frequency separation $\delta \nu_{0,2}$. Figure adapted from \citet{2015EAS....73..193G}.}
\end{figure}

\modifeng{Examining the frequency range near the p-mode envelope, one may see that} it is composed of a sequence of peaks following a repetitive pattern as shown in Fig.~\ref{fig:Separation} for the \emph{Kepler} target 16~Cyg~A \citep{2012ApJ...748L..10M,2015MNRAS.446.2959D}. Two main regularities can then be extracted: the large frequency separation, $\Delta\nu$, and the small frequency separation, $\delta\nu$, or simply the large and small separations (see Sec~\ref{sec:freqSep} for more details). A third global seismic parameter, $\nu_\mathrm{max}$, can also be defined as the frequency at maximum power of the p-mode envelope \citep{1991ApJ...368..599B,1995A&A...293...87K}.

\section{Theory of oscillations}
\label{sec:theory}

After describing the different observations, we present in this section the theory developed to interpret the stellar oscillation spectra. We focus here on theoretical concepts that are useful for this review. More detailed descriptions may be found in, e.g., \citet{1980tsp..book.....C,1989nos..book.....Ub,2002RvMP...74.1073C,2010aste.book.....A}.

\begin{figure}[!htb]
\begin{center}
\begin{tabular}{cc}
	\includegraphics[width=0.5\textwidth]{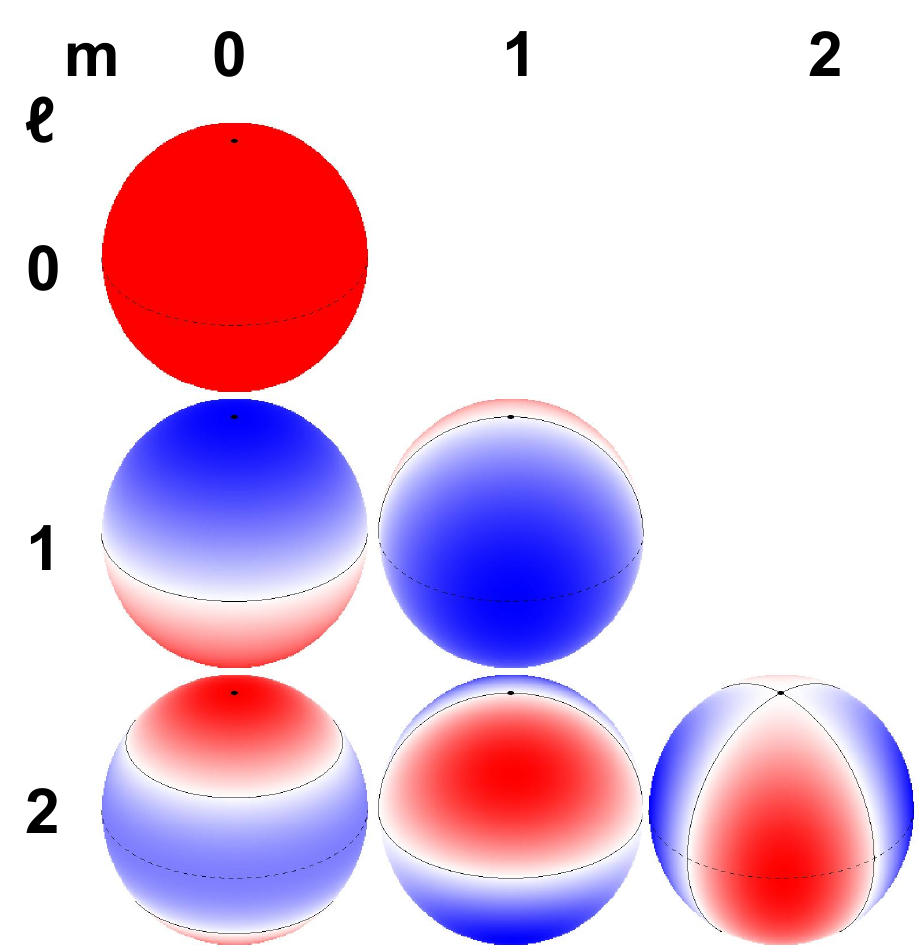} &
	\includegraphics[width=0.5\textwidth]{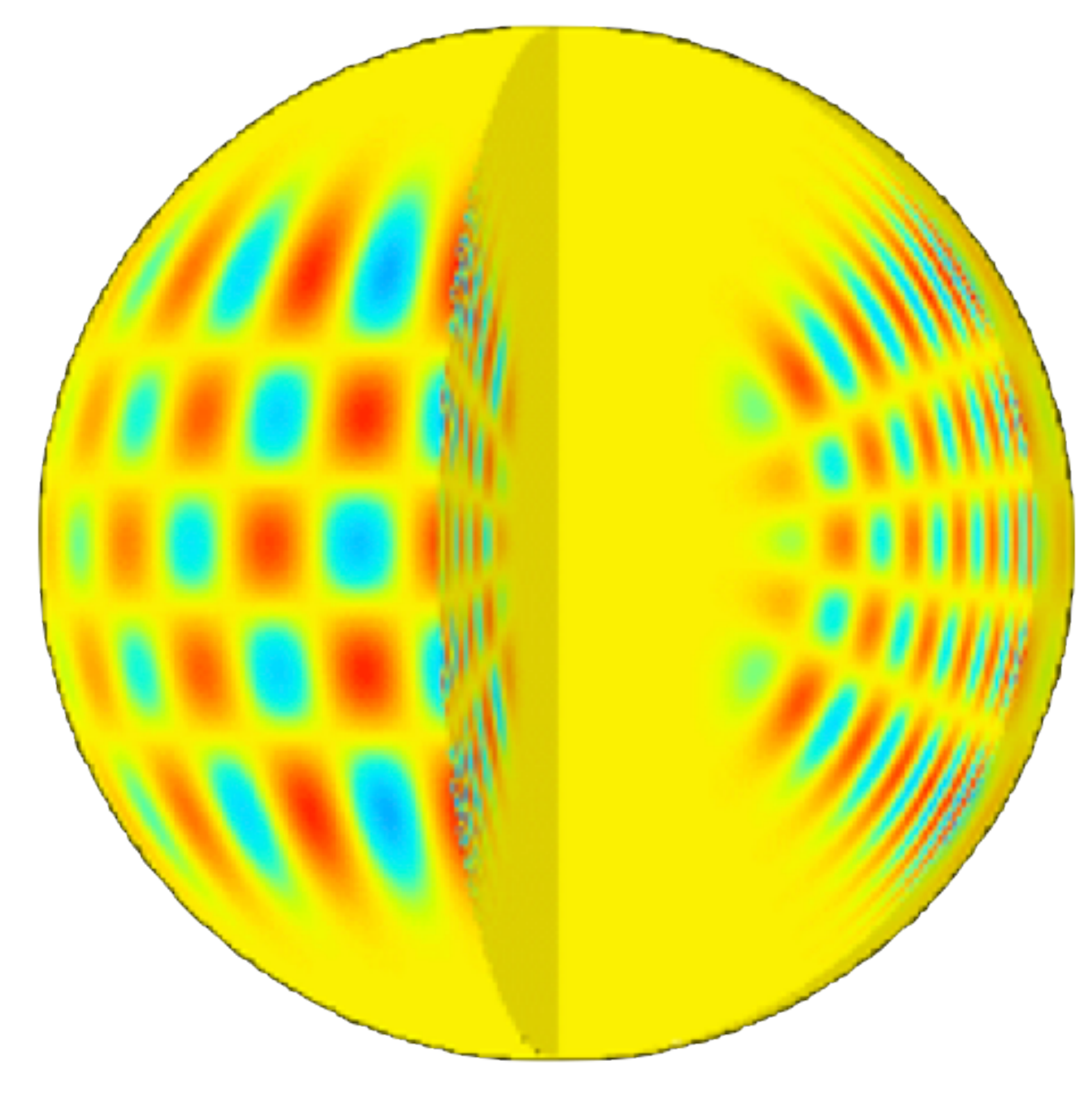}
\end{tabular}
\end{center}
\caption{\label{spher_harm} Left: Example of modes of degree $\ell$=0, 1, 2 and azimuthal order $m$=0, 1, 2. Modes are represented by spherical harmonics. The blue regions are those \modifeng{moving toward} the observer, while the red regions represent those that are moving away; half a period latter this is reversed. Right: Mode $\ell$=20, $m$=16 and $n$=14.
}
\end{figure}

Solar-like oscillations are resonant modes, resonances occurring at specific frequencies. These oscillations are small enough to be considered as linear perturbations around the equilibrium state of the star. Thus, studying the oscillations boils down to an eigenvalue problem; by solving it, we get a discrete set of \textit{eigenfrequencies}, each one being associated with an \textit{eigenmode} describing the distribution of the perturbation inside the star. Stellar oscillation modes are standing waves in a meridional plane and may be propagating in the azimuthal direction, as in a waveguide. Assuming that solar-like stars are purely spherically symmetrical objects, i.e. all of the quantities describing the equilibrium depend on the radius $r$ only, the problem is separable and the horizontal part of the modes are described by spherical harmonics (denoted $Y_\ell^m$); only the radial part depends on the stellar structure. A mode is then fully characterized by three integers:
\begin{itemize}
 \item the radial order, $n$, indicates the number of nodes along the radius. By convention, we denote the p modes by positive numbers and g modes by negative ones (see Sect. \ref{ssec:pgmixeddesc}).
 \item The angular degree, $\ell$, is a non-negative integer denoting the number of nodal lines at the surface of the sphere. Thus, modes with $\ell$=0 are radial modes while those with $\ell \ge 1$ are the non-radial modes. More specifically $\ell$=1 are called dipole modes, those with $\ell$=2 are the quadrupole modes, the $\ell$=3 are the octupole modes, etc.
 \item The azimuthal order, $m$, gives the number of nodal lines passing through the poles. It can take values from $-\ell$ to $+\ell$ including zero. Positive and negative values corresponding to retrograde and prograde waves respectively. When $m=0$, modes are axisymmetric. These modes are usually called zonal modes; modes with $|m|=\ell$ are called sectoral modes.
\end{itemize}
In Fig.~\ref{spher_harm}, a few modes are represented: low-degree modes as well as a higher degree mode.

We usually denote the frequency $\nu_{n,\ell,m}=\omega_{n,\ell,m} / 2 \pi$, expressed in Hz, where $\omega_{n,\ell,m}$ is the angular frequency in $\mathrm{rad\,s^{-1}}$. Due to the spherical symmetry, modes are degenerate relative to $m$, making the frequencies independent on $m$. Thus, we can write $\nu_{n,\ell,m}=\nu_{n,\ell}$. By considering such a symmetry, we implicitly neglect the impact of stellar rotation. We will introduce rotation in Sect.~\ref{ssec:osc_rot}.

\subsection{Acoustic, gravity and mixed modes}
\label{ssec:pgmixeddesc}
\begin{figure}[!htp]
 \begin{tabular}{cc}
	\includegraphics[width=0.5\textwidth]{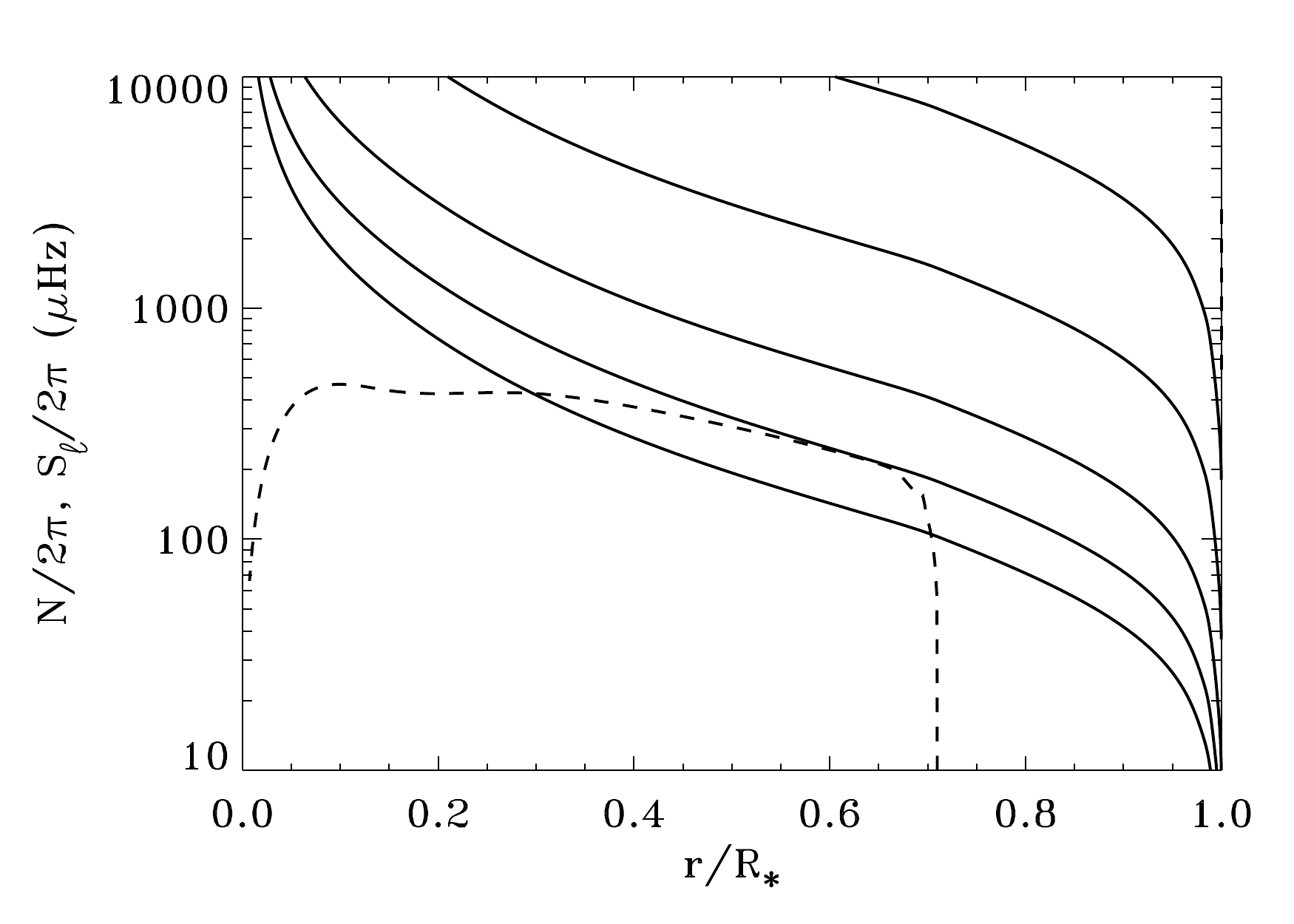} &
	\includegraphics[width=0.5\textwidth]{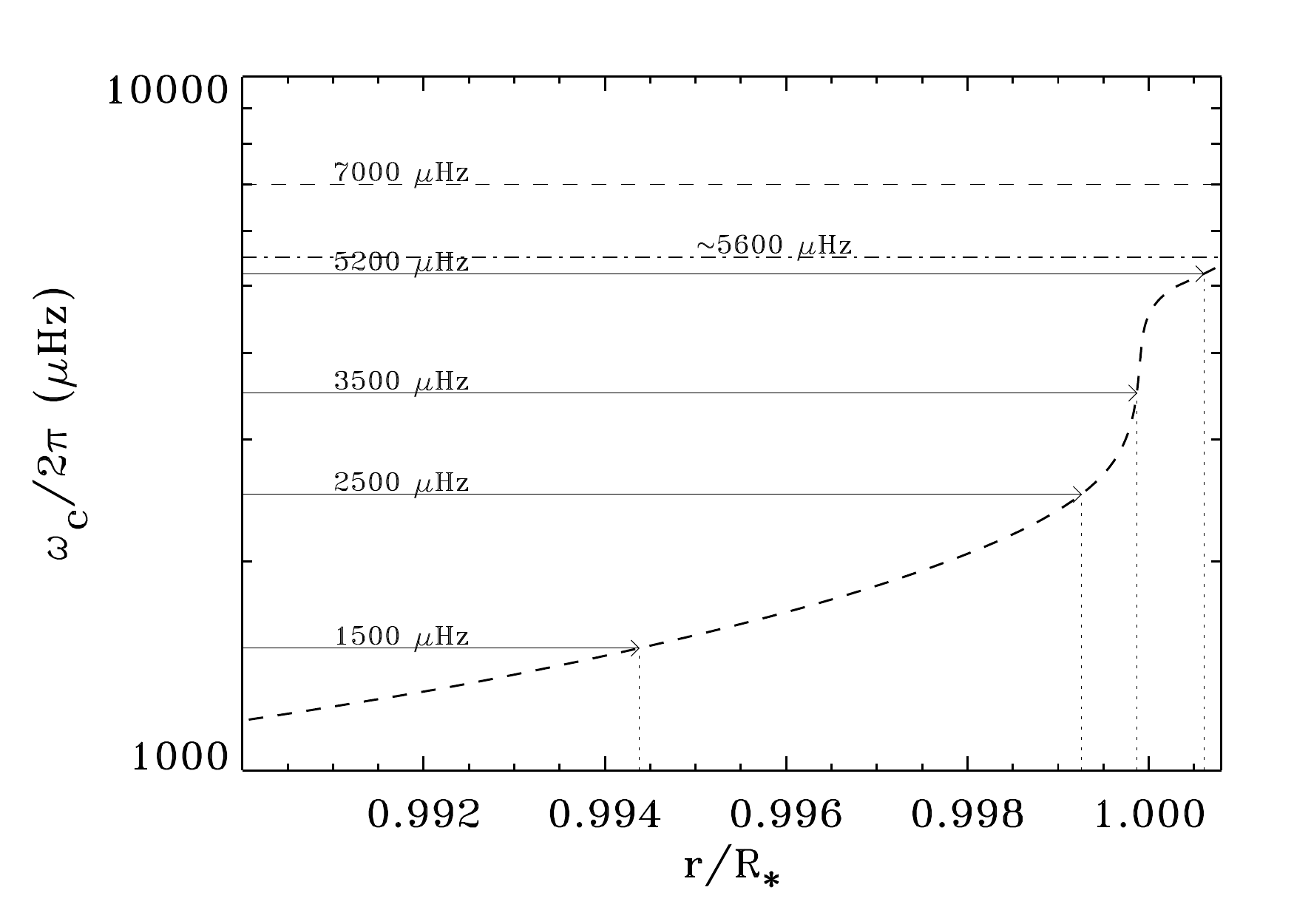}
\end{tabular}
\caption{\emph{(Left)} Brunt--V\"ais\"al\"a frequency (dashed line) and Lamb frequencies for $\ell=1,2,5,20$ and 100 (solid lines) for a typical solar-type star. \emph{(Right)} The cutoff-frequency profile $\omega_c$ \modifeng{near the solar photosphere,} computed within the isothermal approximation, is plotted with a dashed line. The horizontal dashed--dot line depicts the cutoff frequency of modes ($\sim$5600 $\mu$Hz). At higher frequency, waves are not trapped, whereas at lower frequency, waves are trapped below an external turning point (indicated with vertical dotted lines for different frequencies shown with horizontal solid lines). \label{fig:propag}}
\end{figure}

The equations for oscillations in non-rotating spherical stars lead to a collection of 1D eigenvalue problems that can be independently solved for each value of $\ell$. Assuming adiabatic perturbations, these equations are fourth order and may be solved numerically given a radial model of the stellar structure. Various codes have been developed, such as ADIPLS, LOSC, PULSE, POSC, GraCo, NOSC, OSCROX, FILOU, LNAWNR, or GYRE \citep[e.g.][]{2008Ap&SS.316..113C,2008Ap&SS.316..149S,2008Ap&SS.316..107B,2008Ap&SS.316..121M,2008Ap&SS.316..129M,2008Ap&SS.316..135P,2008Ap&SS.316..141R,2008Ap&SS.316..155S,2008Ap&SS.316..163S,2013MNRAS.435.3406T}. However, to understand the nature of the oscillations, \modifeng{we make} some approximations. By assuming that the modes vary much more rapidly than the equilibrium structure, we may crudely reduce the problem to a classical turning-point wave equation
\begin{equation}\label{eq:turning}
 \frac{\textrm{d}^2\xi_r}{\textrm{d}r^2}+K(r)\xi_r = 0 \quad\mbox{with}\quad
 K(r)=\frac{\omega^2}{c^2}\left(\frac{N^2}{\omega^2}-1\right)\left(\frac{S_\ell^2}{\omega^2}-1\right) \; .
\end{equation}
In this equation $\xi_r$ is the amplitude of the radial displacement, $\omega$ the angular frequency of the wave, $c$, $N$ and $S_\ell=\sqrt{\ell(\ell+1)}c/r$ are the sound speed, the Brunt--V\"ais\"al\"a frequency and the Lamb frequency. Within this approximation, waves can only propagate in the stellar interior where $K(r)>0$, that occurs where either (i) $\omega > N$ and $\omega > S_\ell$ or (ii) $\omega < N$ and $\omega < S_\ell$. The first case corresponds to acoustic or pressure (p) modes and the second one to gravity (g) modes. P modes are then high-frequency modes whereas g modes are low-frequency (long-period) modes. In a main-sequence solar-like star these two frequency domains are well separated. Figure \ref{fig:propag} shows the profiles of $N$ and $S_\ell$ in a solar-type star. We deduced from this plot that g modes are confined in the internal radiative region of stars whereas p modes are mainly confined in the outer part of the star.

Modes are also characterized by their inertia ${\cal I}$ defined from the eigenmodes as
\begin{equation}
 {\cal I}_{n,\ell,m} = \int_V \rho |\vec{\xi}_{n,\ell,m}|^2 \textrm{d}V \; ,
\end{equation}
where $\vec{\xi}_{n,\ell,m}$ \modifeng{denotes the total mode displacement normalized by its value at the stellar surface. Thus, more energy is required to excite higher inertia modes}. Since the g modes \modifeng{are largely restricted to the deeper layers} of the stars, their inertia is far larger than that of p modes.

\subsubsection{p modes}
\label{sssec:pmodes}
As indicated by their name, p modes are pressure, or acoustic, waves for which the restoring force arises from the pressure gradient. They are the most important modes in solar-like star seismology since they are by far the most observed ones, with periods of several minutes. Normally stable regarding non-adiabatic processes such as $\kappa$ mechanism, these modes are stochastically excited by the turbulent convective envelope. 
Observed modes correspond to high-order low-degree modes. The asymptotic (in the sense that $n\gg\ell$) theory of p modes has initially been developed by \citet{1967AZh....44..786V} and continued in the 80's by \citet{1980ApJS...43..469T} and \citet{1984ARA&A..22..593D} for example.
Asymptotic theory described also the structure of p-mode spectra. The theory predicts regular patterns: modes are organized in a comb structure as observed. These different regularities are detailed in Section~\ref{sec:freqSep}.

\begin{figure}[!htp]
\begin{center}
\includegraphics[width=0.7\textwidth]{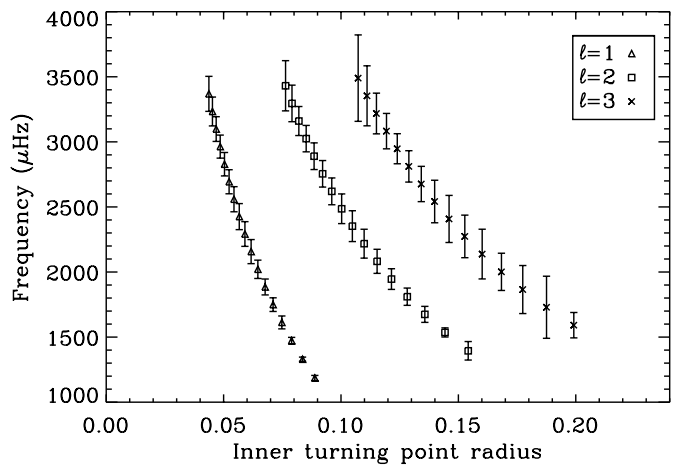}
\end{center}
\caption[Inner turning points of low-degree versus central frequency of low-degree modes observed by GOLF]{\label{Fig_turningPoints} Fractional radii of the inner turning point for low-degree acoustic modes obtained from GOLF \citep{2008SoPh..251..119G}. The error bars of the modes have been multiplied by 5000 to be visible.}
\end{figure}

P modes are confined in the outer region of the stars. Using Equation \ref{eq:turning}, their resonant cavity is limited at the top by the stellar photosphere and at the bottom by an \emph{internal turning point} $r_{\rm{t}}$ defined such as $S_\ell(r_{\rm{t}}) = \omega_{n,\ell}$, i.e.
\begin{equation}\label{eq:intTP}
 r_{\rm{t}} = c(r_{\rm{t}}) L / \omega_{n,\ell} \; ,
\end{equation}
where considering $L=\ell+1/2$ is a better approximation \citep[e.g.][]{1991SvA....35..400V,1994A&A...290..845L}.
As a consequence, we see that for the same radial order, lower-degree modes propagate deeper into the interior of the star, while for the same degree, higher frequency modes, which correspond to higher order modes, also penetrate deeper than lower-order, lower-frequency modes (see Fig.~\ref{Fig_turningPoints}).

The \emph{outer turning point}, close to the photosphere, also depends on the frequency. It corresponds to the radius $r_{\rm{e}}$ where an acoustic wave propagating outwards is reflected due to the brutal drop in density and pressure. A wave propagates as long as $\omega>\omega_{\rm{c}}$ where $\omega_{\rm{c}}$ is the cut-off frequency. We then define $r_{\rm{e}}$ such that $\omega_{n,\ell}=\omega_{\rm{c}}(r_{\rm{e}})$. In an isothermal atmosphere,
\begin{equation}\label{eq:cutoff_freq}
 \omega_{\textrm{c}} = \frac{c_{\textrm{s}}}{2 H} \; ,
\end{equation}
where $H$ is the classical density scale height, that is equal to the pressure scale height $H_{p}=(-\rm{d}\ln p / \rm{d}r)^{-1}$ in an isothermal atmosphere. \citet{1984ARA&A..22..593D} proposed a more detailed expression that is well approximated by this one.  Figure \ref{fig:propag} shows the profile of $\omega_{\rm{c}}$ for the Sun.
In the upper layers of solar-like stars, the profile $\omega_{\rm{c}}(r)$ increases with $r$ and reaches a maximum. This has two consequences: (i) above a cutoff frequency ($\sim 5600\:\mathrm{\mu Hz}$ for the Sun) acoustic waves are no more reflected, then mode trapping is no longer possible. Nevertheless, we may observe a mode-like pattern in the oscillation spectrum above the cutoff frequency \citep[][]{1988ESASP.286..279J,1988ApJ...334..510L,1991ApJ...373..308D}. These so-called pseudo-modes or high interference peaks (HIPS) were first observed in Sun-as-a-star solar observations using GOLF by \cite{1998ApJ...504L..51G} and then confirmed by BiSON \citep{2003ESASP.517..247C}. A few years later, in 2007, there were also found in solar-like stars from ground-based observations \citep{2007MNRAS.381.1001K} and from space using \emph{Kepler} \citep{2015A&A...583A..74J}.
From disk-integrated observations, these HIPS may be explained as an interference of high-frequency waves partially reflected at the \modifeng{unobserved far side} of the observed star \citep{1998ApJ...504L..51G}.
(ii) Higher frequency modes reach higher layers in the atmosphere and this does not depend on $\ell$; as a consequence any perturbation in the outer layers of the star affects similarly low-degree modes with very close frequencies. \modifeng{These layers near the photosphere} are known to be poorly modelled with 1D stellar evolution codes, since they are highly turbulent, non-adiabatic and with a low plasma beta. All of these phenomena generate the so-called near-surface effects visible as a departure between computed and observed frequencies. Such effects have to be taken into account when analysing Solar-type stars. We can also build diagnosis tools that cancel out this effect (see Sect \ref{sec:freqSep}).

\subsubsection{g modes}
At the \modifeng{low-frequency} part of the spectrum of solar-like stars, we find g modes, for which the restoring force is \modifeng{the buoyancy due to density fluctuations and gravity}. These low-frequency modes are confined in the radiative interior of solar-like stars since gravity waves can only propagate in a region where $N^2>0$ (see Eq.~\ref{eq:turning}), i.e. in non-convective zones by definition. They are thus evanescent in solar-like envelopes and reach the surface with \modifeng{very low amplitudes relative to the p modes}. For the Sun, expected periods for g modes are of hours and longer than 35 minutes for the shortest period. The g-mode spectrum also has a regular pattern: g modes of the same degree $\ell$ are evenly spaced in period, the period increasing when the absolute value of the radial order increases. Moreover, \modifeng{radial gravity modes do not exist}.
Up to now, the only claims of g-mode detection in solar-like stars concerns the Sun \citep[e.g.][]{STCGar2004,2007Sci...316.1591G,2017A&A...604A..40F}.

\subsubsection{Mixed p/g modes}
When solar-like stars reach the end of the main sequence, due to the build-up of strong density gradients in the core, the Brunt--V\"ais\"al\"a frequency increases there. As a consequence there exists a frequency range where g modes in the core and p modes in the envelope may coexist. If the evanescent region between the p- and g-mode cavities is small enough, a coupling between them occurs. In such a case we get mixed modes with both p and g characteristics. Their properties were initially discussed theoretically by \citet{1974A&A....36..107S} and the first observations of mixed modes in solar-like stars were reported from ground-based observations of $\eta$ Bootis, by \citet{1995AJ....109.1313K} and confirmed later by \citet{2003AJ....126.1483K}, and \citet{2005A&A...434.1085C}. They were in very good agreement with theoretical predictions by, e.g., \citet{1995ApJ...443L..29C}.
More recently, many observations of mixed modes have been made from ground-based observations but also from space thanks to CoRoT  \citep[e.g.][]{2010A&A...515A..87D} and \emph{Kepler} observations \citep{2010ApJ...713L.169C,2011A&A...534A...6C,2011ApJ...733...95M}. \citet{2011Sci...332..205B} first reported the existence of mixed modes in red-giant stars. Later, \citet{2011Natur.471..608B} and \citet{2011A&A...532A..86M} showed the power of these modes to measure the evolutionary status of red giants, with a clear difference between stars ascending the red-giant branch (RGB) and those in the so-called ``clump''.

Mixed modes are very useful to put constraints on the internal structure and dynamics of stars since they are very sensitive to the core due to their g-mode behaviour, whereas their p-mode properties  make their surface amplitudes high enough to be detected.

\subsection{Effects of rotation}\label{ssec:osc_rot}
In previous sections, we have assumed a perfect spherical symmetry implying that modes are degenerate in $m$.
Everything that breaks this symmetry lifts this degeneracy. Rotation is the most common phenomenon breaking the symmetry. When the rotation rate is slow enough, as it is in most solar-type stars, it may be treated as a perturbation of the non-rotating case. Modes are thus split into $2\ell+1$ $m$-components forming a multiplet. Each component has a frequency $\omega_{n,\ell,m}=\omega_{n,\ell} + \delta\omega_{n,\ell,m}$ where the second term is a small perturbation usually called a rotational splitting (or just a splitting). This quantity will directly depend on the rotation rate of the star. To \modifeng{first order, in the perturbative expansion, the splittings may be expressed as:}
\begin{equation}
 \delta\omega_{n,\ell,m} = m \iint K_{n,\ell,m}(r,\theta) \rm{ \Omega } (r,\theta) \textrm{d} r \textrm{d}\theta \; ,
\end{equation}
where $K_{n,\ell,m}$ is the rotational kernel depending on the eigenmode and $\rm{ \Omega} $ is the rotation profile in the star \citep{1977ApJ...217..151H}. Thus, the kernel indicates how the splitting is sensitive to the rotation inside the star. By assuming the rotation is solid in the mode cavity, this equation is simplified \citep{1951ApJ...114..373L}:
\begin{equation}
 \delta\omega_{n,\ell,m} = m (C_{n,\ell}-1) \rm{ \Omega }  \; ,
\end{equation}
where $C_{n,\ell}$ is named the Ledoux (or Coriolis) coefficient. For p modes $C_{n,\ell}\approx 0$ for high-order modes (for the Sun, with $n\gtrsim 10$ we get $C_{n,\ell} \lesssim 10^{-2}$), whereas for g modes $C_{n,\ell}\approx 1/[\ell(\ell+1)]$.

As a consequence, p modes of solar-like stars are generally modelled as symmetrical multiplets of frequencies
\begin{equation}
\nu_{n,\ell,m}=\nu_{n,\ell}-m\nu_{\rm{s}} \label{eq:split} \; ,
\end{equation}
where $\nu_{\rm{s}}=\rm{ \Omega } /(2\pi)$ is simply called the splitting. If the rotation is nearly solid and observed modes are pure p modes, $\nu_{\rm{s}}$ is the same for all of them. If the rotation changes with the stellar radius, the splittings depend on the cavities probed by the modes and they will be different. The presence of mixed modes in sub-giant and red giant stars gives opportunities to measure their core rotation.

Equation \ref{eq:split} induces symmetrical multiplets. However asymmetrical splittings may be observed if differential rotation in latitude is strong enough \citep[e.g.][]{2004SoPh..220..169G}, or if the star is oblate due to fast rotation, or in mixed modes due to near-degeneracy effects \citep{2017A&A...605A..75D}.

\modif{It is important to note that magnetic fields can also produce asymmetries in the multiplets in both amplitudes and frequencies \citep[e.g.][]{1990MNRAS.242...25G,GooTho1992,1993ASPC...40..563S,2018ApJ...854...74K,2018sf2a.conf..113A}}.

\subsection{Mode visibility}
\begin{figure}[!htp]
\begin{center}
 \includegraphics[width=0.7\textwidth]{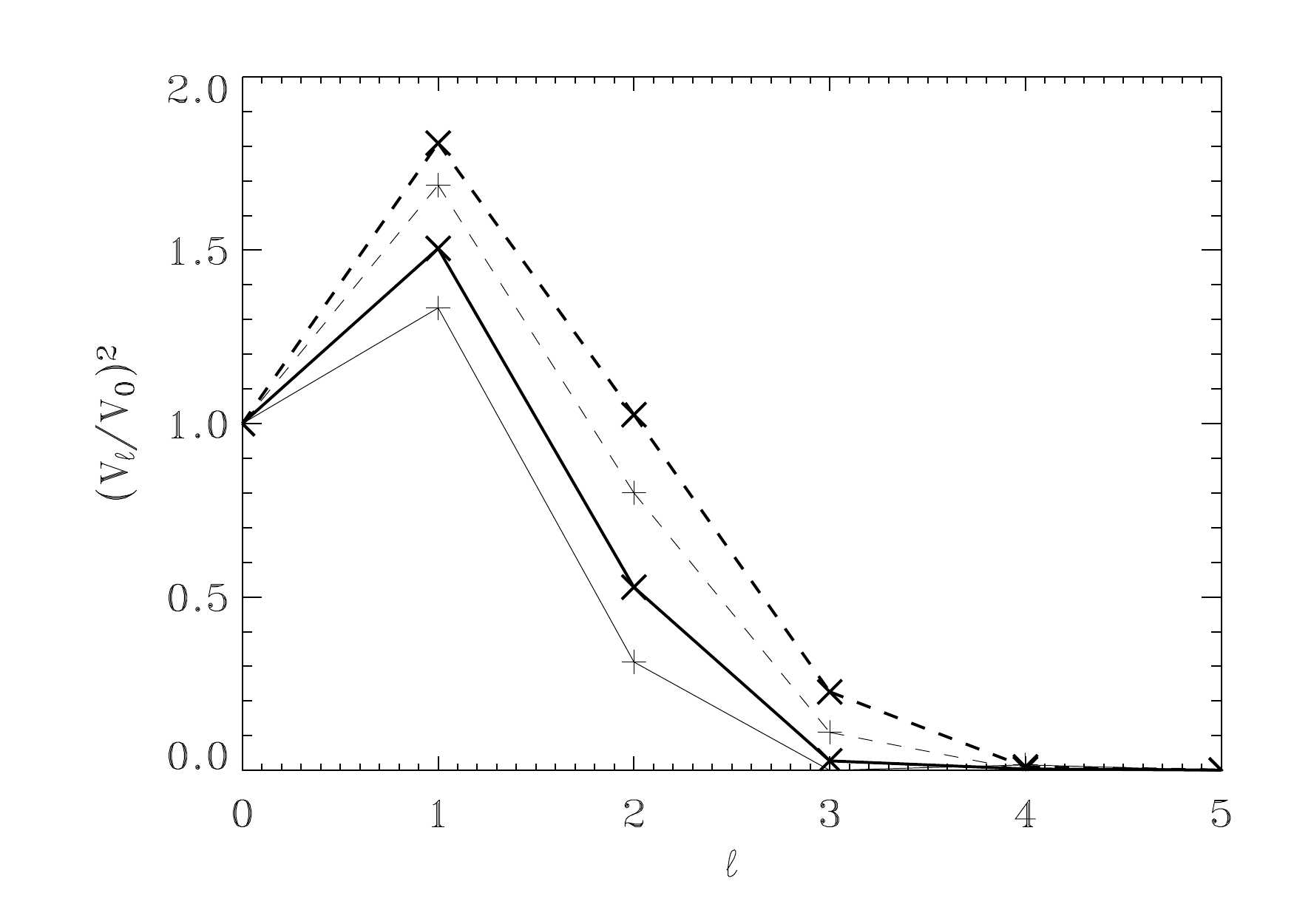}
 \end{center}
 \caption{Mode visibilities $V_\ell$ as a function of the degree $\ell$. Solid lines correspond to intensity measurements ($W(\mu)=L(\mu)$) whereas dashed lines correspond to velocity measurements ($W(\mu)=\mu L(\mu)$). Faint lines are obtained by ignoring the limb darkening ($L(\mu)=1$). Thick lines are obtained within the Eddington approximation ($L(\mu)=0.4+0.6\mu$).\label{fig:vis_theo}}
\end{figure}

We focus in this review on low-degree modes, because they are the only ones that we observe in stars without any spatial resolution due to cancellation effects. Indeed, since all of the information (intensity or velocity) is integrated over the visible stellar disc, the contribution of small-scale modes, i.e. high-degree modes, vanishes. We can only measure high-degree modes in the Sun because \modifeng{its surface is resolved}. If we denote the fluctuation induced by a mode $f_{n,\ell,m}(\theta,\phi)=A Y_\ell^m (\theta, \phi)$, then we may show \citep{1977AcA....27..203D,1993A&A...268..309T,2003ApJ...589.1009G} that the observed amplitude can be expressed as:
\begin{equation}\label{eq:ampl_vis}
 a_{n,\ell,m}= r_{\ell,m}(i) V_\ell  A \; .
\end{equation}
$V_\ell$ is called the mode visibility, $r_{\ell,m}(i)$ is the relative amplitude of mode inside a multiplet and depends only on the inclination angle $i$ between the rotation axis and the line of sight. \modifeng{Mode visibilities may be written as:}
\begin{equation}\label{eq:Vl_exp}
 V_\ell = \sqrt{(2\ell+1)\pi}\int_0^1 P_\ell(\mu) W(\mu) \mu \rm{d}\mu \; ,
\end{equation}
where $P_\ell$ is the $\ell$-th order Legendre polynomial, $W(\mu)$ a weighting function depending only on the distance to the limb $\mu$. It indicates the contribution of each surface element over the stellar disc. For intensity measurements $W(\mu)$ is mainly the limb darkening function $L(\mu)$. Limb darkening depends on atmosphere properties (effective temperature, surface gravity, metallicity,...) and the observed wavelength. For velocity observations, since the motion induced by low-degree p modes are mainly radial at the surface, we may approximate the weighting function as $W(\mu)\approx\mu L(\mu)$. Mode visibilities are plotted in Fig.~\ref{fig:vis_theo} for low degree. We see that they dramatically drop for $\ell>2$. In practice, we observe mainly modes with $\ell=0$, 1 and 2 and some $\ell = 3$. $V_\ell$ mainly depends on the atmospheric properties and the mode physics in the upper layers. We notice that $\ell=3$ modes are only visible thanks to limb-darkening effects. Specific derivations of $V_\ell$ for CoRoT and \textit{Kepler} observations have been carried out by \citet{2009A&A...495..979M} and \citet{2011A&A...531A.124B}.

\begin{figure}[!htp]
 \begin{tabular}{cc}
	\includegraphics[width=0.5\textwidth]{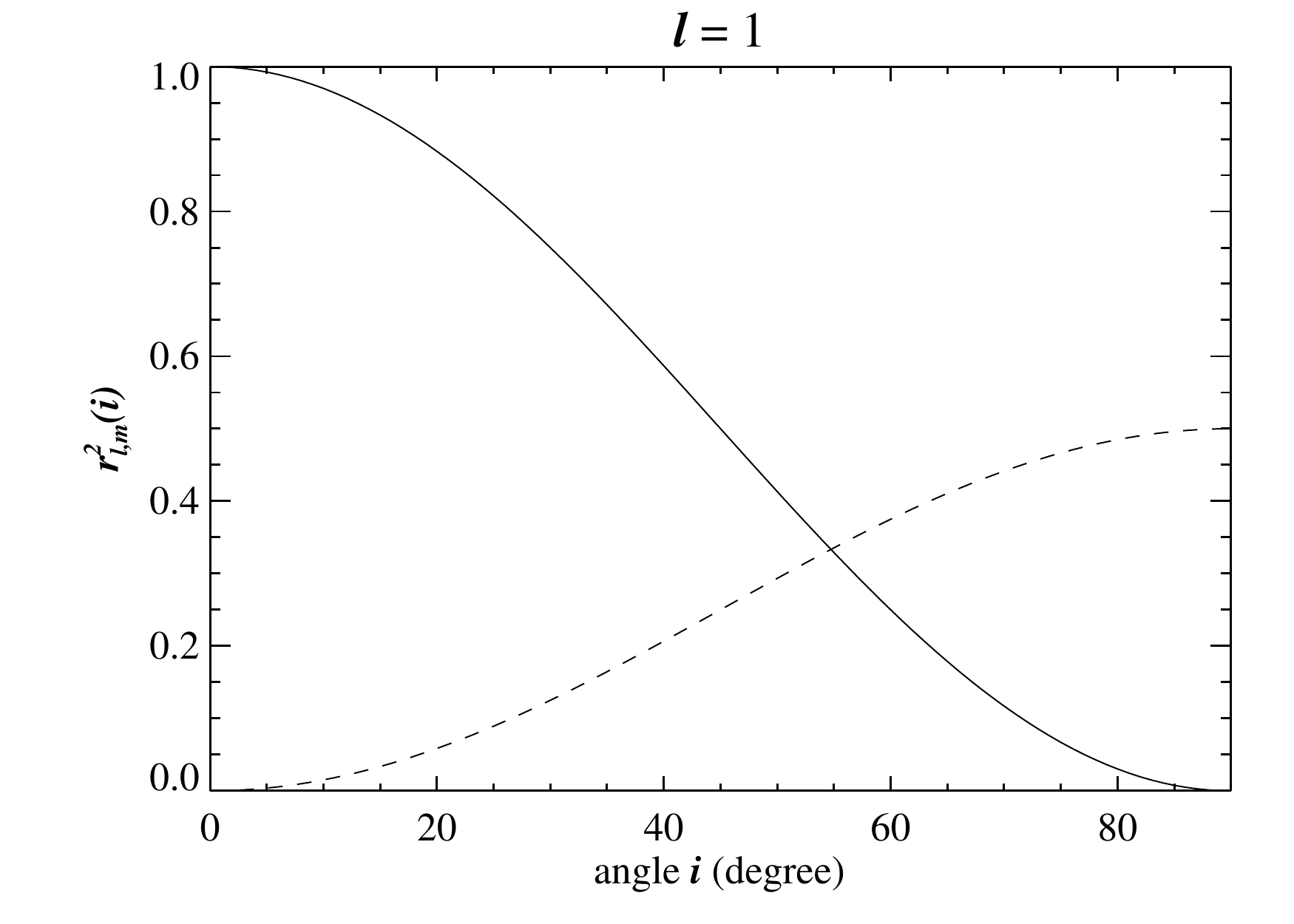} &
	\includegraphics[width=0.5\textwidth]{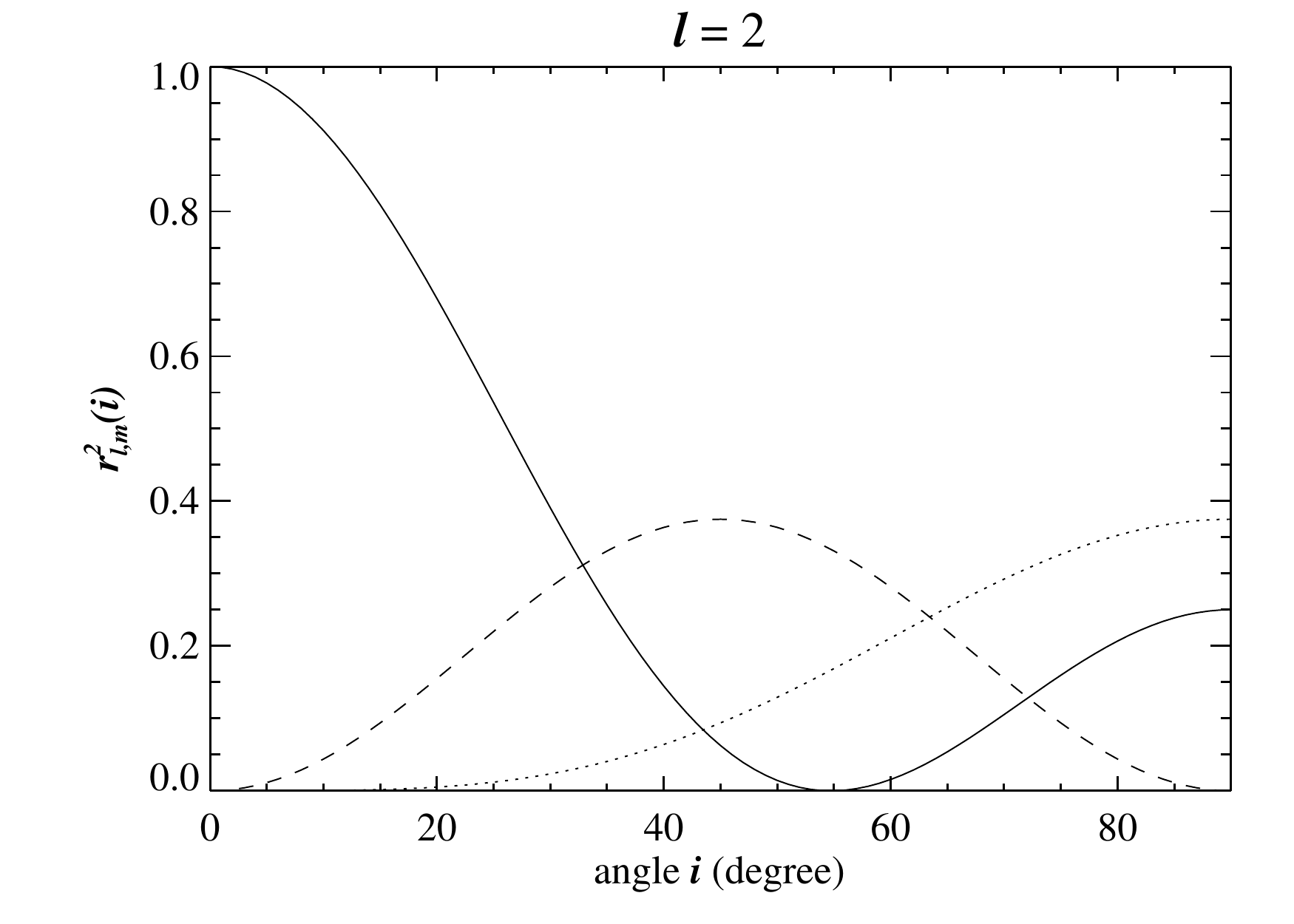}
\end{tabular}

 \caption{\modif{Relative power $r^2_{\ell,m}$ of modes in a multiplet} as a function of the inclination angle $i$ for $\ell=1$ \emph{(left)} and $\ell=2$  \emph{(right)}. Solid, dashed and dotted lines correspond to $|m|=0$, 1 and 2, respectively.\label{fig:ampl_theo}}
\end{figure}

When $V_\ell$ depends on the stellar atmosphere and the instrument, the factor $r_{\ell,m}(i)$ is purely geometric and reads
\begin{equation}\label{eq:alm_exp}
 r^2_{\ell,m}(i) = \frac{(\ell-|m|)!}{(\ell+|m|)!}[P_\ell^{|m|}(\cos i)]^2 \; ,
\end{equation}
where $P_\ell^m$ is the associated Legendre functions. Figure \ref{fig:ampl_theo} shows the variation of these coefficients with $i$ for $\ell=1$ and 2. \modif{The squared factor $r^2_{\ell,m}(i)$ represents the relative \emph{power} of modes in a multiplet.} We notice that $\sum_m r_{\ell,m}^2(i)=1$. For stars observed pole-on ($i=0\degr$) only axisymmetric ($m=0$) modes are visible; it is then not possible to infer any information about rotation. For stars observed equator-on ($i=90\degr$), only components with even $\ell+m$ are visible.

To be able to separate $r_{\ell,m}$ and $V_\ell$ factors in Eq.~\ref{eq:ampl_vis}, $W$ must depend on $\mu$ only. It is, for example, well known that the amplitude ratio in $\ell=2$ and 3 multiplets in solar data observed, for instance, by GOLF does not follow Eq.~\ref{eq:alm_exp} since the spatial response of the instrument does not depend on $\mu$ only. Recent detailed measurements have been presented and discussed in \citet{2011A&A...528A..25S}.

\subsection{Frequency separations}
\label{sec:freqSep}

\begin{figure}[!htp]
\begin{center}
\includegraphics[width=0.9\textwidth]{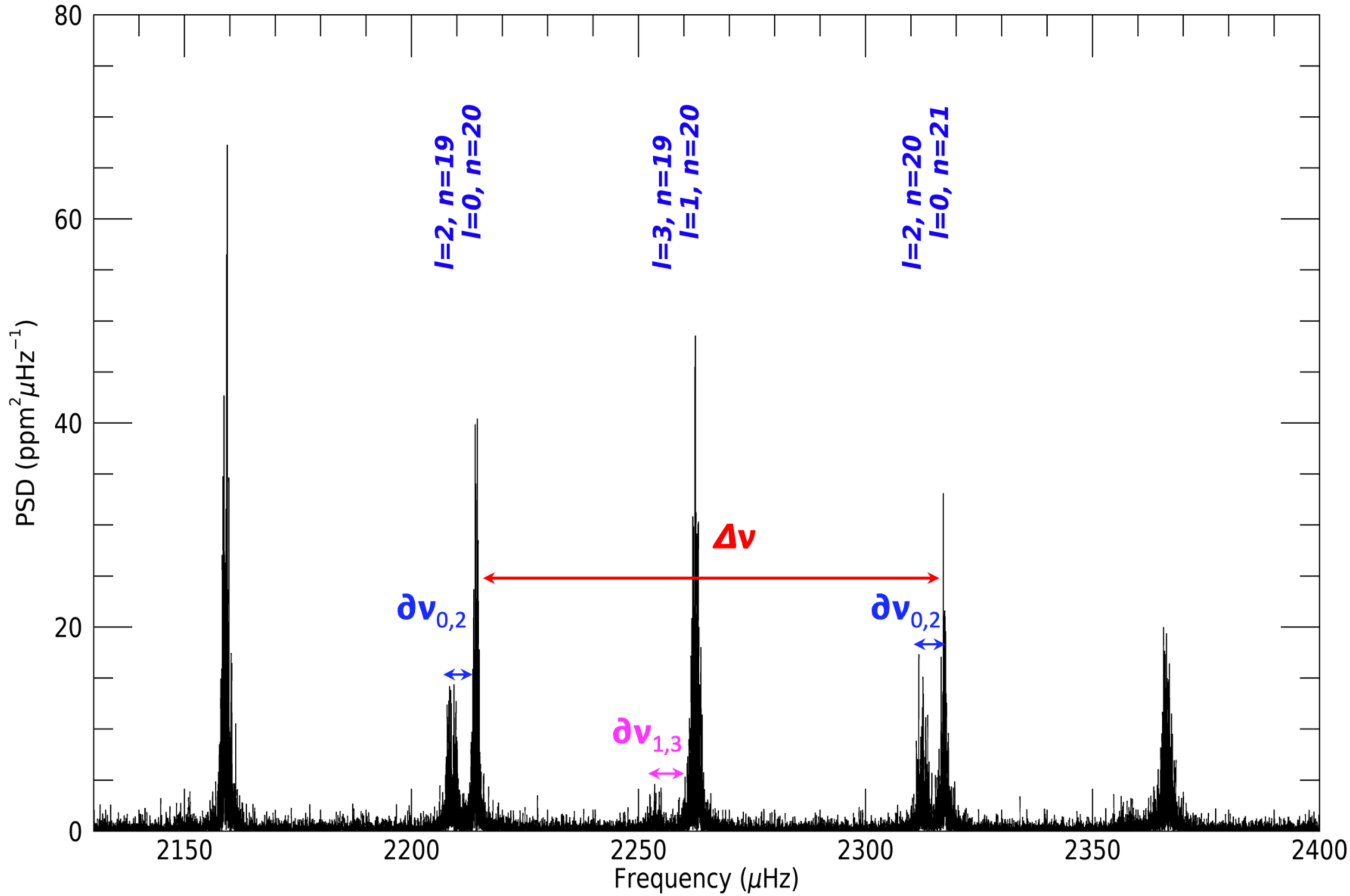}
\end{center}
\caption{\label{Separation} \modif{Region of the power spectrum of 16~Cyg A (shown in Fig.~\ref{fig:Separation} between 2130 and 2400 $\mu$Hz. Some modes $\ell=$0, 1, 2 and 3 are labelled with radial orders between 19 and 21. The horizontal lines ending with arrows indicate the large spacing, $\Delta\nu$, and the small spacings, $\delta_{0,2}$ and $\delta_{1,3}$.}
}
\end{figure}

The regular structure of p-mode spectra predicted by asymptotic theory is well observed  as we can see in Fig.~\ref{Separation} for the solar case. A sequence of modes of degrees 2, 0 and 3, 1 is repeated. We then may define two important seismic variables: the large- and the small-frequency separations or simply large and small separations.

The large separation of low-degree p modes is given by (see also Fig.~\ref{Separation}):
\begin{equation}
\Delta \nu_{\ell} (n)= \nu_{n,\ell}-\nu_{n-1, \ell} \; .
\end{equation}
This regularity is well captured by their first-order asymptotic expansion \citep{1980ApJS...43..469T}:
\begin{equation}
 \nu_{n,\ell} \approx \Delta\nu \left(n+\frac{\ell}{2}+\frac{1}{4}+\epsilon \right) \label{eq:nup_as}\; ,
\end{equation}
where the asymptotic large spacing $\Delta\nu$ depends inversely on the sound-travel time between the centre and the surface of the star, which is also called acoustic radius:
\begin{equation}
\Delta \nu=\left[2 \int_0^R \frac{{\rm{d}}r}{c}\right]^{-1}, \label{eq:expr_dnu}
\end{equation}
where $R$ is the stellar radius, and $c$ is the sound speed.

The small separation of low-degree p modes is given by (see also Fig.~\ref{Separation}):
\begin{equation}
\delta \nu_{\ell, \ell+2} (n)= \nu_{n,\ell}-\nu_{n-1, \ell+2} \; .
\end{equation}
The small separation is the difference of two modes with nearly identical eigenfunctions at the surface (e.g. those with almost the same frequency, and thus similar outer turning points) and being only different in the deeper layers, with different inner turning points.
Using second-order asymptotic theory \citep{1990ApJ...358..313T,1991SvA....35..400V} it can be shown that:
\begin{equation}
\delta \nu_{\ell, \ell+2} (n) \simeq -(4\ell + 6) \frac{\Delta \nu_\ell (n)}{4 \pi^2 \nu_{n,\ell}}\int_0^R \frac{{\rm{d}}c}{{\rm{d}}r} \frac{{\rm{d}}r}{r} \;\;\;.
\end{equation}
This asymptotic expression shows that the small separation is dominated by the sound-speed gradient near the core (the integral is weighted by $1/r$) and, therefore, it is sensitive to the chemical composition in the central regions.
In solar-like stars, generally only $\delta \nu_{0,2}$ is directly observable.
As the frequencies of both modes are very close, they have similar near-surface effects. Hence, the small separation is less affected by such effects than the large separation. However, some residuals can still remain. Therefore, it has been demonstrated that the ratio of the small separation to the large separation, defined as $r_{0,2}  \equiv r_{0,2} (n) = \delta \nu_{0,2} (n) / \Delta \nu_\ell (n)$, can exclude such near-surface effects to a great extent \citep[for more details see][]{2003A&A...411..215R}.

By extension, we also define small separations between radial and dipole modes, $\delta_{0,1}$ and  $\delta_{1,0}$, as the amount by which the modes $\ell$=1 (resp. $\ell=0$) are offset from the midpoint of the modes $\ell$=0 (resp. $\ell=1$):
\begin{equation}
\delta \nu_{0,1} (n)=\frac{1}{2} ( \nu_{n,0}+\nu_{n+1,0})-\nu_{n,1} \;\;\; ,
\end{equation}
\begin{equation}
\delta \nu_{1,0} (n)=\frac{1}{2} ( \nu_{n-1,1}+\nu_{n,1})-\nu_{n,0} \;\;\; .
\end{equation}
These quantities are also very sensitive to stellar cores and may be used, for example, to probe for the presence of a convective core (see Sect.~\ref{ssec:intern_struct}).

\section{Spectral analysis}

In this section, we introduce various practical tools that are used to analyze seismic observations of solar-like stars. We do not deal with the computation of a power spectrum density from a velocity curve or a light curve.

\subsection{The \'echelle diagram}
\begin{figure}[!htp]
\begin{center}
\includegraphics[width=0.85\textwidth]{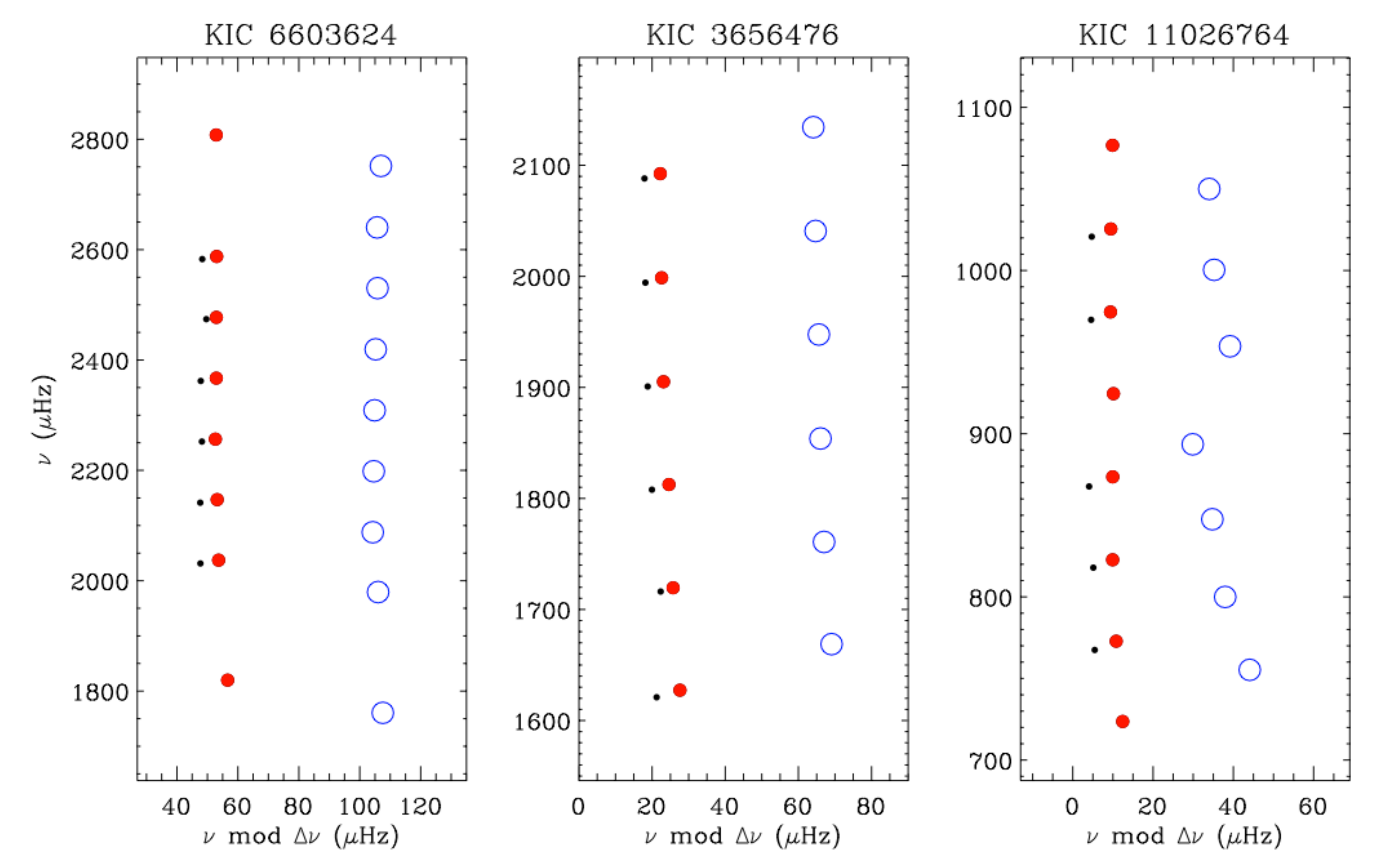}
\end{center}
\caption[\'Echelle diagram of 3 solar-like stars measured by {\it Kepler}]{\label{ED169392} \'Echelle diagrams of 3 solar-like stars observed by {\it Kepler}, showing the $\ell$ = 0 (filled red symbols), $\ell$ = 1 (open blue symbols), and $\ell$ = 2 (small black symbols) ridges. Extracted from \citet{2010ApJ...713L.169C}.
}
\end{figure}

A common way to represent the oscillation spectrum is the \'echelle diagram. It consists of plotting the mode frequencies as a function of the frequencies modulo the large frequency separation. Similarly, an \'echelle diagram may be built from an observed spectrum by cutting the spectrum into segments of multiples of the large-frequency spacing, stacking them one on top of the next, and making then a 2D map. Doing so, modes with the same degree $\ell$ are aligned along almost vertical ridges. It was first used in helioseismology by \citet{1980Natur.288..541G} to identify the modes in the solar oscillation spectrum observed from the South Pole. It is now commonly used in asteroseismology to correctly identify the degree and the order of the modes. Any departure from the first-order asymptotic relation will produce curvature and/or wiggles in the vertical ridges. For example, the presence of mixed modes in evolved solar-like stars is clearly visible and easy to identify: some dipole modes are displaced, or bumped, from their original position due to the presence of a mixed mode in the vicinity. Figure~\ref{ED169392} shows the \'echelle diagram of three solar-like stars observed by {\it Kepler}. The ridges corresponding to the even ($\ell=0$ and 2) modes are clearly shown on the left-hand side of the diagrams, while the ridge of the $\ell$=1 is visible onto the right. ($\ell$=3 mode amplitudes are too small to be unambiguously observed). From left to right, stars are increasingly evolved. Indeed the \'echelle digram of KIC~11026764 (right-hand panel) shows a bumped $\ell$=1 mode (a mode that has been displaced from its original position by the presence of a mixed mode or by the mode immediately below) at $\sim$ 900 $\mu$Hz. This is a clear signature of a more evolved star, probably a sub-giant star.

\subsection{Modelled spectrum}
The observed power spectrum density is the superimposition of several components (see Fig. \ref{fig:PSD}) composing a background on top of which the oscillation modes are visible.
We denote $B(\nu)$ the background and $P(\nu)$ the p-mode spectrum. The full model is thus
\begin{equation}
 S(\nu)=B(\nu)+P(\nu).
\end{equation}
This section provides models for $B$ and $P$ that are commonly used in the literature.

\subsubsection{Background model}
\label{sssec:bgmodel}
At high frequency, the spectrum is dominated by the photon noise $W$; this is a white noise, independent of the frequency. At low frequency, the spectrum is dominated by long-term trends, including both instrumental (drifts, thermal variations, etc.) and stellar effects (especially activity and rotation). Between these two extreme ranges, the background originates from the surface convection, which is dominated by its granular scales. Low-frequency trends and convective components may be described by empirical laws -- initially proposed by  \citep{harvey85} -- corresponding to exponentially decaying temporal variations. The background is then modelled as
\begin{equation}
 B(\nu) = \sum_i H_i(\nu) + W \; ,
\end{equation}
where the different components are
\begin{equation}
 H_i(\nu) = \frac{\zeta_i\sigma_i^2\tau_i}{1+(2\pi\nu\tau_i)^{\alpha_i}} \; .
\end{equation}

Here, $\sigma_i$ is the rms amplitude of the component, $\tau_i$ its characteristic time scale, $\zeta_i=2\alpha_i \sin (\pi/\alpha_i)$ is a normalisation constant. The exponent $\alpha_i$ measures the amount of memory in the physical process responsible for the component. A larger exponent means less memory in the process. Exponential decay gives $\alpha=2$. In this case, $\zeta_i=4$. Sometimes, following \citet{2011ApJ...741..119M}, authors define a new timescale, $\tau_{\mathrm{eff}}$, as the e-folding time of the autocorrelation function of the signal; for $\alpha=2$ both timescales are identical ($\tau=\tau_\mathrm{eff}$).
In practical cases, one granulation component may be enough to model the convective contribution, but a second one, possibly due to faculae, is sometimes required \citep{2013ApJ...767...34K}.

This model $B(\nu)$ is the background limit spectrum that would be obtained after an infinite observing time, which would average all statistical fluctuations. The observed background is then this limiting spectrum multiplied by a random noise following a two-degree-of-freedom (2-dof) $\chi^2$ statistic. \modif{Random processes in time series tend to produce normal (Gaussian) noises in the Fourier domain, both for the real and imaginary parts of the Fourier transform, due to the  central limit theorem. Thus, a power spectrum being the sum of squared real and imaginary parts follows a 2-dof $\chi^2$ statistic. It is  true for the background but also for stochastically excited modes (see next section). This statistical distribution has been well verified in observations.}

\subsubsection{Mode model}
The last components of the spectrum are the modes themselves. As previously mentioned, solar-like oscillations are not unstable modes but are excited stochastically and damped by turbulence in the outer layers of the convection zone \citep[e.g.][]{1977ApJ...212..243G,1988ApJ...326..462G}.
Following these authors, each mode can be simply modelled as a randomly excited and damped harmonic oscillator following the equation
 \begin{equation}
 \frac{{\rm{d}}^2\xi}{{\rm{d}}t^2}+2\eta \frac{{\rm{d}}\xi}{{\rm{d}}t}+\omega_o^2\xi=f(t) \; ,
 \end{equation}
where $\xi(t)$ is the displacement of the oscillator, $\eta$ its damping rate, $\omega_o=2\pi\nu_o$ the frequency of the undamped oscillator and $f(t)$ the random forcing function. Assuming that the mean value of the square of the Fourier transform of $f(t)$, $\langle|F(\nu)|^2\rangle$, is a slowly-varying function of $\nu$, and that $\eta \ll \nu_{o}$, then the power spectrum density of a mode is modelled as a Lorentzian profile multiplied by a stochastic noise following a 2-dof $\chi^2$ statistic. More specifically, the limit Lorentzian profile reads
\begin{equation}
 L(\nu; \nu_o, \Gamma, H) = \frac {H}{1+\left(\frac{2(\nu-\nu_o)}{\Gamma}\right)^2} \; ,
\end{equation}
where $H$ is the mode height and $\Gamma$ is the width at half-height. They are expressed as 
 \begin{equation}
 H= \frac{\langle |F(\nu)|^2 \rangle}{16\pi\eta^2\nu{_o}^2},\label{eq:modeH} 
 \end{equation}
and
 \begin{equation}
 \Gamma=\frac{\eta}{\pi} \; .
 \end{equation}
 
 For a single mode, the integrated power, $P$, is given by:
  \begin{equation}
 P = \frac{\pi}{2} H\Gamma \; ,
 \end{equation}
which corresponds to the mean square of the mode amplitude \citep[see also the discussion in][]{2003ApJ...595..446J}.
 
The total energy, $E$, is taken to be the sum of both the kinetic and the potential energy:
 \begin{equation}
 E ={\cal I} P \; ,
 \end{equation}
where ${\cal I}$ is the corresponding mode inertia defined in Sect.~\ref{ssec:pgmixeddesc}. The rate at which energy is dissipated in the modes \citep[e.g.][]{1999A&A...351..582H} can be derived by using the harmonic damped oscillator analogy \citep[e.g.][]{2000MNRAS.313...32C}:
 \begin{equation}
 \frac{{\rm{d}}E}{{\rm{d}}t}=\dot{E}=2\pi H \Gamma^2 \; .
 \end{equation}

From the study of these equations, the linewidth provides a direct measurement of the damping rate. The mode power (or mode energy) provides a measure of the balance between the excitation and the damping of the modes. Finally, the energy supply rate provides information about the excitation or the forcing of the oscillator.

Such a Lorentzian model is sufficient to reproduce the observed modes, even if asymmetries were reported for the Sun since \citet{1993ApJ...410..829D}. These asymmetries are small enough for low-degree modes -- typically of the order of a few percent \citep[e.g.][]{1998ApJ...506L.147T,1999MNRAS.308..424C,ThiBou2000} -- to consider such Lorentzian description as sufficiently accurate for analysing observations shorter than a year. For longer time series, including asymmetries may be relevant \citep{2018ApJ...857..119B}.

This Lorentzian model is also valid as long as the modes are resolved, i.e. the spectra resolution is finer than the mode width. It means that the observing time is longer than the mode lifetime. This point is verified for observed p modes in solar-like stars, but it may be invalidated at very low frequency.

The oscillation spectrum is  the sum of a full sequence of modes:
\begin{equation}
 P(\nu) = \sum_{n,\ell} M_\ell (\nu; H_{n,\ell},\, \Gamma_{n,\ell}, \nu_{n,\ell},\, \nu_{\rm{s}}, i) \; ,
\end{equation}
where $M_\ell$ is the profile of a multiplet
\begin{equation}
 M_\ell(\nu; H_{n,\ell},\, \Gamma_{n,\ell}, \nu_{n,\ell},\, \nu_{\rm{s}}, i) = \sum_{\ell=-m}^{m} a_{\ell,m}(i) L(\nu; \nu_{n,\ell}-m\nu_{\rm{s}},\, H_{n,\ell},\, \Gamma_{n,\ell}) \; ,
\end{equation}
where $i$ is the inclination angle and $\nu_{\rm{s}}$ the rotation splitting. The coefficients $a_{\ell,m}$ are given by Eq.~\ref{eq:alm_exp} (see. Sect.~\ref{sec:theory}). To obtain this expression, we assume that the intrinsic mode heights and widths inside a multiplet are the same. Since the components have similar frequencies ($\nu_{\rm{s}} \ll \nu_{\rm{o}}$), the energy injected by stochastic excitation is the same (see Eq.~\ref{eq:modeH}), and we also assume that damping processes only depend smoothly on frequency.

\subsection{Maximum Likelihood Estimators}

\subsubsection{Fitting spectra}
This model $S(\nu; \vec{p})$ described in the previous section is parametrized and we aim in determining the most likely values for these parameters $\vec p$ (e.g. mode frequencies, widths, heights, splittings...), given an observed spectrum. The observed spectrum $\vec Y=\{Y_{i}\}$ is sampled at $n$ frequencies $\nu_i$ and we assume that the realization noise for each point is independent. This last point is absolutely verified when the spectrum is estimated through a Fast Fourier Transform of an evenly spaced time series without gaps. Since the observed power density is distributed around the limit spectrum with 2-dof $\chi^2$ statistics \citep[e.g.][]{1984PhDT........10W,AppGiz1998}, the likelihood is
 \begin{equation}
 {\cal L}(\vec Y;{\vec p}) = \prod_{i=1}^n \frac{1}{S(\nu_{i},\vec{p})}
 \exp\left[-\frac{Y_i}{S(\nu_{i},\vec{p})}\right]. 
 \label{dpf} 
 \end{equation}
Given a model $S$, one seeks to find, for a spectrum $\vec Y$ observed over a chosen frequency interval, the parameters $\vec{p}$ that
maximize ${\cal L}(Y;{\vec p})$. The way to choose the frequency interval to consider and the details of the models depend on the fitting strategy and will be discussed in \ref{ssec:fitstrat}.
Maximum likelihood estimators (MLE) are frequently used to analyse seismic data of solar-like stars. This is an inheritance of helioseismology where this approach was intensively used \citep{AppGiz1998}.
\modif{In practice, the negative logarithm of the likelihood function $-\ln \cal L$ is minimized with a standard algorithm such as a modified Newton's method.}

\subsubsection{Errors and correlations}
To estimate the covariance matrix of parameters $\vec{p}$, a typical method is to approximate it by the inverse of the Hessian matrix. The uncertainties on fitted parameters are therefore taken as the square roots of the diagonal elements of the inverted matrix. These estimates are based on the Kramer--Rao theorem. By using them, we have to keep in mind a few crucial points: (i) these error estimates are only lower limits of the statistical errors and (ii) it is only asymptotically valid: the statistical distribution of parameters are assumed to be normal, that is not necessarily the case \citep[e.g.][]{2010AN....331..933B}. Thus, the variables are to be carefully chosen, for example $h=\ln H$ and $\gamma=\ln \Gamma$ are more suitable variables than $H$ and $\Gamma$ to estimate the errors through the Hessian \citep[see discussion in][]{1994A&A...289..649T}. The prefered way to estimate the errors remains in performing Monte Carlo simulations.

Nevertheless, the Kramer--Rao theorem may be used to understand the evolution of errors with various parameters. For example, following  \citet{1990ASSL..159..253D}, \citet{1992ApJ...387..712L} showed that the lower limit for the error of the frequency of a radial mode is
\begin{equation}
 \sigma_\nu = \sqrt{f(\beta)\frac{\Gamma}{4\pi T}}\label{eq:err_lib} \; ,
\end{equation}
where $T$ is the observation time, $\beta=B/H$ is the local background to mode height ratio, and $f(\beta)=(1+\beta)^{1/2}[(1+\beta)^{1/2}+\beta^{1/2}]^3$. More general expressions have been proposed for multiplets in the solar case ($i\approx 90\degr$) by \citet{1994A&A...289..649T} and for any inclination angle by \citet{2008A&A...486..867B}.
Such relations show that for resolved modes, the precision of frequencies varies with $T^{-1/2}$: to reduce the error by a factor of 2, we need observations 4 time longer.

\begin{figure}[!htp]
\begin{center}
\includegraphics[width=0.8\textwidth]{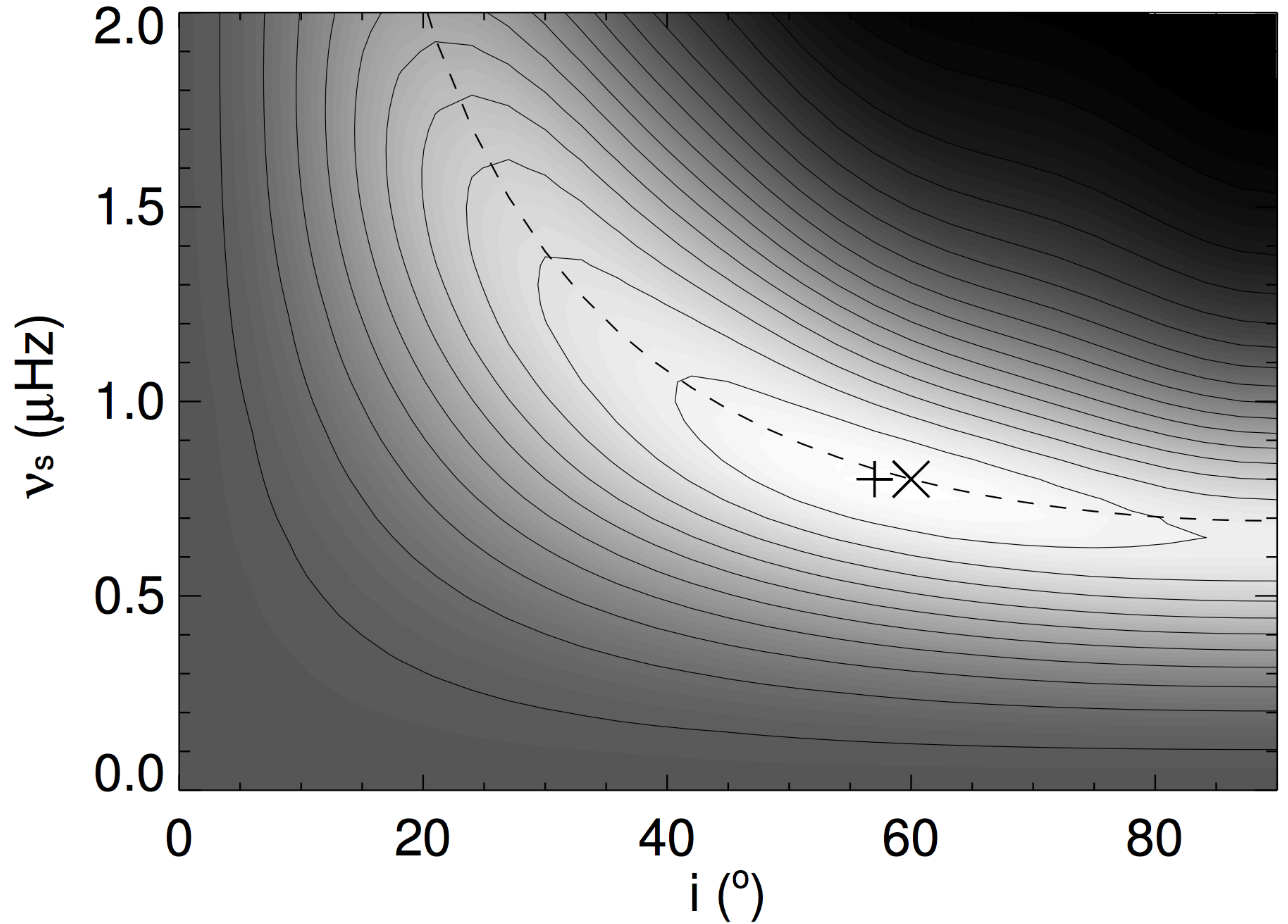}
\end{center}
\caption{\label{fig:banane} Likelihood for a simulated spectrum in the  
$(i,\nu_{\rm{s}})$ plane of the parameter space. All of the other parameters are 
fixed to their simulated value.
The white area corresponds to the highest likelihoods and the black 
to the lowest. The $\times$ is the simulated value $(i_0,\nu_{s,0})$ 
and the $+$ is the maximum of the likelihood. The dashed line
follows a constant projected splitting $\nu_s \sin i$. Extracted from \citet{2006MNRAS.369.1281B}.
}
\end{figure}

It is also important to notice that some parameter estimates are strongly correlated. It is specially the case between the mode height $H$ and width $\Gamma$, which are strongly anticorrelated, making the quantity $H\Gamma$, hence the mode energy $P$, better determined than $H$ and $\Gamma$ individually \citep[e.g.][]{1994A&A...289..649T}. Similar correlations exist between the inclination angle $i$ and the splitting $\nu_{\rm{s}}$ when $\nu_{\rm{s}}$ is not significantly larger than the mode width. The correlation is well visible in a likelihood mapped in the $(\nu_{\rm{s}},i)$ plane. The likelihood maximum show a banana shape structure (Fig.~\ref{fig:banane}). In this case, the projected splitting $\nu_{\rm{s}}\sin i$ is better determined than the splitting itself, even when the inclination is poorly measured \citep[see][]{2006MNRAS.369.1281B,2008A&A...486..867B}.

\subsection{Bayesian methods}

\subsubsection{Bayesian inference}
During the last decade, ``Frequentist'' approaches based on MLE have been completed with Bayesian approaches to analyse oscillation spectra. First introduced for the seismology of solar-like stars to correctly interpret the first CoRoT data \citep{2009A&A...506...15B,2009A&A...506....7G,2009A&A...506.1043G}, it has been extensively developed and used ever since \citep[e.g.][]{2011A&A...527A..56H,2013MNRAS.435..242G,2013MNRAS.430.2313C,2016MNRAS.456.2183D,2017ApJ...835..172L}.
The likelihood is nothing but the probability to observe the data set $\vec Y$ assuming a parameter set $\vec p$, under given a priori information $I$ (including for example a spectral model $S$), we denote it $p(\vec Y|\vec p,I)(\equiv {\cal L}(\vec Y;{\vec p}))$. We maximize this probability with MLE. However, we would prefer to assign a probability to a parameter set $\vec p$ for the observation $\vec Y$, i.e. $p(\vec p|\vec Y,I)$. The is called the posterior probability. It is linked to the likelihood through Bayes' theorem:
\begin{equation}
 p(\vec p|\vec Y,I)=\frac{p(\vec p|I)p(\vec Y|\vec p,I)}{p(\vec Y|I)},
\end{equation}
$p(\vec p|I)$ is the prior probability, i.e. the probability of getting these parameters before looking in the data; it includes our current knowledge (physical properties, information coming from other data or other parts of the spectrum) and current ignorance. The probability $p(\vec Y|I)$ is called the global likelihood:
\begin{equation}
p(\vec Y|I)=\int p(\vec p|I)p(\vec Y|\vec p,I) \,{\rm{d}}\vec p.
\end{equation}
Using a marginalization procedure, we can derived the marginal posterior probability distribution for a subset of parameters of interest $\vec p_{\rm{I}}$ by integrating out the remaining parameters $\vec p_{\rm{N}}$ ($\vec p=\{\vec p_{\rm{I}},\vec p_{\rm{N}}\}$), called nuisance parameters:
\begin{equation}
p(\vec p_{\rm{I}}|\vec Y,I)=\int p(\vec p|\vec Y,I) \,{\rm{d}}\vec p_{\rm{N}}.
\end{equation}
Using Bayesian methods, we then obtain the statistics for all of the parameters of our model: not only an estimated mean and variance, but the full distribution. For spectra with low signal to noise ratio, \cite{2009A&A...506....7G} showed that MLE may be biased and Bayesian approaches are more robust.

\subsubsection{Priors: knowledge and ignorance}
Bayesian methods sample the function $p(\vec p|\vec Y,I)$ to give a global picture of the problem. Moreover, Bayesian approaches allow us to include relevant priors in fitting procedures. Even more, priors are needed: posterior probabilities only make sense when prior probabilities are set. Sometimes, when our knowledge is limited we look for priors that take that ignorance into account. As an example, let us consider the prior for the inclination angle $i$. We may certainly consider our prior independent of the other stellar quantities. When we do not have any complementary observations (or we do not want to use them), we would naturally assumed that all orientations are evenly probable. This does not mean that prior probability is uniform between $0\degr$ and $90\degr$. If we assume isotropy, the prior probability distribution for $i$ is $p(i)\,{\rm{d}}i = \sin i\,{\rm{d}}i$. A uniform distribution for $i$ would favour a rotation axis oriented toward us. Ignorance priors for frequency (or splitting) are uniform probability distributions whereas it is uniform in logarithm for heights and widths \citep[see, e.g.,][ for more detailed discussions]{2009A&A...506...15B,2011A&A...527A..56H}.

\subsubsection{Markov Chain Monte Carlo}
The most common method to sample the posterior probability is to use a Markov Chain Monte Carlo (MCMC) approach. Even if they are rather slow, their implementation is easy and flexible. The Metropolis--Hastings algorithm \citep{1953JChPh..21.1087M,Hast70} is the most commonly used method. Its application to solar-like star oscillation spectra are detailed in \citet{2009A&A...506...15B} and \citet{2011A&A...527A..56H}. The distribution that we want to sample, called the target distribution, is $p(\vec p|\vec Y,I)$. To do so, we construct a pseudo-random walk in the parameter space such that the density of drawn points in a given region of the parameter space is proportional to $p(\vec p|\vec Y,I)$. A Markov Chain achieves such a random walk. The chain is built such that a new point $\vec p_{t+1}$ is added to the chain depending on the previous point $\vec p_{t}$ according to a time-independent transition kernel $p(\vec p_{t+1}|\vec p_{t})$. In the Metropolis--Hastings algorithm, a new point $\vec p_\mathrm{new}$ is randomly drawn from a proposed probability distribution centred on $\vec p_{t}$, $p_p(\vec p_\mathrm{new}|\vec p_{t})$ (In practice we consider multivariate normal distributions). We then accept $\vec p_\mathrm{new}$ as a new point ($\vec p_{t+1}=\vec p_\mathrm{new}$) with an acceptance probability
\begin{equation}
 \alpha(\vec p_\mathrm{new},\vec p_t) = \min \left[1, \frac{p(\vec p_\mathrm{new}|\vec Y,I)}{p(\vec p_t|\vec Y,I)}   \right]. \label{eq:mcmc_accept}
\end{equation}

If we reject the new point, then $\vec p_{t+1}=\vec p_{t}$. The expression for $\alpha(\vec p_\mathrm{new},\vec p_t)$ is simpler than the ones found in literature, because we have taken into account the symmetry of normal distributions that we used as proposed distributions ($p_{\rm{p}}(\vec p_1|\vec p_2)=p_{\rm{p}}(\vec p_2|\vec p_1)$).

From a mathematical point of view, the Markov chain asymptotically represent the target function (i.e. $p(\vec p|\vec Y,I)$) independently from the choice of the proposal probability distribution $p_p$. However, since we will get a chain with a limited size, the $p_p$ law must be chosen to ensure an efficient sampling of $p(\vec p|\vec Y,I)$ in reasonable computing time. As we consider $p_p$ as a multivariate normal law, we must find an suited covariance matrix for this law. Different automated algorithms are used to constructed such covariance matrices, for example in \citet{2009A&A...506...15B} and \citet{2011A&A...527A..56H}.

Finally, improved versions of MCMC using parallel tempering \citep{2005PCCP....7.3910E} are often implemented in asteroseismology. It is very similar, except that we do not use only one chain, but several ones corresponding to different ``temperature'' (by analogy with statistical physics), characterised by a parameter $\beta$. Each chain samples the following probability:
\begin{equation}
 p_\beta(\vec p|\vec Y, I) \propto p(\vec p|I)p(\vec Y|\vec p,I)^\beta
\end{equation}
for $\beta=1$ (the ``cool'' distribution) we recover the usual posterior probability. By decreasing $\beta$ (i.e., by using ``hotter and hotter''), the function is flatter and flatter, and is reduced to the prior distribution for $\beta=0$. Doing so, we understand that it is easier to explore regions of a hotter distribution that would never been visited for a cooler one. The different chains are independent, but randomly we swap the position of the two chains. It is very useful, for example, when the posterior distribution possesses several well separated local maximums. With a classical MCMC, the chain may get stuck in one of them and it may take a very long time to sample the different maxima. Parallel tempering is an efficient way to avoid this \citep[e.g.][]{2011A&A...527A..56H}.

MCMC is not the only sampling method developed for seismology of solar-like stars. For example, \citet{2014A&A...571A..71C} proposed a spectrum analysis method based on a nested sampling Monte Carlo algorithm.

\subsection{Fitting strategy: local vs. global fits}
\label{ssec:fitstrat}
Several strategies can be used to fit an oscillation spectrum. Historically, to analyse helioseismic data of the Sun seen as a star, a local strategy was developed: modes are fitted within windows narrow enough to isolate a single mode or a pair of modes $\ell=0,2$ or $\ell=1,3$. For intensity measurements, $\ell=3$ modes are generally small enough to be ignored, and $\ell=1$ modes are often fitted alone. A second approach is to perform a global fit of all modes above a given amplitude threshold around the maximum of the p-mode hump, simultaneously. This approach, pioneered by \citet{1999ESASP.448..135R} on solar data, became more popular in asteroseismology when first CoRoT data had to be analysed \citep{2008A&A...488..705A}.

\begin{figure}[!htp]
\center
 \includegraphics[width=1\textwidth]{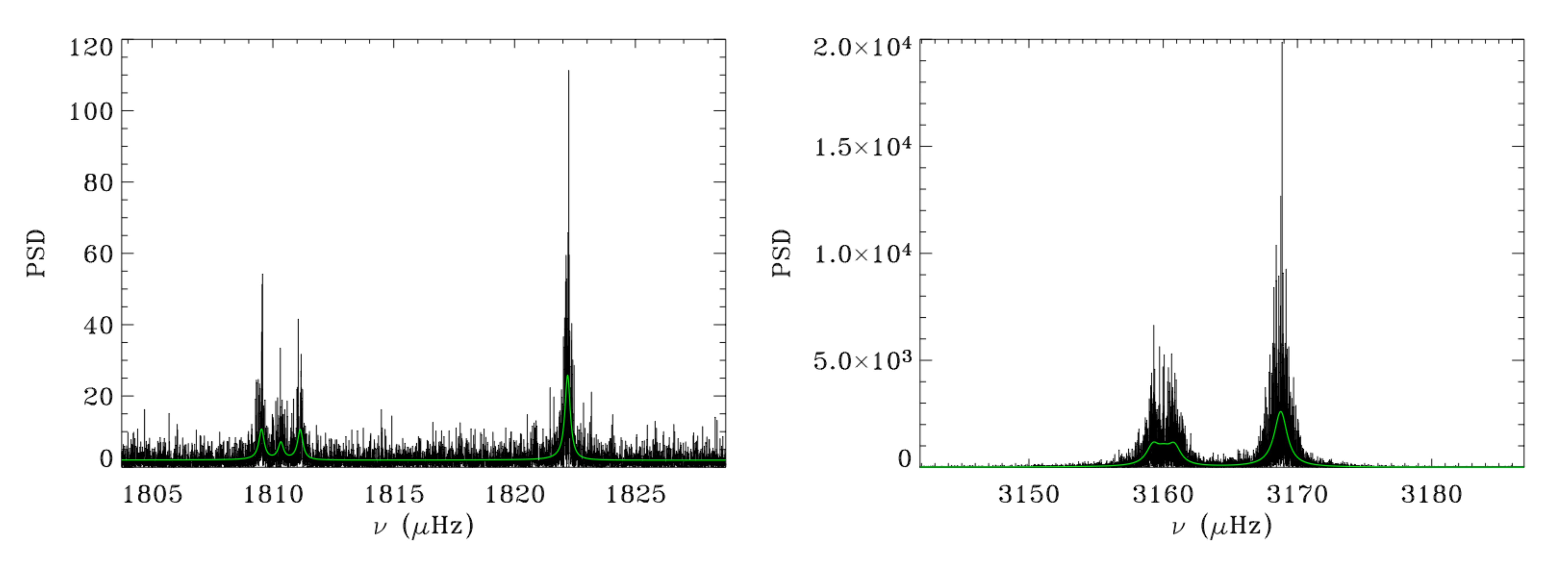}\\
 \includegraphics[width=1\textwidth]{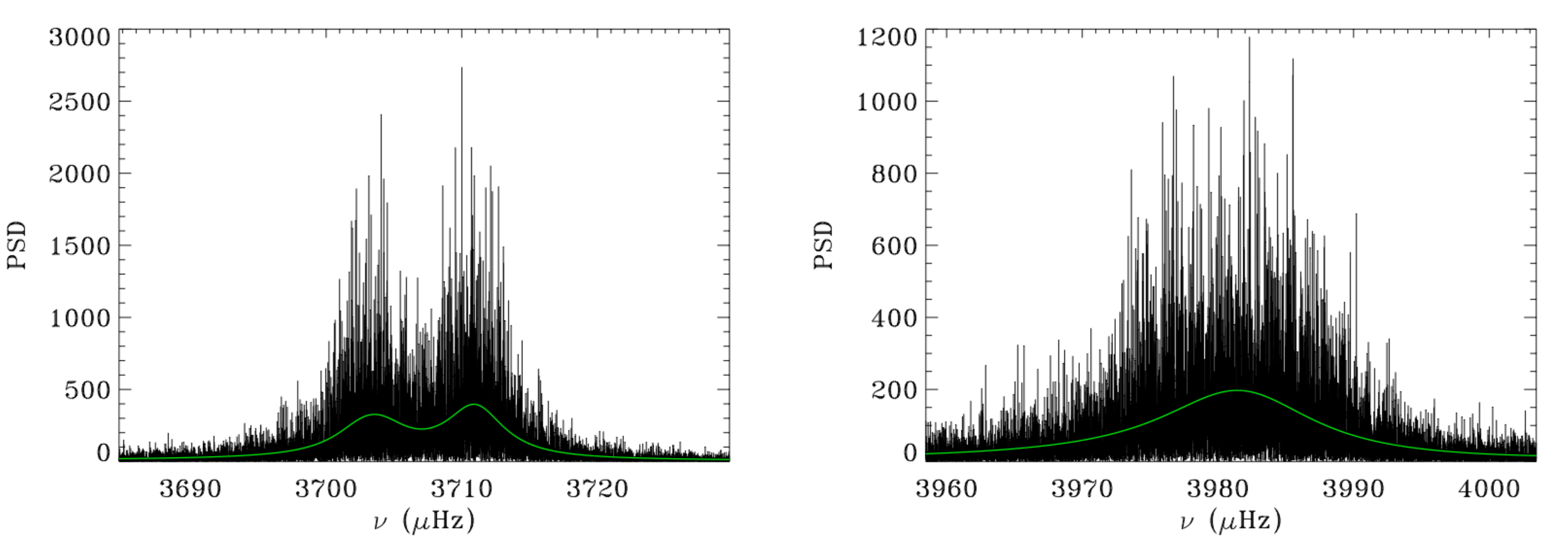}
  \caption[Examples of fits of low-order $\ell$=2, 0 and $\ell$=3, 1 modes at low and high frequency]{\label{fit_mp_lf} Example of fits (green line) $\ell$=2, 0 and $\ell$=3, 1 modes, left- and right-hand panels respectively at low (top) and high (bottom) frequencies of a typical GOLF spectrum. In the first case, the lifetimes of the modes are longer and therefore the linewidths are smaller than for the modes at high frequency.}
\end{figure}

Local fitting can be performed when mode pairs are well separated from each other, i.e. when the large separation is large enough to ensure that the wings of the Lorentzian profiles of other modes do not contaminate the fitting window. In a local approach, we assume the background is flat and is only parametrised with a constant. We use a single linewidth for the two modes ($\ell=0$ and 2), and we also may use a single free height, $H_0$, which is the height of the $\ell=0$ mode, and impose $H_2=V_2^2/V_0^2H_0$ for the $\ell=2$ mode by fixing the visibilities from theory. This last constraint is useful when $\ell=2$ and 0 modes are blended.

\begin{figure}[!htp]
  \centering
   \includegraphics[width=0.95\textwidth]{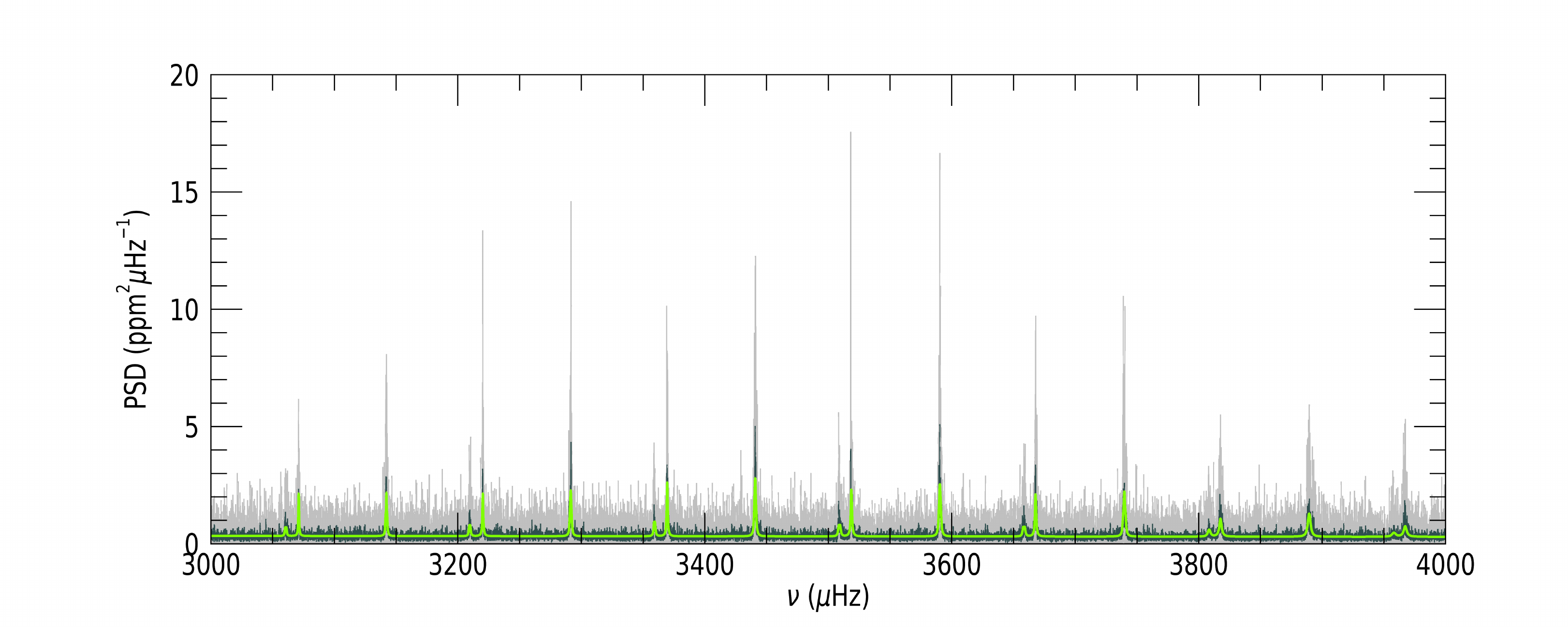}
      \caption[Example of a global fitting done in KIC~8006161]{\modif{Expanded view of the PSD of KIC~8006161 in the p-mode region at full resolution (light grey) and smoothed over 11 bins wide boxcar (dark gray).  The green line corresponds to the global fitting performed over 7 orders.}}
         \label{fpsd}
\end{figure}

Even if this local fitting scheme is easier to implement, with a reduced number of free parameters for each fitting procedure, it has been proven that a global approach is generally better suited \citep{2008A&A...488..705A,2014aste.book..123A}. An example of such global fit is presented in Fig.~\ref{fpsd} corresponding to the analysis of 11 orders of the CoRoT target HD~169392 \citep{2013A&A...549A..12M}.
 A main advantage of this approach is to impose that some parameters be the same for all modes, especially the inclination angle $i$, and for some cases the splitting $\nu_{\rm{s}}$. Due to the correlation previously mentioned between $i$ and $\nu_s$, $i$ may be poorly determined for each mode independently, we thus increase its precision by using a common value for all modes. Since all parameters are correlated, a better determination of $i$ improve the determination of $\nu_s$, hence linewidths, hence heights... However, by doing this, we assume that all of the stellar layers rotate with the same axis. This reasonable assumption may be questioned for the deep core \citep[e.g.][for the Sun]{1993ApJ...409..476B}. Free parameters to parametrize widths and heights can also be reduced assuming that they vary only slowly with frequency; one free parameter per large separation may be enough. Mode visibilities can also be fixed or fitted as global parameters. Concerning the background, it cannot be taken as constant in the fitting window, a more complete profile (Sect.~\ref{sssec:bgmodel}) must be used. Its parameters may be fixed by a previous overall fitting or left (partially) free \citep[see, e.g.,][]{2008A&A...488..705A,2011A&A...530A..97B,2011ApJ...733...95M}.

\begin{figure}[!htp]
  \centering
   \includegraphics[width=0.9\textwidth]{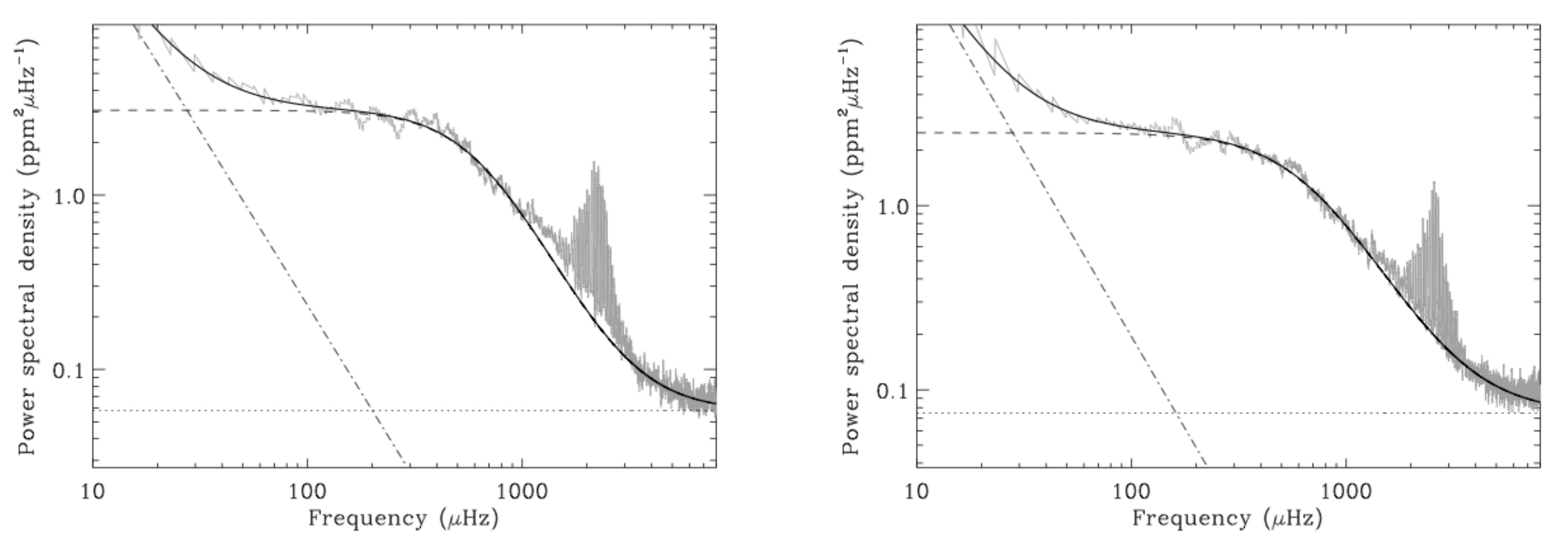}
      \caption[Background fitting of the two solar analogs 16~Cyg~A and B observed by {\it Kepler}]{Background fitting of the two solar analogs 16~Cyg~A (left) and B (right) observed by {\it Kepler}. The PSD has been smoothed by a  20 $\mu$Hz boxcar (grey), with best-fitting background components attributed to granulation (dashed lines), stellar activity and/or larger scales of granulation (dot-dashed lines) and shot noise (dotted lines), with the sum of the background components plotted as solid black lines \citep{2012ApJ...748L..10M}.   \label{16CygBack}          }
        
\end{figure}

A global fit of the background is generally previously performed by ignoring the frequency range of p modes, or modelling the p-mode hump as a Gaussian profile or two Lorentzian profiles \citep[as has been shown for the solar case by][]{2008A&A...490.1143L}. An example of such a fit for the two solar analogs 16~Cyg~A and B observed by {\it Kepler} is shown in Fig.~\ref{16CygBack} \citep{2012ApJ...748L..10M}.

\subsection{Hypothesis tests and model comparisons}

The fitting techniques described in the previous section provide the best parameters of a model, assuming that the hypotheses are correct. Hypotheses include especially the choice of the model, and some a priori information such as the mode identification. It is however frequently necessary to test several competiting hypotheses to make a final decision. The most outstanding one is to test the mode identification. The first solar-like star observed by CoRoT was the F-type star HD49933. Early-type solar-like stars have broad modes due to strong mode damping, thus mode widths are not smaller than the small separation, making the even modes $\ell=0,2$ overlap. As a consequence, identifying the $\ell=1$ ridge from the $\ell=0,2$ ridge is not obvious in an \'echelle diagram. Both hypotheses have to be tested with a frequentist  \citep{2008A&A...488..705A} or Bayesian approach \citep{2009A&A...506...15B}. Nevertheless to complement statistical approaches, \citet{2012ApJ...751L..36W} propose a method based on physical properties of the stars to disentangle both scenarios using the variation of $\epsilon$ (see Eq.~\ref{eq:nup_as}) with effective temperature.

It is also very useful to use model comparisons to validate the significance of a splitting. Thus, we can verify if a mode is significantly better fitted with a multiplet than with a single Lorentzian profile \citep[e.g.][]{2015A&A...580A..96D}.

When the two hypotheses $H_0$ and $H_1$ have the same number of free parameters (typically when we want to compare the two different mode identifications), the ratio of their likelihoods ${\cal L}_0$ and ${\cal L}_1$ gives a direct comparison of the two hypotheses. The $p$-value of favouring $H_0$ over $H_1$ is then $p=(1+{\cal L}_1/{\cal L}_0)^{-1}$. However, as already mentioned, since MLE may be biased for low SNR, such a comparison may be skewed.

Assuming now that the two hypotheses $H_0$ and $H_1$ have respectively $n_0$ and $n_1$ free parameters ($n_1>n_0$), we want to determine whether $H_1$ may be favoured. We thus need to assess the likelihood improvement of using more free parameters. \citet{wilks1938} showed that the quantity $\Lambda = 2(\ln {\cal L}_1 - \ln {\cal L}_0 )$ follows a $\chi^2$ distribution with $n_1-n_0$ degrees of freedom. The $p$-value is thus the probability
\begin{equation}
 p = P(\chi^2[n_1-n_0 \mathrm{\ dof}] > \Lambda) \, .
\end{equation}
A low $p$-value rejects the simplest model in favour of the most complex.

Using a Bayesian approach to compare two different competing models $M_i$ and $M_j$ (including different set of priors), we need to compute the evidence of the model 
\begin{equation}
 p(\vec Y|M_i,I)=\int p(\vec p|M_i,I)p(\vec Y|\vec p,M_i,I) {\rm{d}}\vec p \; .
\end{equation}
We then introduce the odds ratio
\begin{equation}
 O_{ij} = \frac{p(M_i|\vec Y,I)}{p(M_j|\vec Y,I)}
 = \frac{p(M_i|I) p(\vec Y | M_i,I)}{p(M_j|I) p(\vec Y | M_j,I)}\; .
\end{equation}
By assuming that the two models are equally probable ($p(M_i|I)=p(M_j|I)=1/2$), odds ratios reduces to Bayes factor
\begin{equation}
 O_{ij} = B_{ij} = \frac{p(\vec Y | M_i,I)}{p(\vec Y | M_j,I)}\; .
\end{equation}
The relative probability to favour model $M_j$ over model $M_i$ is $p=(1+O_{ij})^{-1}$. It may be generalized to more competing models.
The main difficulty is to compute the evidences $p(\vec Y | M_i,I)$. Parallel tempering MCMC allows us to compute them using the different chains with different temperatures. It reads \citep[e.g.][]{Gregory2005}
\begin{equation}
 p(\vec Y | M_i,I) = \int_0^1 \langle \ln p(\vec Y|M_i,I)\rangle {\rm{d}}\beta \; ,
\end{equation}
where the term $\langle\cdots\rangle$ is the average of the logarithm of the likelihood for all points of a given tempered chain, characterised by $\beta$.
This shows another interest to perform parallel tempering MCMC.

\subsection{Global seismic parameters}
We mainly detailed in the previous sections how individual mode properties can be extracted. Of course, global parameters, especially the mean large separation $\Delta\nu$, the frequency at maximum amplitude $\nu_\mathrm{max}$ and the maximum amplitude of radial mode $A_{\rm max}$ can be derived from a detailed fit. However, there exist quick methods to recover these main features. Such methods are useful when we must deal with a large number of stars. This is the reason why they have been massively used to analyse several thousand red giants observed by \emph{Kepler}, but they have also been applied to main-sequence stars. Various pipelines have been developed during the last decade by several teams around the world \citep[e.g.][]{2009A&A...506..435R,2010MNRAS.402.2049H,2010ApJ...723.1607H,2010A&A...522A...1K,2010A&A...511A..46M,2011A&A...525L...9M}.

The most common technique to measure $\nu_\mathrm{max}$ is to fit a bell-shape curve to the p-mode hump and define its maximum as $\nu_\mathrm{max}$. $A_\mathrm{max}$ can be derived from the total mode power around the maximum. By measuring the total power of the modes over a large separation interval around $\nu_{\rm max}$, we measure the total power of one mode of each degree. Using mode visibilities and assuming that all modes in the interval have the same intrinsic amplitude, we recover the power of a radial mode. The square root of this quantity provides the rms amplitude \citep[see][]{2008ApJ...682.1370K}. Generally, we convert the observed amplitude into a bolometric amplitude using bolometric corrections derived for each instrument \citep[see, e.g., for CoRoT and \emph{Kepler}][]{2009A&A...495..979M,2011A&A...531A.124B}.

To determine $\Delta\nu$, two techniques may be used to recover the regular pattern of p modes. This first one is to find a maximum of the autocorrelation of the spectrum in the p-mode region. The autocorrelation lag providing the largest peak is the large separation, another peak (generally slightly smaller) occurs at $\Delta\nu/2$ when the $\ell=1$ modes coincide with the $\ell=0,2$ modes. The second technique, which has appeared to be more robust in practice, is to consider the Fourier transform of the power spectrum in the p-mode region. Doing so, the largest peak of the Fourier transform is $\tau=2/\Delta\nu$, which corresponds to the main regularity visible in spectra produced by the alternation of even and odd modes. Since products in Fourier domain are convolutions in the time domain, we can easily show that this technique is equivalent to looking at the autocorrelation of the time series. We can refer to \citet{2006MNRAS.369.1491R} for detailed discussions of its use and \citet{2009A&A...508..877M} for discussions on measurement errors with this technique.

\newpage

\section{Inferences on stellar structure}
\subsection{Scaling relation for masses and radii}

If detailed modelling of a star to reproduce the observed frequencies is the most accurate way to determine its mass and radius, a simpler approach using scaling relations from solar values have been massively used these last few years. These scaling relations link the mass $M$ and radius $R$ of a star to the large separation $\Delta\nu$, the frequency at maximum amplitude $\nu_\mathrm{max}$, and the effective temperature  $T_{\rm{eff}}$.

A scaling relation for $\Delta\nu$ can easily be derived assuming an homology relation between stellar structures \citep{2013ASPC..479...61B}. Assuming two homologous stars (with radii $R$ and $R'$ and masses $M$ and $M'$) means that $m(r)/M=m'(r')/M'$ for all $r/R=r'/R'$, where $m(r)$ is the mass of the star encompassed within the radius $r$. \modif{In homologous stars, \citet[][\S 20.1]{1990sse..book.....K} show that the density and pressure profiles from one star is deduced from the other through the following scaling relations: $\rho'(r')=(M'/M)(R'/R)^{-3} \rho(r)$ and $p'(r')=(M'/M)^2(R'/R)^{-4} p(r)$. Hence we deduce for the sound speed profiles that $c'(r')=(M'/M)^{1/2}(R'/R)^{-1/2} c(r)$.}
Using Eq.~\ref{eq:expr_dnu}, we then \modif{show} that
\begin{equation}
 \frac{\Delta\nu'}{\Delta\nu} =\left(\int_0^R \frac{{\rm{d}}r'}{c}\right)\left(\int_0^{R'} \frac{{\rm{d}}r}{c'}\right) ^{-1}= 
 \left(\frac{M'}{M}\right)^{1/2}
 \left(\frac{R'}{R}\right)^{-3/2}.
\end{equation}
The large separation then scales with the square root of the mean stellar density. Of course, real stars are not homologous, however, it is well verified with models \citep[e.g.][]{1986ApJ...306L..37U}. \citet{2011ApJ...743..161W} show that the agreement is better than 5\%. Fixing the Sun as a reference, any star can be scaled from it as \citep{1995A&A...293...87K}:
\begin{equation}
\label{dnuMR}
   \Delta \nu \approx \Delta \nu_\odot  \left( \frac{M}{{\rm{M}}_\odot} \right)^{1/2} \left( \frac{R}{{\rm{R}}_\odot}\right)^{-3/2},
\end{equation}
where $\Delta \nu_\odot$=135.1 $\pm$ 0.1 $\mu$Hz, as derived by \citet{2011ApJ...743..143H} using 111 subseries of 30-day each collected by the \emph{Variability of solar IRradiance and Gravity Oscillations} (VIRGO) instrument \citep{1995SoPh..162..101F} aboard the \emph{Solar and Heliospheric Observatory spacecraft}  \citep[SoHO,][]{DomFle1995} spanning from 1996 to 2005 and analyzed them in the same way as asteroseismic data. Nevertheless, the choice of the reference values is not that straight forward and may be the subject of debate. One thing is clear, each pipeline producing $\Delta \nu$ and  $\nu_\mathrm{max}$ to be used in these scaling relations has their own solar reference values computing strictly with the same procedure as the stellar values.

The second quantity scaling with stellar global parameter is $\nu_{\rm{max}}$. \citet{1991ApJ...368..599B} suggested that this quantity scales as the acoustic cutoff frequency $\nu_{\rm{c}}$ (see Sect.~\ref{sssec:pmodes}). This assumption has been justified by \citet{2011A&A...530A.142B}. Using Eq.~\ref{eq:cutoff_freq} and noting that in the stellar atmosphere, \modif{approximated by an isothermal atmosphere of temperature $T_{\rm eff}$,} $H_p \propto T_{\rm eff}/g$ and $c^2 \propto T_{\rm eff}$, we deduce that $\nu_c \propto g/\sqrt{T_{\rm eff}}$.
Hence, $\nu_{\rm{max}}$ can be scaled from the solar value as follows \citep{1995A&A...293...87K}:
\begin{equation}
\label{numaxMRT}
   \nu_{\rm{max}} \approx  \nu_{\rm{max}, \odot}   \left(\frac{M}{{\rm{M}}_\odot}\right) \left(\frac{R}{{\rm{R}}_\odot}\right)^{-2} \left(\frac{T_{\rm{eff}}}{{\rm{T}}_{\rm{eff},\odot}}\right)^{-1/2},
\end{equation}
where ${\rm{T}}_{\rm{eff},\odot}$ = 5770 K, and $\nu_{\rm{max}, \odot}$ = 3090 $\pm$ 30 $\mu$Hz \citep{2011ApJ...743..143H}.

\begin{figure}[!htp]
\begin{center}
\includegraphics[width=0.7\textwidth]{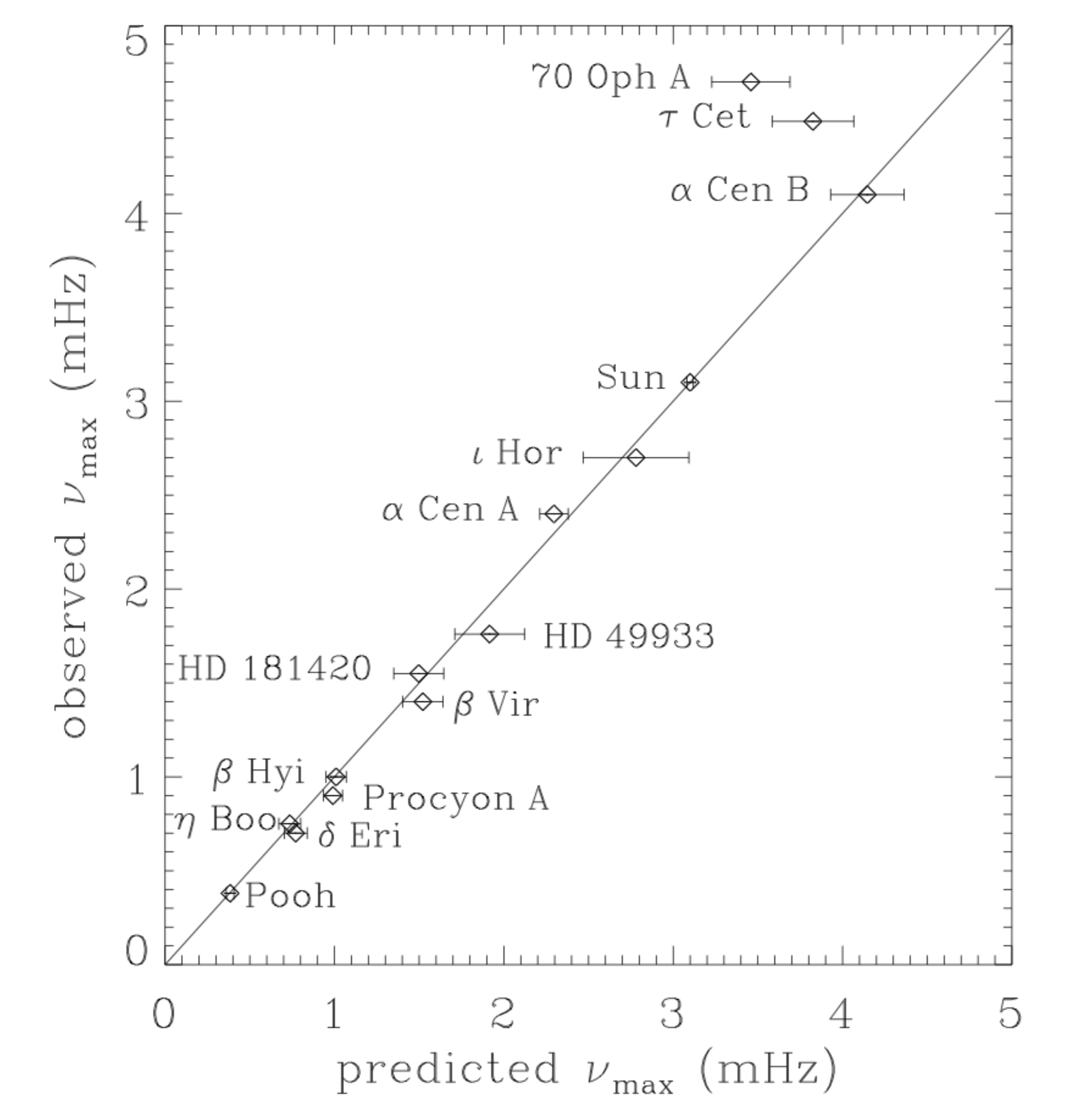}
\end{center}
\caption{\label{nmaxtest} Expected and observed $\nu_{\rm{max}}$ from \citet{2014aste.book...60B}. These stars have very well-known stellar parameters as well as accurate parallaxes.
}
\end{figure}

Scaling relations from solar values have been extensively tested, in particular,  by comparing for example the expected and observed values of $\nu_{\rm{max}}$ for some well-studied stars with accurate parallaxes \citep{2003PASA...20..203B}. Figure~\ref{nmaxtest} shows the comparison of the observed and predicted values of $\nu_{\rm{max}}$ for 14 stars. The general agreement is very good excepting the two low-mass stars $\tau$~Cet and 70~Oph~A.  From a theoretical point of view, some predictions of lighter 0.7 and 0.9 ${\rm{M}}_\odot$ models show a double bump in the p-mode hump, which complicates the correct extraction of this parameter \citep[e.g.][]{2008A&A...485..813C}. However, one of the maxima usually lies close to the $\nu=\nu_{\rm{max}}$ line. Moreover, ground-based observations of Procyon
\citep{2008ApJ...687.1180A}, as well as other CoRoT \citep[e.g.][]{2010A&A...511A..46M} and {\it Kepler} stars have shown that more massive stars could also present a double bump in the p-mode envelope. The question of the definition for $\nu_\mathrm{max}$ arises in this case. 

\begin{figure}[!htp]
\begin{center}
\includegraphics[width=0.7\textwidth]{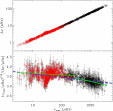}
\end{center}
\caption{\label{nmaxdnu} $\Delta\nu$ as a function of $\nu_{\rm{max}}$ (top panel) for 1700 stars observed by {\it Kepler} from \citet{2011ApJ...743..143H}. The red triangles show stars observed in long cadence (red giants), while black diamonds are stars observed in short cadence (solar-like stars). In the bottom panel we have shown the same relation but after removing the luminosity dependence by raising $\nu_{\rm{max}}$ to the power of 0.75. Green lines show power-law fits to the $\Delta\nu-\nu_{\rm{max}}$ relation for two different intervals of $\nu_{\rm{max}}$. The dashed blue line shows the relation of Eq.~\ref{dnu_numax} derived using both red giants and main-sequence stars \citep{2009MNRAS.400L..80S}.
}
\end{figure}

Combining Eqs.~\ref{dnuMR} and \ref{numaxMRT}, we deduced that
\begin{equation}
 \Delta\nu \propto M^{-1/4} T_{\rm eff}^{3/8} \nu_{\rm max}^{3/4}.
\end{equation}
Because of the relatively weak change in $T_{\rm eff}$ among solar-like stars and the weak dependency on $M$, we recover that solar-like oscillations in main-sequence stars to a good approximation follow:
\begin{equation}
\label{dnu_numax}
\Delta\nu \approx \Delta\nu_\odot \left(\frac{\nu_{\rm{max}}}{\nu_{\rm{max},\odot}}\right)^b,
\end{equation}
with $b\approx 0.75$, as shown, for example, by \citet{2009MNRAS.400L..80S}, \citet{2009A&A...506..465H}, \citet{2010A&A...517A..22M} , and \citet{2011A&A...525A.131H}. Figure~\ref{nmaxdnu} (top) shows the relation $\Delta\nu$ versus $\nu_{\rm{max}}$ computed using 1700 stars from the main sequence (black diamonds) completed with red clump (red triangles appearing in a diagonal branch between 20 to 50 $\mu$Hz in the bottom panel) observed by the {\it Kepler} mission. Although the relation appears to be constant for all of these stars, several authors \citep{2010A&A...517A..22M,2010ApJ...723.1607H} have suggested that the slope is different for red-giant and main-sequence stars. To enhance such a difference, we have subtracted the luminosity dependence by raising $\nu_{\rm{max}}$ to the power of $b=0.75$. A fit to the residuals below and above $\nu_{\rm{max}}$= 300 $\mu$Hz --  which roughly marks the transition from low-luminosity red giants to sub-giants -- enhances a steeper slope as $\nu_{\rm{max}}$ increases. It is important to note that, for $\nu_{\rm{max}}$  close to the solar value for example, the use of a power-law relation calibrated to red-giant stars would lead to an underestimation in $\Delta\nu$ by $\approx 10\%$. Recent comparisons of different global seismic pipelines (mostly for red giants) can be found in \citet{2017ApJS..233...23S} and \citet{2018arXiv180409983P}.

In a similar way, it is possible to derive a scaling relation from the solar values for the maximum oscillation \modif{full -- not RMS -- intensity amplitude} of radial modes, $A_{\rm{max}}$, of the radial modes \citep[following, e.g.,][]{1995A&A...293...87K,2007A&A...463..297S}:

\begin{equation}
A_{\rm{max}}={\rm{A}}_{\rm{max},\odot} \; \beta \left(\frac{L/{\rm{L}}_\odot}{M/{\rm{M}}_\odot}\right)^s \left(\frac{T_{\rm{eff}}}{{\rm{T}}_{\rm{eff},\odot}}\right)^{-2} \;\; ,
\end{equation}
where the exponent $s$ usually has values between 0.7 and 1.0. The $\beta$-coefficient has been introduced  by \citet{2011ApJ...732...54C} because without any further correction the above relation is known to overestimate the amplitudes for the hottest solar-like stars \citep{2006ESASP.624E..28H}. A$_{\rm{max},\odot}$ is 4.4 $\pm$ 0.3 ppm for the VIRGO green channel ($\lambda$ = 500nm) as calculated by \citet{2011ApJ...743..143H}. The inferred bolometric amplitude of $\rm{A}_{\rm{max},\odot,bol}$ = 3.5 $\pm$ 0.2 ppm is in good agreement with the value of 3.6 ppm \citep{2009A&A...495..979M} commonly adopted in asteroseismology.

\subsection{Model-independent determination of masses and radii}

Combining Eq.~\ref{dnuMR} and \ref{numaxMRT}, we derive masses and radii of stars as a function of the seismic variables $\nu_{\rm{max}}$ and $\Delta_\nu$, assuming that $T_{\rm{eff}}$ is known from photometry or spectroscopy:

\begin{equation}
{R} \approx {{\rm{R}}_\odot} \left( \frac{{\rm{\Delta\nu}}_\odot}{\Delta \nu}\right)^2 \left(\frac{\nu_{\rm max}}{{\rm{\nu}}_{\rm{max},\odot}} \right) \left(\frac{T_{\rm eff}}{{\rm{T}}_{\rm{eff,\odot}}} \right)^{1/2} ,
\end{equation}

\begin{equation}
M \approx {\rm{M}}_\odot \left(\frac{{\rm{\Delta\nu}}_\odot}{\Delta \nu} \right)^4 \left(\frac{\nu_{\rm max}}{{\rm{\nu}}_{\rm{max},\odot}} \right)^3 \left(\frac{T_{\rm eff}}{{\rm{T}}_{\rm{eff,\odot}}} \right)^{3/2}.
\end{equation}

\begin{figure}[!htp]
\begin{center}
	\includegraphics[width=0.76\textwidth]{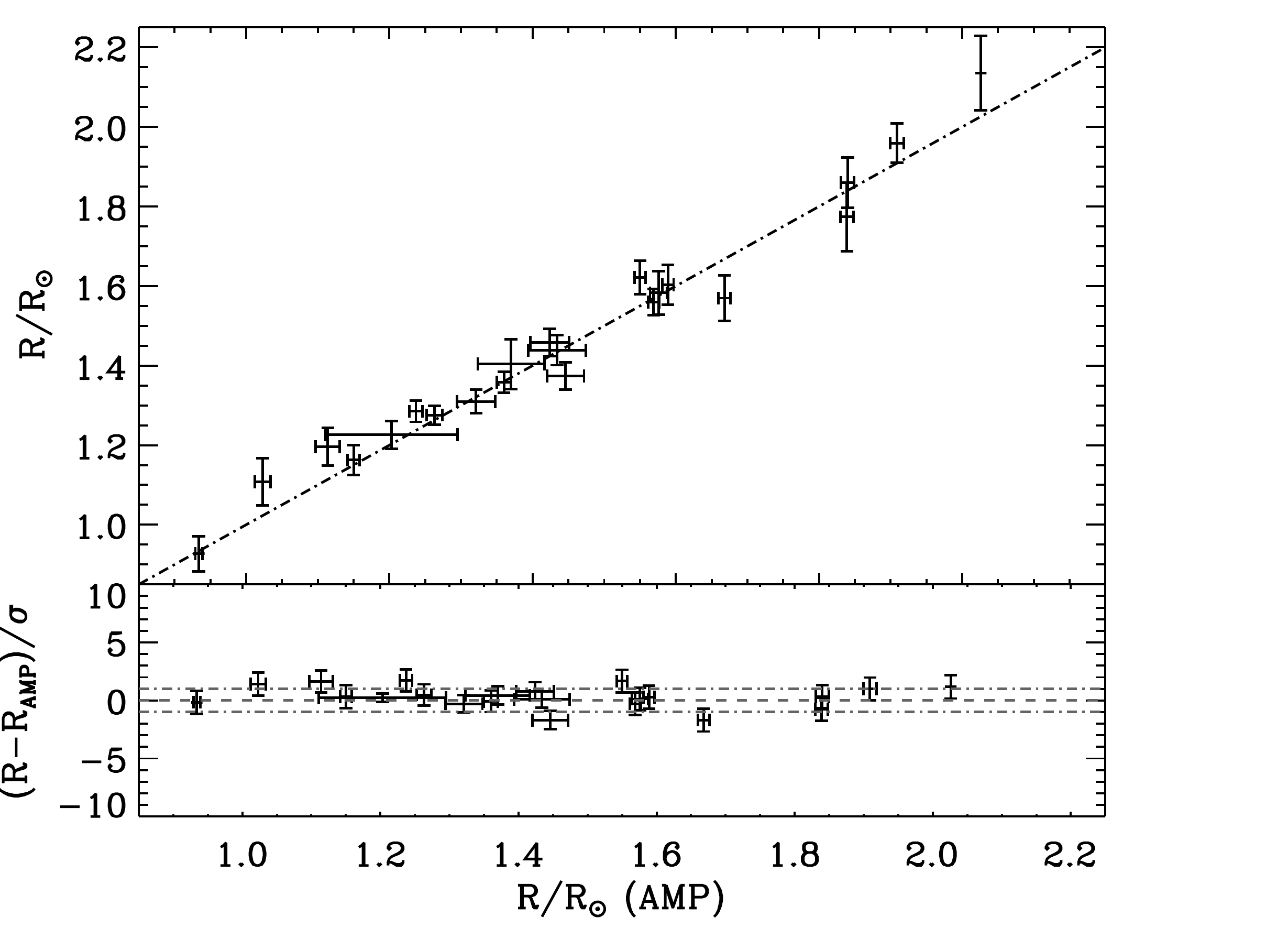}  \\ 
	\includegraphics[width=0.76\textwidth]{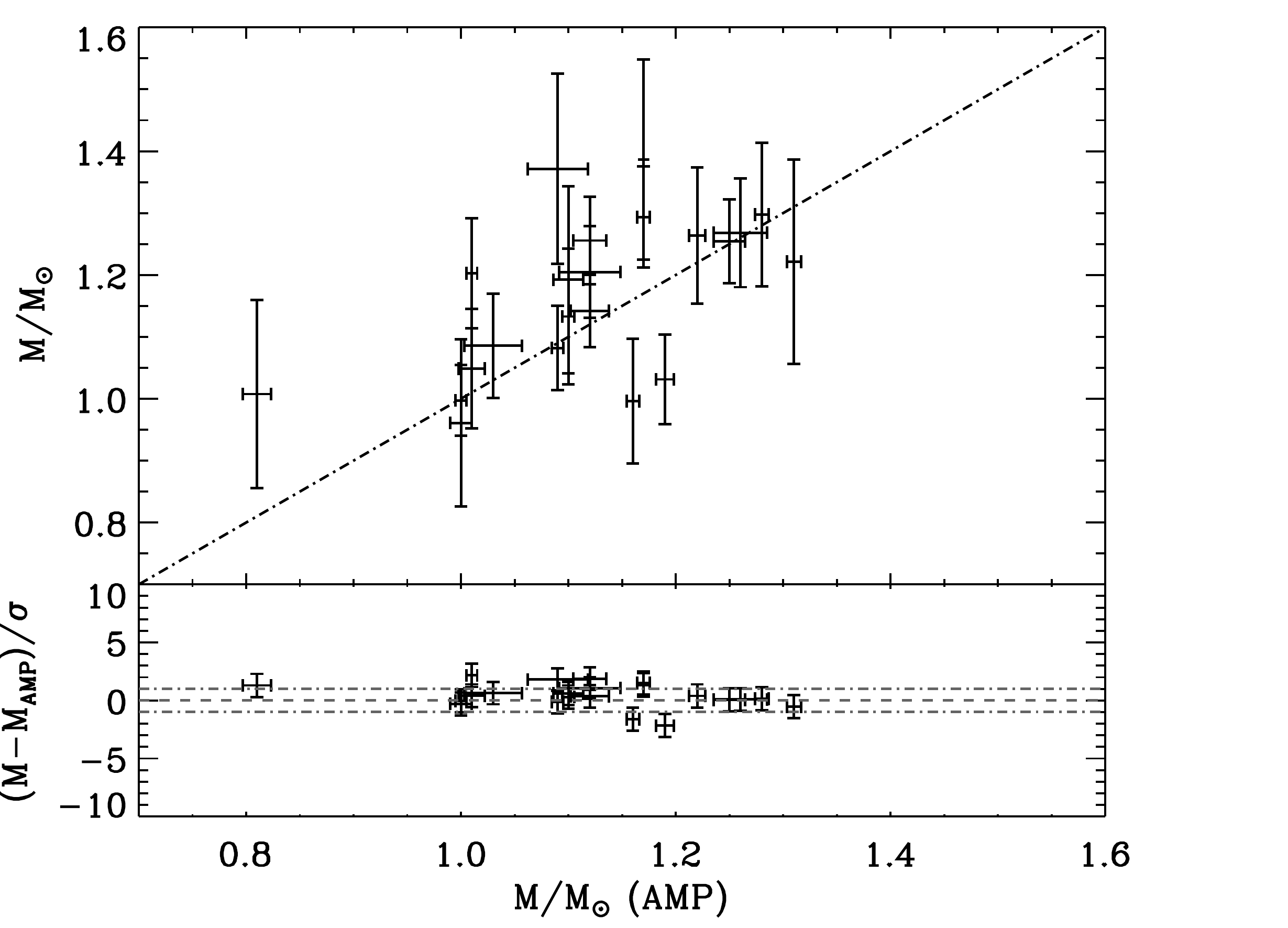}
\end{center}
\caption[Comparison of the Masses and Radii computed form the scaling relations and AMP]{\label{amp_RM} Comparison of stellar radii (top) and masses (bottom) computed using the Asteroseismic Modeling Portal (AMP) with those computed from the empirical scaling relations for 22 solar-like stars observed by {\it Kepler} from \citet{2012ApJ...749..152M}. The top of each panel compares the actual values, while the bottom shows the relative differences in units of the statistical uncertainty ($\sigma$). See the text for details.
}
\end{figure}

These scalings provide a quick way to measure mass and radius without any models. The errors on these measurements will depend first on the precision of $\Delta\nu$ and $\nu_{\rm{max}}$ determination. For \emph{Kepler} data, global methods such as the A2Z or the Octave pipelines \citep{2010A&A...511A..46M,2010MNRAS.402.2049H} \modif{provide precisions around 10\% and 20\% for masses and radii according to their authors. However, using individual frequencies to compute averaged $\Delta\nu$ and $\nu_{\rm max}$ gives uncertainties of about 3 $\%$ in radius and 9 $\%$ in mass  \citep{2012ApJ...749..152M}.}
Nevertheless, these relations are only approximations and to test their accuracy we can compare the values inferred from the scaling relations with the best modeling performed on 22 {\it Kepler} targets for which we have been able to determine the individual frequencies of a few dozens modes \citep{2012ApJ...749..152M}. The modeling was done using the Asteroseismic Modeling Portal (AMP), which uses a parallel genetic algorithm \citep{2003JCoPh.185..176M} to optimize the match between the model output and the observational constraints. In Fig.~\ref{amp_RM} we have compared the estimation of both parameters. Stellar radii from both methods are in a very good agreement, while we have a higher scatter in the comparison of the masses, as was expected because the exponents involved on the seismic parameters were 2 to 3 times bigger. According to this work, the scaling relations tend to overestimate the radius by +0.3$\sigma$  and the mass by +0.4$\sigma$ relative to the values from AMP ($\sigma^2$ being the quadratic sum of the uncertainties from the two methods). \modif{This corresponds to radii overestimated by about 1\% and masses by about 4\% typically}. These results suggest that observations of the global oscillation properties combined with an effective temperature through the scaling relations from the solar values can provide reliable estimates of the radius and mass but with a lower precision than AMP when individual frequencies are available. Indeed, the precision reached in that latter case by AMP, in the mass and radius, is much better: 0.8 $\%$ in radius and 1.1$\%$ in the mass.

It has also been possible to test the global seismic scaling relations using independent measurements of radii for main-sequence and giant stars using interferometry. \citet{2012ApJ...760...32H} performed such an analysis comparing asteroseismic radii obtained with \emph{Kepler} with interferometric radii obtained with the CHARA array and found an agreement of around 4\% \citep[see also][]{2013MNRAS.433.1262W}.  Unfortunately, because main-sequence stars have small angular diameters, it is extremely difficult to properly extract their radius, these stars being more prone to systematic errors in the adopted calibrator diameters than sub-giants and red giants. This is the reason why it has been recommended to restrict any comparison of asteroseismic diameters with interferometry to stars with angular diameters larger than $> 0.3$ mas excluding many main-sequence solar-like stars \citep{2017ApJ...844..102H}.

Another independent validation of the asteroseismic radius can be done using astrometric results \citep[e.g.][]{2012ApJ...757...99S} from Hipparcos \citep{2007ASSL..350.....V} or Gaia \citep{2001A&A...369..339P}. Indeed, \cite{2017ApJ...844..102H} compared asteroseismic radii of 2200 oscillating stars observed by {\it Kepler} \citep[including 440 main-sequence stars and sub-giants from ][]{2011Sci...332..213C}  with Gaia DR1 results included in the Tycho-Gaia Astrometric Solution \citep[TGAS,][]{ 2015A&A...574A.115M}.  The overall agreement found was excellent, which helped to empirically demonstrate that asteroseismic radii computed using global seismic scaling relations were accurate to $\approx 10 \%$ for stars ranging from $\approx$0.8 to 10 R$_\odot$ without any visible offset between the two radii determinations for main-sequence stars  (1 to 1.5 R$_\odot$). Moreover, no significant trends were found with metallicity  [Fe/H] = -0.8 to +0.4 dex.

\subsection{Model dependent determination of masses and radii}

\begin{figure}[!htp]
\begin{center}
 \includegraphics[width=0.45\textwidth]{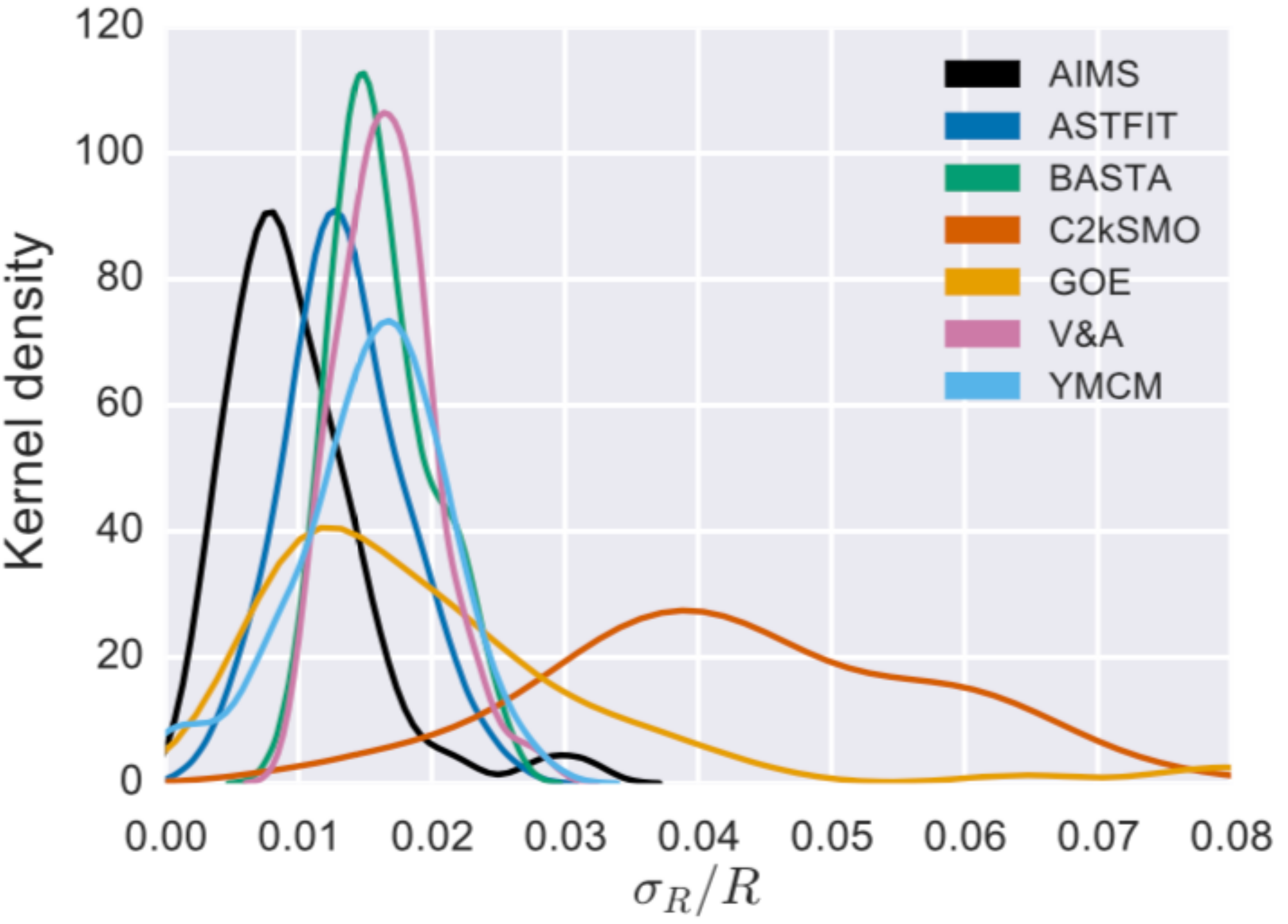} 
 \includegraphics[width=0.45\textwidth]{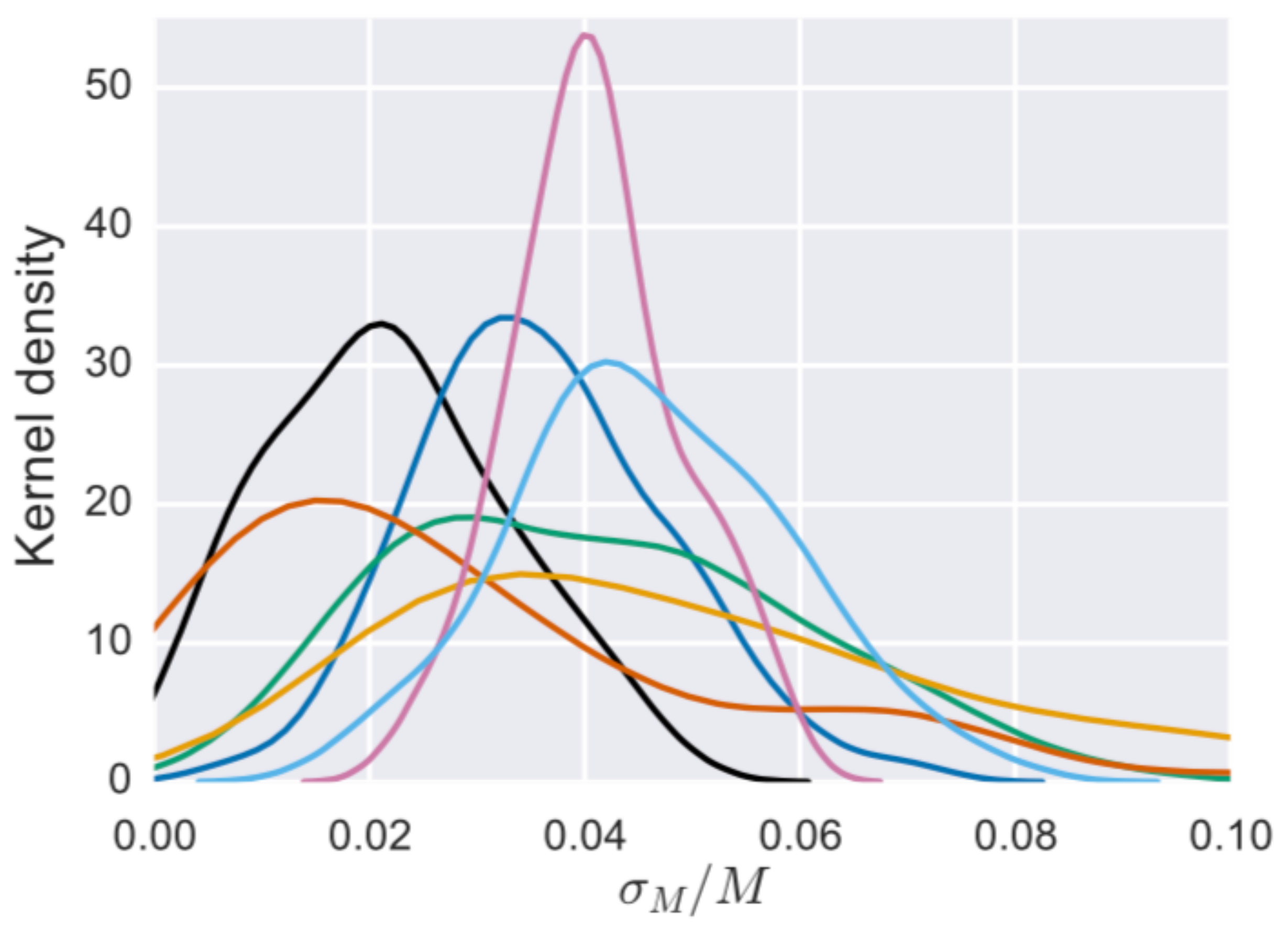}  
 \end{center}
\caption[Distribution of masses and Radii from Silva-Aguirre et al. 2017]{\label{distrib_silva} Distribution of radius and mass uncertainties of 66 solar-like and sub-giant stars analyzed by \cite{2017ApJ...835..172L}. Asymmetric error bars are added in quadrature for the plots. The different pipelines are described in detail in \cite{2017ApJ...835..173S}. Figure adapted from \cite{2017ApJ...835..173S}.
}
\end{figure}

The ultimate precision of asteroseismology can only be reached when individual mode frequencies are combined with spectroscopy and astrometric observations to infer the best stellar model of each star. Tens of individual mode frequencies of the best 66 main-sequence and sub-giant solar-like stars seismically characterized by \cite{2017ApJ...835..172L} were distributed among seven different modeling teams to determine radii, masses, and ages for all the stars in this sample \citep{2017ApJ...835..173S}. As it is shown in Fig.~\ref{distrib_silva},  masses and radii agree within the error bars for most of the cases with average uncertainties better than $\sim 2 \%$  and $\sim$ 4$\%$ in radius and mass respectively. The differences found between the different results could be explained in terms of the different physics used in the codes and the way in which the uncertainties were computed during the minimization process (that explains the large uncertainties found in radius computed by Cesam2k Stellar Model Optimization -- C2kSMO, \cite{2014A&A...569A..21L}).  For further details see \cite{2017ApJ...835..173S}. 

\subsection{Ensemble observational asteroseismology}

\begin{figure}[!htp]
\begin{center}
\includegraphics[width=0.90\textwidth]{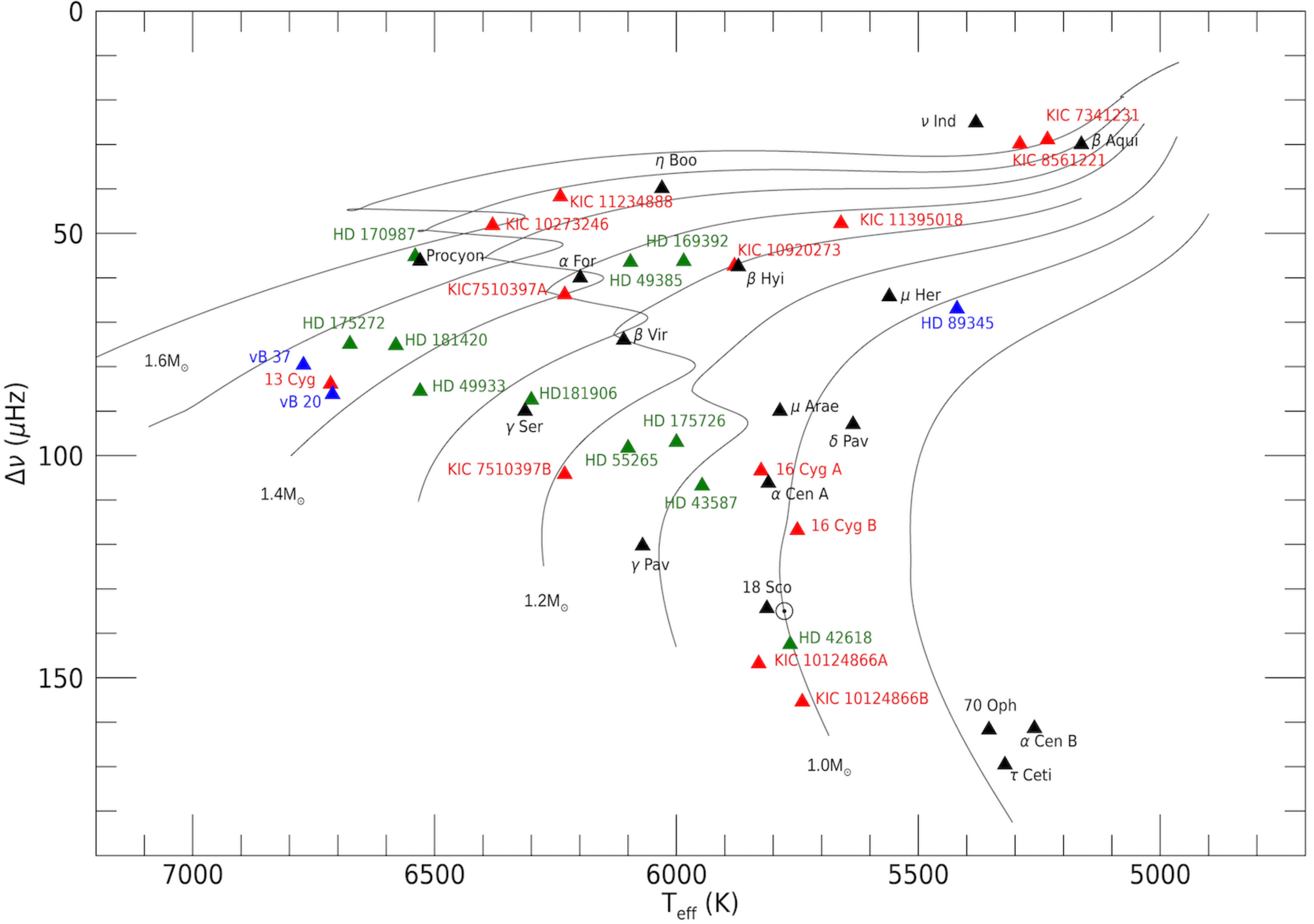}
\includegraphics[width=0.90\textwidth]{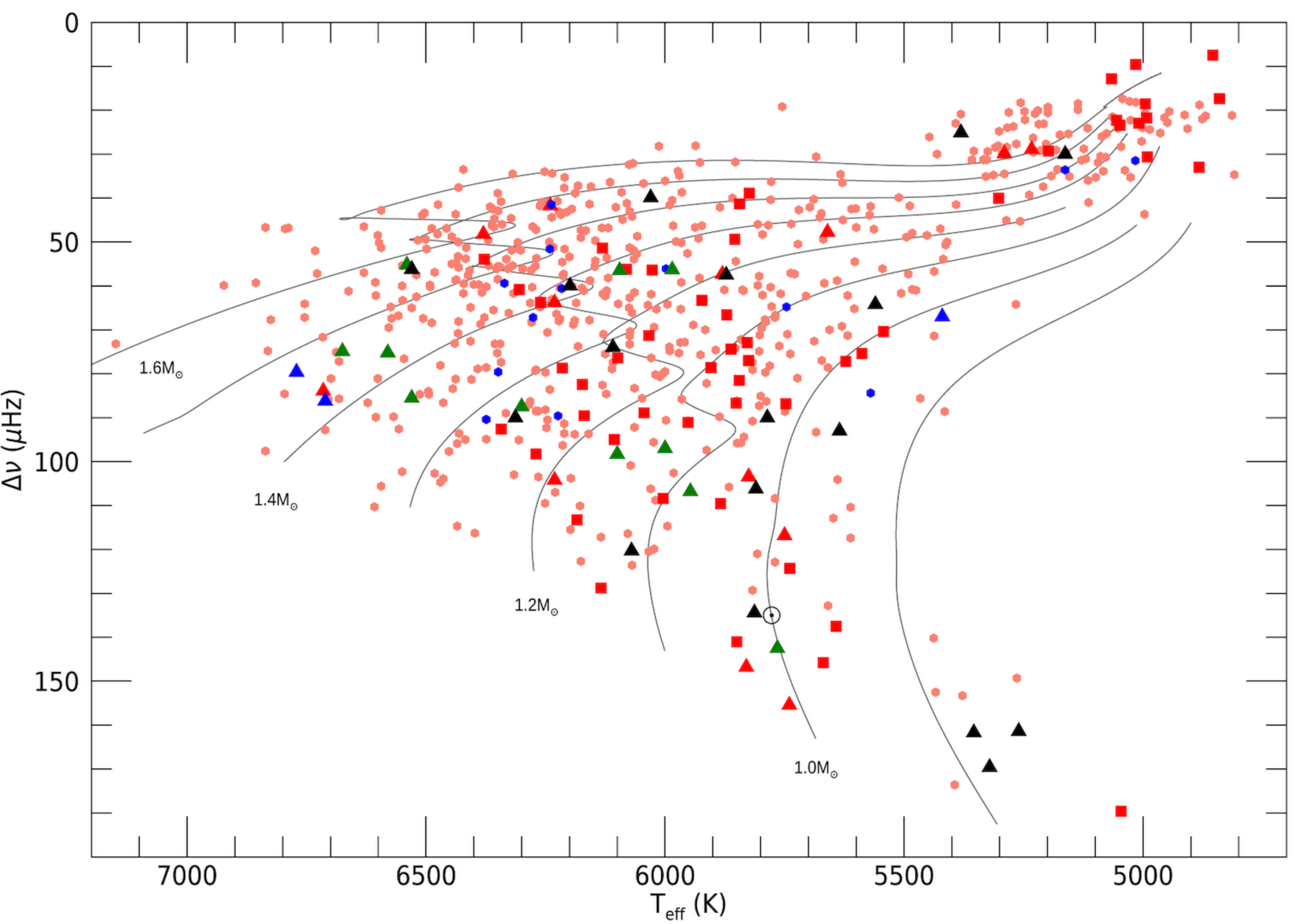}
\end{center}
\caption[Seismic HR diagram]{\label{Seis_HR_simple} Seismic Hertzsprung--Russell diagram showing $\Delta\nu$ as a function of $T_{\rm eff}$ \modif{of main-sequence, sub-giants, and some early red giant stars}. In the top panel a selection of stars is represented while in the bottom panel stars observed \modif{in short-cadence} by \emph{Kepler} and K2 have been added. Triangles represent stars analyzed individually or in groups of a few stars. Circles represent stars studied as an ensemble. Colors represent observations made from the ground (black) or with spacial missions:  green for CoRoT, red or magenta for \emph{Kepler} and blue for K2. WIRE observations are included as ground-based because these targets were also observed from ground. The Sun is represented with its usual label ($\odot$). Evolutionary tracks, computed with the ASTEC code \citep{2008Ap&SS.316...13C}, are shown for masses ranging between 0.9 and 1.6 M$_\odot$ at solar composition (Z$_\odot$ = 0.0246). The values used to plot these figures are listed in Table~\ref{tab1}.}
\end{figure}

We can take advantage of the ensemble analysis of several hundreds MS solar-like pulsating stars to reveal general trends on the properties of stellar pulsations. 
In Fig.~\ref{Seis_HR_simple} the ensemble of solar-like stars with measured pulsations is shown in a seismic HR diagram, i.e., plotting the large frequency separation as a function of the effective temperature. In the top panel are represented ground-based Doppler-velocity observations (black triangles) \citep[][and references therein]{2002A&A...390..205B,2003AJ....126.1483K,2004ESASP.559..563M,2005A&A...440..609B,2005ApJ...635.1281K,2006ApJ...647..558B,2006A&A...450..695C,2007ApJ...663.1315B,2008ApJ...682.1370K,2008A&A...488..635M,2008ApJ...676.1248B,2009A&A...494..237T,2010ApJ...713..935B,2011A&A...526L...4B}, CoRoT observations (green triangles) \citep{2008A&A...488..705A,2009A&A...507L..13B,2009A&A...506...51B,2009A&A...506...41G,2009A&A...506...33M,2010A&A...515A..87D,2010A&A...518A..53M,2011A&A...530A..97B,2013A&A...549A..12M,2013A&A...558A..79O,2013JPhCS.440a2030B,2014A&A...564A..34B}, \emph{Kepler} observations \citep{2011ApJ...733...95M,2011A&A...534A...6C,2012ApJ...756...19D,2012ApJ...748L..10M,2014A&A...563A..84G,2015A&A...582A..25A,2016ApJ...831...17G,2017A&A...601A..82W}, and K2 observations \citep{2016MNRAS.463.2600L,2018arXiv180501860V}  not included in ensemble analyses. 
In the bottom panel ensemble analyses of \emph{Kepler} field stars \citep[magenta circles,][]{2014ApJS..210....1C}, \emph{Kepler} stars hosting planets \citep[red squares,][]{2016MNRAS.456.2183D}, and K2 stars of campaigns 1 to 3 \citep[blue circles,][]{2016PASP..128l4204L} have been added. 

\begin{figure}[!htp]
\begin{center}
\includegraphics[width=0.8\textwidth]{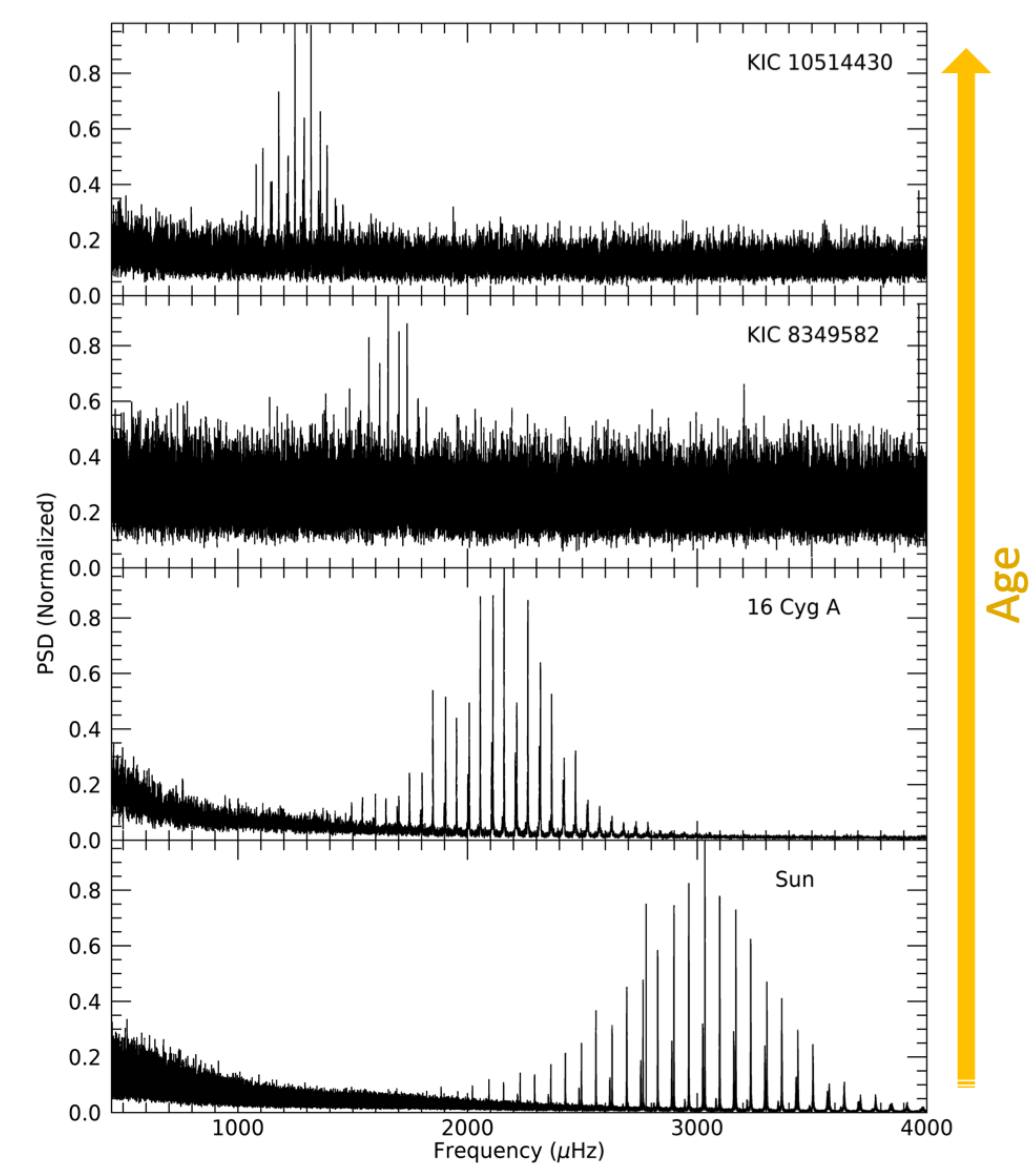}
\end{center}
\caption[Stellar Evolution at constant Mass during the MS]{\label{1mass_evol} Example of an evolutionary sequence of solar analogue stars exhibiting acoustic modes along the MS starting with the Sun. The Sun is the youngest star of the sequence. From bottom to top, the Sun observed by the green channel of the VIRGO/SPM instrument onboard SoHO, 16 Cyg A, KIC~8349582, and KIC~10514430 with masses of 1, 1.011 $\pm$ 0.02, 1.068 $\pm$ 0.02, and 1.059 $\pm$ 0.04 ${\rm{M}}_\odot$ respectively \citep[see for the details][]{2012ApJ...748L..10M,2015MNRAS.452.2127S}.}
\end{figure}

Using different solar-like stars located along the evolutionary tracks traced in Fig.~\ref{Seis_HR_simple}, it is today possible to build observational evolutionary sequences of dozens of stars of similar masses (and whenever possible, similar metallicities), using scaling relation discussed in previous sections. Therefore, ensemble asteroseismology studies open new ways of performing differential analysis on field stars along evolutionary sequences and thus constraining how the internal properties of stars change with mass, metallicity, and evolution \citep[e.g.][]{2011ApJ...740L...2S,2013A&A...558A..79O}. An example of an observational evolutionary sequence of solar analogues is shown in Fig.~\ref{1mass_evol}. 

Mode amplitudes, heights, and linewidths are more difficult to measure. \cite{2014A&A...566A..20A} showed that systematic effects between 8 different groups of ``fitters'' were mainly due to the way that the convective background was treated, as well as on the fitted values of the rotational splittings and inclination angles. Following a correction scheme based on the one-fit approach of \cite{2005A&A...433..713T}, they demonstrated that the systematic effects could be reduced to less than $\pm$ 15 $\%$ for the linewidths and heights, and to less than $\pm$ 5 $\%$ for the amplitudes. It is worth mentioning that different convective background models introduce frequency-dependent systematic errors that could bias any comparison with theoretical predictions and between different groups of fitters. Once frequencies, amplitudes, and linewidths are measured with precision and accuracy, it is possible to look for global trends with, for example, age, effective temperature, or mass. 

\begin{figure}[!htp]
\begin{center}
\includegraphics[width=0.95\textwidth]{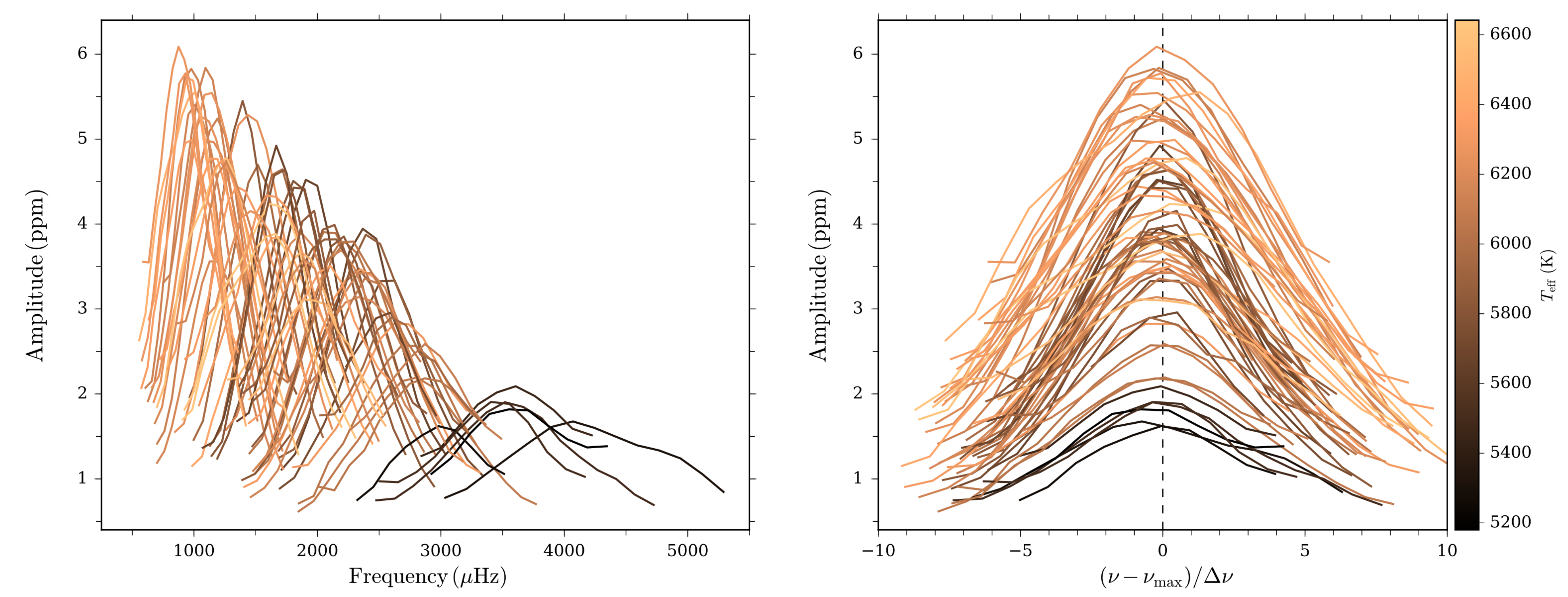}
\end{center}
\caption[Radial p-mode amplitude of MS solar-like stars]{\label{FIG_amps} Radial p-mode amplitude envelopes as a function of frequency (left panel) for 66 MS solar-like stars observed by \emph{Kepler}. The color scale represents the stellar effective temperature. In the right panel, the same envelopes are represented as a function of a proxy of the radial order $(\nu-\nu_{\rm{max}})/\Delta\nu$. For clarity, the envelopes have been smoothed with a five-point Epanechnikov filter. Figure obtained from \cite{2017ApJ...835..172L}.}
\end{figure}

During the MS evolution, acoustic-mode frequencies  -- and hence $\nu_{\rm{max}}$ -- decrease (see Fig.~\ref{1mass_evol}) while their amplitudes increase progressively as shown in Fig.~\ref{FIG_amps} \citep[see also][]{2005ApJ...635.1281K,2008ApJ...682.1370K,2008ApJ...687.1180A}. Moreover, p-mode amplitudes also scale with increasing effective temperature \citep[e.g.][]{1995A&A...293...87K,2011A&A...529L...8K,2014A&A...566A..20A,2017ApJ...835..172L}, which includes a variation with mass as explained by \cite{2011ApJ...743..143H}. 

\begin{figure}[!htp]
\begin{center}
\includegraphics[width=0.95\textwidth]{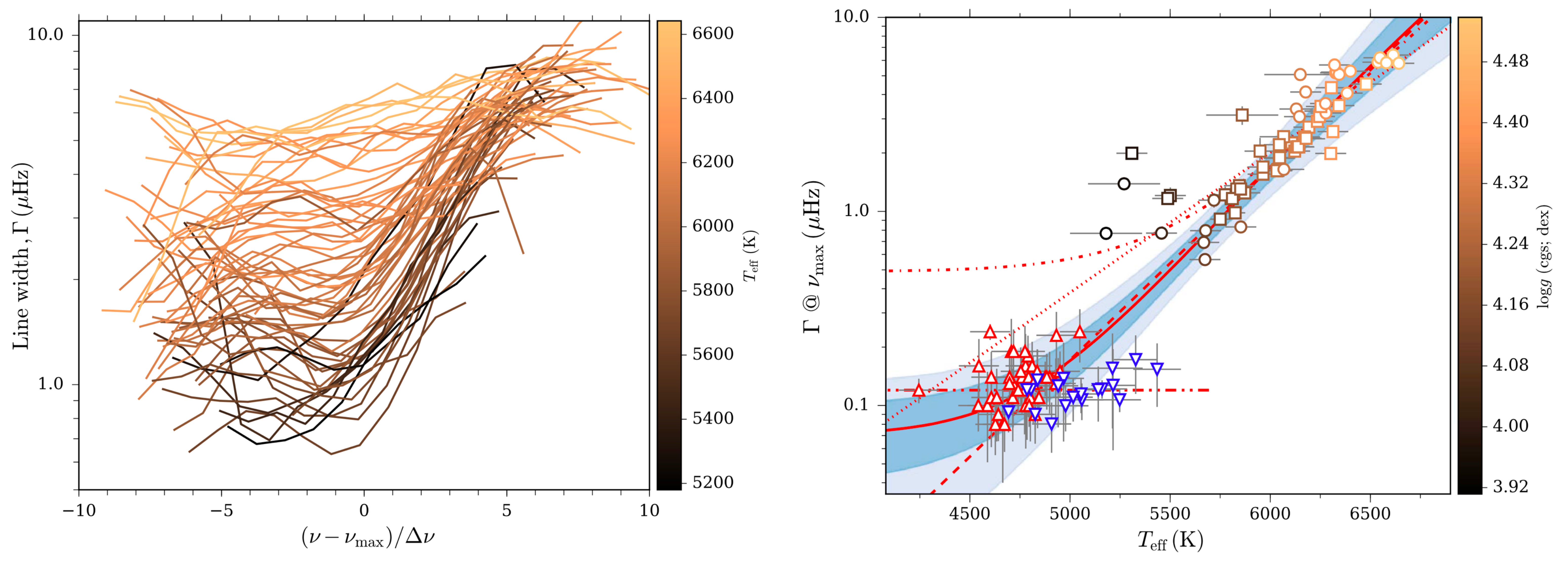}
\end{center}
\caption[radial p-mode amplitude of MS solar-like stars]{\label{FIG_widths} Left: Radial p-mode linewidths as a function of a proxy of the radial order of the same 66 MS solar-like stars observed by \emph{Kepler} of the previous figure. For clarity, the linewidths have also been smoothed with a five-point Epanechnikov filter. Right: Linewidth at $\nu_{\rm{max}}$ as a function of the effective temperatures (color-coded  with $\log g$) for these 66 MS stars (open circles and squares).  Upward red triangles correspond to linewidths of 42 giants in NGC 6819 from \cite{2017MNRAS.472..979H} while downward blue triangles are the linewidths of 19 red giants from \cite{2015A&A...579A..83C}. The full red line and the dashed line are the fits from \cite{2017ApJ...835..172L} with the shaded-dark and light-blue regions indicating the 1- and 2-$\sigma$ intervals of the fit; the dashed--dotted line gives the fit proposed by \cite{2012A&A...537A.134A}; the dotted line gives the fit proposed by \cite{2012ApJ...757..190C}; the dashed--dotted-dotted line gives the constant fit to only the red giants. Figures from \cite{2017ApJ...835..172L}.}
\end{figure}

Linewidths also depend strongly on effective temperature. They increase with temperature as is shown in Fig.~\ref{FIG_widths}. Several scaling relations of the linewidths at $\nu_{\rm{max}}$ with the effective temperature have been proposed in the literature \citep{2009A&A...500L..21C,2011A&A...529A..84B,2012ApJ...757..190C,2012A&A...537A.134A,2014A&A...566A..20A,2016A&A...595C...2A}. These relations are qualitatively in good agreement with theoretical predictions, although it would be necessary to introduce other dependencies such as with $\log g$ \citep{2012A&A...540L...7B}. The main differences arises at low temperatures where it is necessary to introduce sub-giants and giants to improve the fit. Moreover, to go further in the comparison with the theoretical predictions, it would be necessary to improve the precision in the calculation of the effective temperatures whose large errors limit the final precision of the fitted scaling relation \citep[][]{2012A&A...537A.134A}.

\subsection{The C-D diagram}

\begin{figure}[!htp]
\begin{center}
\includegraphics[width=0.85\textwidth]{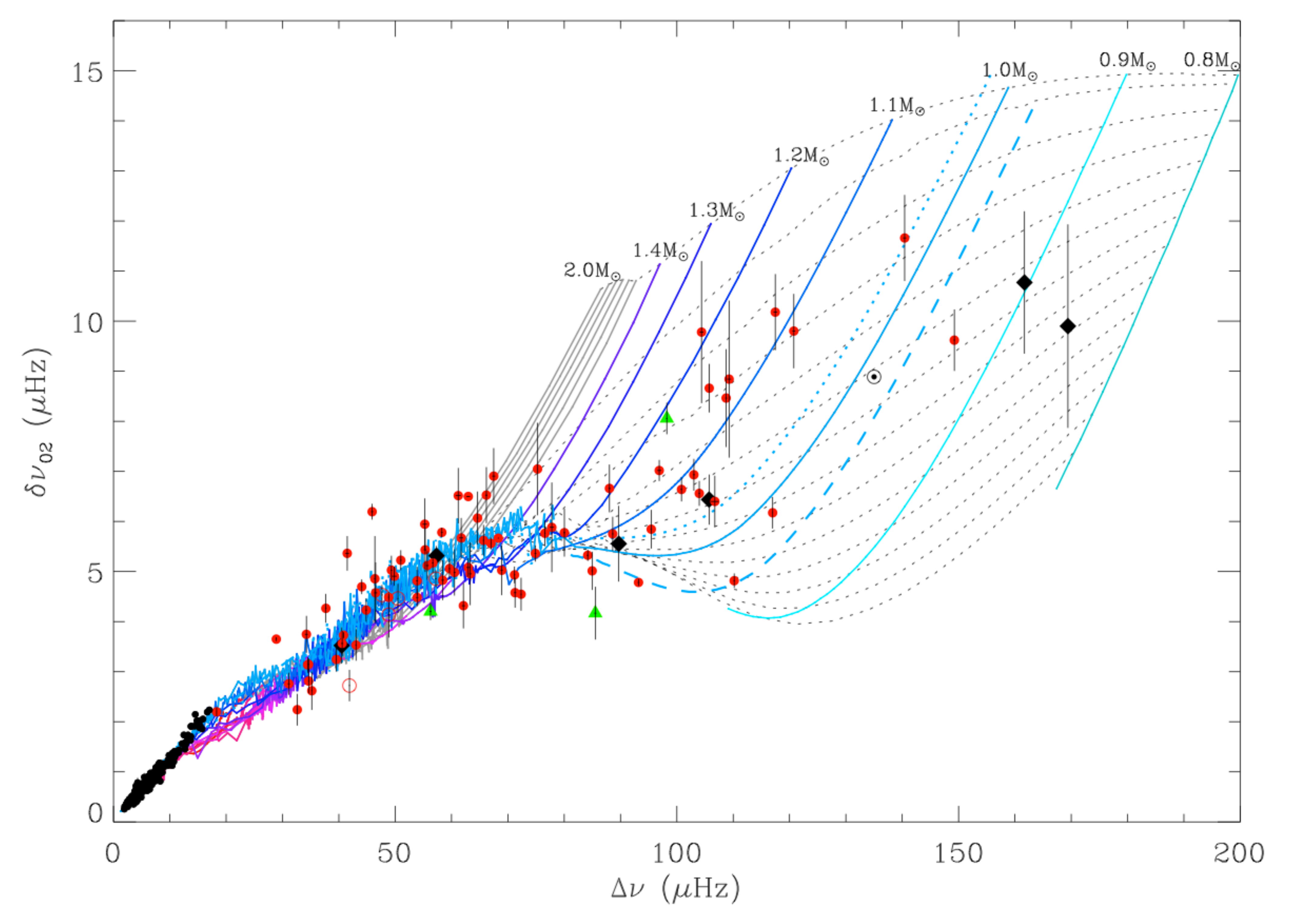}
\end{center}
\caption[The C-D diagram]{\label{CDdiag} C-D diagram, showing $\delta \nu_{0,2}$ {\it versus} $\Delta \nu$ from \citet{2011ApJ...742L...3W}. The filled red circles are the 76 \emph{Kepler} stars used in that work, while open red circles are {\it Kepler} stars previously published. Also shown are {\it Kepler} red giants \citep[black circles; from][]{2010ApJ...723.1607H}, and main-sequence and sub-giant stars from CoRoT (green triangles) and ground-based observations (yellow diamonds). The Sun is marked by its usual symbol. Error bars show the uncertainties derived from the linear fits for both $\Delta \nu$ and $\delta \nu_{0,2}$, although those in $\Delta \nu$ are generally too small to be visible. Model tracks for near-solar metallicity ($Z_0$ = 0.017) increase in mass by 0.1 ${\rm{M}}_\odot$ from 0.8 ${\rm{M}}_\odot$ to 2.0 ${\rm{M}}_\odot$ (light blue to red lines). Also shown are tracks for metal-poor ($Z_0$ = 0.014; dotted) and metal-rich ($Z_0$ = 0.022; dashed) solar-mass models. The section of the evolutionary tracks where the models have a higher $T_{\rm{eff}}$ than the approximate cool edge of the classical instability strip \citep{1998ApJ...498..360S} are gray: they are not expected to show solar-like oscillations. Dotted black lines are isochrones, increasing from 0 Gyr (ZAMS) at the top to 13 Gyr at the bottom.}
\end{figure}
The Christensen-Dalsgaard diagram (or simply C-D) is the representation of the small separation $\delta \nu_{0,2}$ as a function of the large separation $\Delta\nu$ \citep{1988IAUS..123..295C}. By placing stars on this diagram (see Fig.~\ref{CDdiag}, it is possible to directly determine their masses and ages. Indeed the tracks for different masses and ages are well separated. However, it is important to note that this diagram depends on metallicity. Nevertheless, if the metallicity of a star is known, such a diagram is an interesting tool to derive its age. A C-D diagram has been constructed for 76 \emph{Kepler} stars and several CoRoT targets by \citet{2011ApJ...742L...3W}. Results are shown in Fig.~\ref{CDdiag}. We can see how evolution tracks are well split during the main sequence but converge at the terminal age main sequence. It also gives an estimate of the mass within 4\,--\,7\% when the metallicity is determined within 0.1 dex. The age-determination accuracy depends strongly on the small-separation determination.



\subsection{Modelling a star and surface effects}

To go beyond global parameters and global diagnostics, we can model a star in detail to match the observed frequencies. Several approaches are then possible. We can used a precomputed grid of models to find the model with the closest frequencies \citep[e.g.][]{2009ApJ...700.1589S,2010ApJ...725.2176Q,2011ApJ...730...63G}, we can also perform on-the-fly modelling to recover the best model. For this kind of direct approach, we need to minimize a cost function, generally a $\chi^2$, between observed and modelled frequencies. This cost function may integrate external constraints such as spectroscopic or photometric determinations of temperature, metallicity, and eventually the mass if the star belongs to a multiple system. 

\begin{figure}[!htp]
\begin{center}
\includegraphics[width=0.5\textwidth]{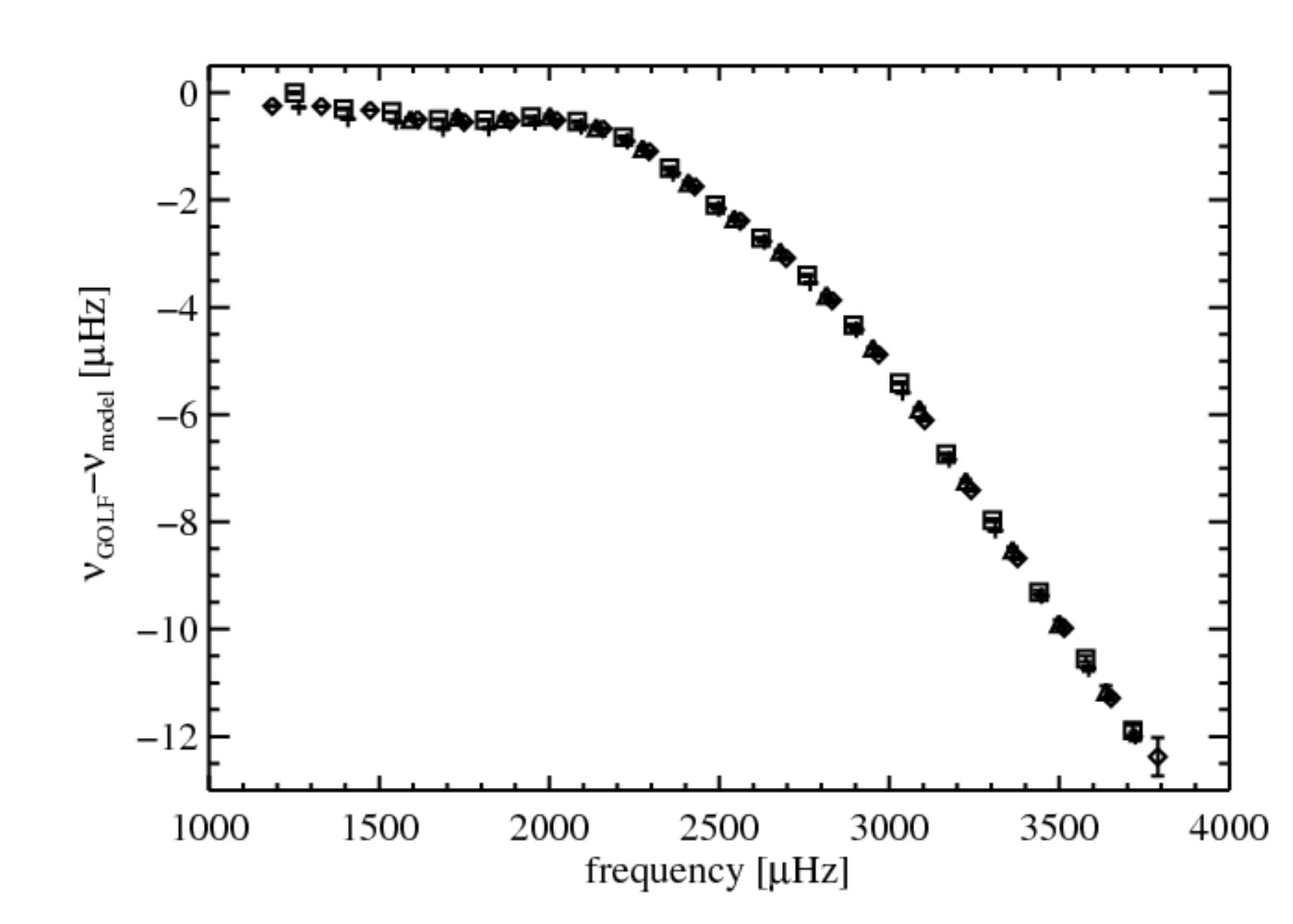}\includegraphics[width=0.5\textwidth]{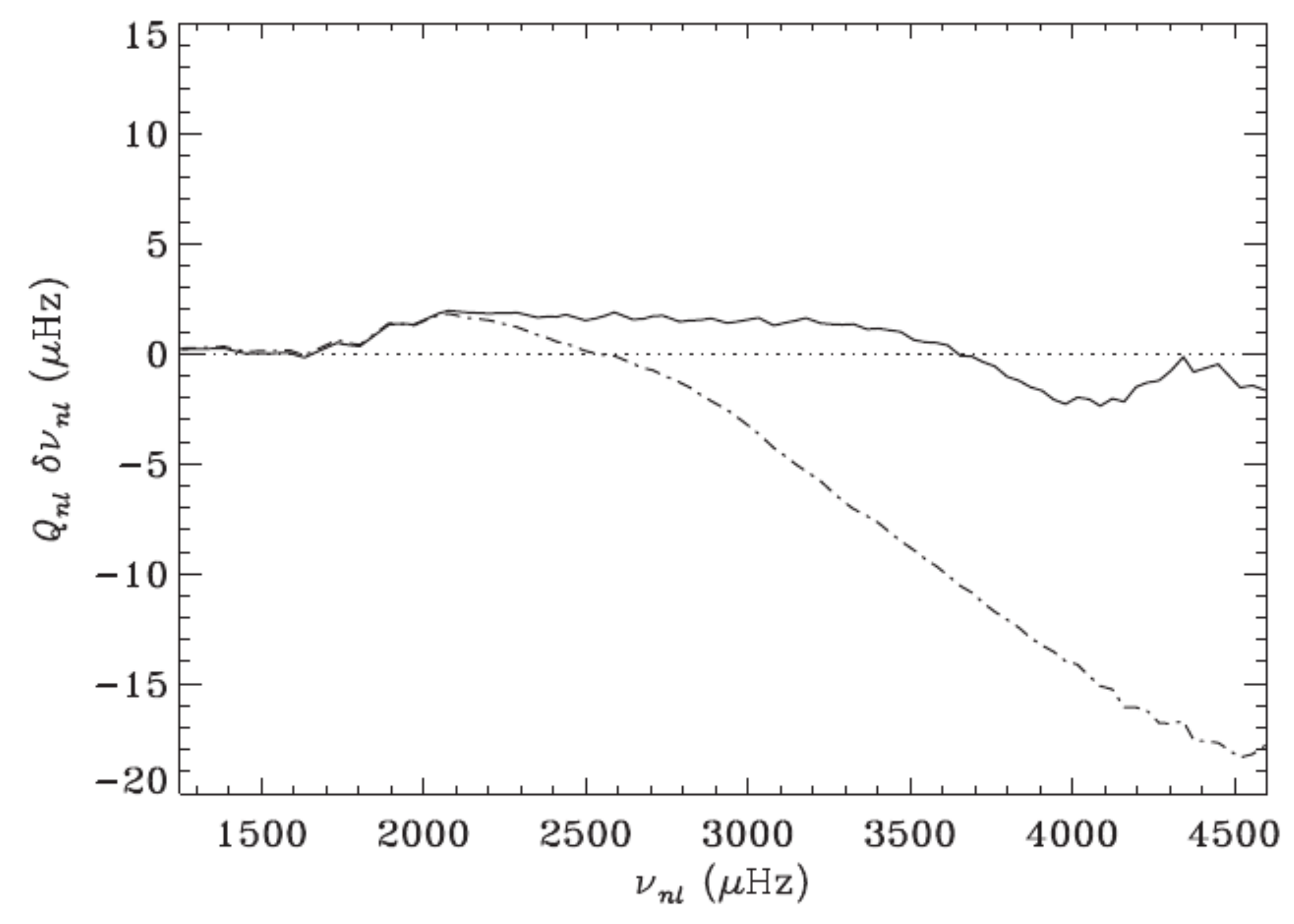}
\end{center}
\caption{\label{fig:surface_sun}Left: Difference between frequencies observed by GOLF and a solar model for $\ell=0\dots 3$ modes. Right: The dashed line is similar to the left panel and shows the difference between frequencies of radial modes observed by MDI and frequency computed in a standard way. To obtain the solid line, a patched model with non-adiabatic effects and time-dependent convection has been used \citep{2017MNRAS.464L.124H}.}
\end{figure}

Nevertheless, the use of frequencies without specific care provides biased results due to surface effects. It is known for the Sun \citep[e.g.][]{1991sia..book..401C} that significant discrepancies between observed frequencies and those computed from 1D model occur above $\approx 2000\:\mu$Hz, as shown in Fig.~\ref{fig:surface_sun}. This is due to the poor modelling of upper layers. Low frequency modes are almost unaffected because their outer turning point are deeper in the star, as discussed in Sect.~\ref{sssec:pmodes} (Fig.~\ref{fig:propag}). Upper layers are affected by dynamical processes missing in 1D modelling:  3D model atmospheres show that turbulent pressure cannot be neglected (at the photosphere it reaches about 15\% of the total pressure) and the structure is also affected by convective back-warming \citep[e.g.][]{2013ApJ...769...18T}. Moreover the $\beta$ plasma becomes small, letting the magnetic field play a role (it is the reason why its variations during stellar cycles are visible on frequencies, see Sect.~\ref{Sec:magnetism}).
There are not only structural effects, but also modal effects: non adiabatic effects must be included as well as fluctuations of the turbulent pressure.

To improve the modelling of outer layers, 1D models patched with 3D simulations have been proposed for decades
\citep[e.g.][]{1991LNP...388..195S,1999A&A...351..689R,2007MNRAS.375..403Y,2014MNRAS.437..164P,2015ApJ...806..246B,2015A&A...583A.112S,2017A&A...600A..31S,2016A&A...592A.159B,2016A&A...592A..24M,2017MNRAS.464L.124H,2017MNRAS.466L..43T}. To compute the oscillations primarily two different assumptions were proposed to model the Lagrangian fluctuations $\delta p_t$ of turbulent pressure. The first one, called the Gas $\Gamma$ model (GGM), consists in assuming that they vary as the total pressure, i.e. $\delta p_{\rm{t}}/p_{\rm{t}} \approx \delta p_{\rm tot}/p_{\rm tot}  \approx \delta p_{\rm gas}/p_{\rm gas} = \Gamma_1 \delta\rho /\rho$. In the second model, called the Reduced $\Gamma$ model (RGM), the turbulent pressure vanishes $\delta p_{\rm{t}}/p_{\rm{t}} \approx 0$, hence $\delta p_{\rm tot}/p_{\rm tot}  \approx (\Gamma_1 p_{\rm gas}/p_{\rm tot}) \delta\rho /\rho$. Rosenthal et al. (1999) shows that GGM gives better results than RGM. Using patched models with GGM improves the computed high frequencies, but a residual of several $\mu$Hz remains. Recently \citet{2017A&A...600A..31S} and \citet{2017MNRAS.464L.124H} used time-dependent convection models to compute the oscillations of patched models. The first team used prescriptions from \citet{2005A&A...434.1055G} and \citet{2005A&A...435..927D}, whereas the second one used a model developed by \citet{1977ApJ...214..196G}. Doing so, the discrepancies do not almost depend on frequency and remain limited to a few $\mu$Hz (Fig.~\ref{fig:surface_sun}).

Whereas it was easy to get rid of surface effects for the Sun because its mass and radius are extremely well known, dealing with surface effects is more difficult for other stars. A solution is to correct the model frequencies $\nu_{\rm model}$ with an ad-hoc term $\delta\nu$ to get corrected frequencies $\nu_{\rm corr}=\nu_{\rm model}+\delta\nu$. The first correction that was broadly used was proposed by \citet{2008ApJ...683L.175K} it takes the form
\begin{equation}
 \delta\nu = a (\nu/\nu_0)^b
\end{equation}
where $\nu_0$ is a reference frequency, $b$ is calibrated from the Sun, and $a$ has to be fitted. This relation has been extensively used during the last 10 years \citep[e.g.][]{2011A&A...526A..35T,2012ApJ...749..152M}. Of course, it assumes that the mode inertia varies slowly with frequency, that is the case for main-sequence stars. For sub-giant stars, corrections must be divided by the mode inertia $\cal I$.
Other prescriptions have been proposed later. Following a suggestion by \citet{1990LNP...367..283G}, \citet{2014A&A...568A.123B} proposed a cubic correction
\begin{equation}
 \delta\nu = a_3 (\nu/\nu_0)^3 / {\cal I}
\end{equation}
where $\nu_0=\nu_c$ and a composite correction with an additional inverse term is
\begin{equation}
 \delta\nu = (a_{-1} (\nu/\nu_0)^{-1} + a_3 (\nu/\nu_0)^3) / {\cal I}.
\end{equation}
These new prescriptions appear to give better fits than the old prescription for main-sequence stars \citep{2014A&A...568A.123B} and sub-giants \citep{2017A&A...600A.128B}. A recent study performed on a large sample of 66 \emph{Kepler} stars shows a significant improvement in the fit by using the composite form including the inverse term \citep{2018MNRAS.479.4416C}.
It  is also worth mentioning an alternative prescription derived from patched models by \citet{2015A&A...583A.112S}:
\begin{equation}
 \delta\nu = a \left(1 - \frac{1}{1+(\nu/\nu_0)^b} \right).
\end{equation}

Finally, another way to get rid of surface effects is to construct seismic variables where surface effects cancel out. \citet{2003A&A...411..215R} proposed to use the ratio of the small separation ($\delta\nu_{02}$ or $\delta\nu_{01}$) to the large separation to remove a large part of the surface effects \citep[see also][]{2005MNRAS.356..671O, 2005A&A...434..665R}. Such a solution has been used, for example, by \citet{2013ApJ...769..141S} on two \emph{Kepler} targets. In such cases, direct modelling (including surface effect corrections) led to several models with frequencies compatible with the observations but significantly different internal structure. Using ratios pins down the number of plausible models, reducing the errors on radius, mass, and age down to 2\%, 4\%, and 10\%. A large sample of 66 solar-like stars observed by \emph{Kepler} has been homogeneously analysed by \citet{2017ApJ...835..172L} and modelled by \citet{2017ApJ...835..173S} in the LEGACY project. The average uncertainties on the model parameters for this sample are again 2\%, 4\%, and 10\% for radius, mass, and age.

\subsection{Constraints on the internal structure}
\label{ssec:intern_struct}

Beyond global parameters, asteroseismology allow us to derive constraints on the internal structure that can be compared with stellar models. Acoustic glitches provide interesting diagnosis. An acoustic glitch is a perturbation in the frequencies due to the presence of discontinuities (or rapid variations) in sound speed (or its derivatives) in the stellar structure. Such discontinuities generate partial reflections of propagating waves, resulting in a change in the resonant frequencies of a cavity (see Gough 1990 for a comprehensive toy model). For the Sun and stars, the impact of glitches have been theoretically studied since late 1980s \citep[see, e.g.,][]{1988IAUS..123..151V,1990LNP...367..283G}. It slightly modifies mode frequencies with an oscillatory behaviour such as
\begin{equation}
 \delta\nu_{n,\ell} \propto \sin (4\pi\tau_{\rm{g}}\nu_{n,\ell} +\phi_{\rm{g}}),
\end{equation}
where $\tau_{\rm{g}}$ is the acoustic depth of the glitch, i.e. the travel time of a sound wave to reach the glitch from the surface:
\begin{equation}
 \tau_{\rm{g}} = \int_{r_{\rm{g}}}^R \frac{{\rm{d}}r}{c},
\end{equation}
$r_{\rm{g}}$ being the radial position of the glitch and $\phi_{\rm{g}}$ a phase shift. For low degree modes, this phase $\phi_{\rm{g}}$ is the same for all modes. The amplitude and the envelope of the perturbation change with the nature and the abruptness of the transition (the shaper, the larger). 

Sharp variations of the sound speed (or derivative) occur in stars where (i) the adiabatic exponent $\Gamma_1$ quickly decreases, which occurs in the ionisation of abundant elements (H, He{\,\sc I}, He{\,\sc II}) (ii) when the thermal gradient changes, especially when energy transport processes change from radiative to convective (or the opposite).
Measuring glitches allow us to probe (i) the helium ionisation zone, possibly giving constraints on the helium abundance in the envelope, (ii) the position and the structure of the base of the convective envelope (BCE). Glitches generated by convective cores are too deep -- in terms of acoustic depth -- to be seen as oscillations in frequency. We will discuss convective cores later on.

To search for these oscillations we can look directly at the frequency \citep[e.g.][]{1994A&A...283..247M,2000MNRAS.316..165M}. It is also possible to look for them in the surface phase shift or its derivatives 
\citep[e.g.][]{1994A&A...290..845L,1994MNRAS.268..880R,1997ApJ...480..794L,2001MNRAS.322...85R}. They are also visible in the large separation, or even better in second differences,
\begin{equation}
 \delta_2\nu_{n,\ell} = \nu_{n+1,\ell}-2 \nu_{n+1,\ell} + \nu_{n-1,\ell},
\end{equation}
or even higher differences \citep[e.g.][]{1994MNRAS.267..209B,1997MNRAS.288..572B, 2001A&A...368L...8M, 2004A&A...423.1051B}.
The second differences should almost vanish without glitches. Variations in $\delta_2\nu_{n,\ell}$ can be modelled as the sum of a small trend and two glitches:
\begin{equation}
 \delta_2\nu_{n,\ell} = a_0 + a_1\nu + b_0\nu {\rm{e}}^{-b_1\nu^2} \sin (4\pi\tau_{\rm He}\nu_{n,\ell} +\phi_{\rm He}) + \frac{c}{\nu^2} \sin (4\pi\tau_{\rm BCE}\nu_{n,\ell} +\phi_{\rm BCE}). \label{eq:d2nugl}
\end{equation}
This parametrization of the glitch envelopes is adapted from \citet{2007MNRAS.375..861H}.

Glitches are also visible in small separations and ratios, but oscillation periods of signatures are not linked to the acoustic depth but to the acoustic radius (time for a sound wave to propagate to the glitch from the stellar centre). As a consequence, the helium ionisation signature almost disappears in these variables \citep[for more details, see][]{2003A&A...411..215R,2005A&A...434..665R,2009A&A...493..185R}.

\begin{figure}[!htp]
\begin{center}
\includegraphics[width=\textwidth]{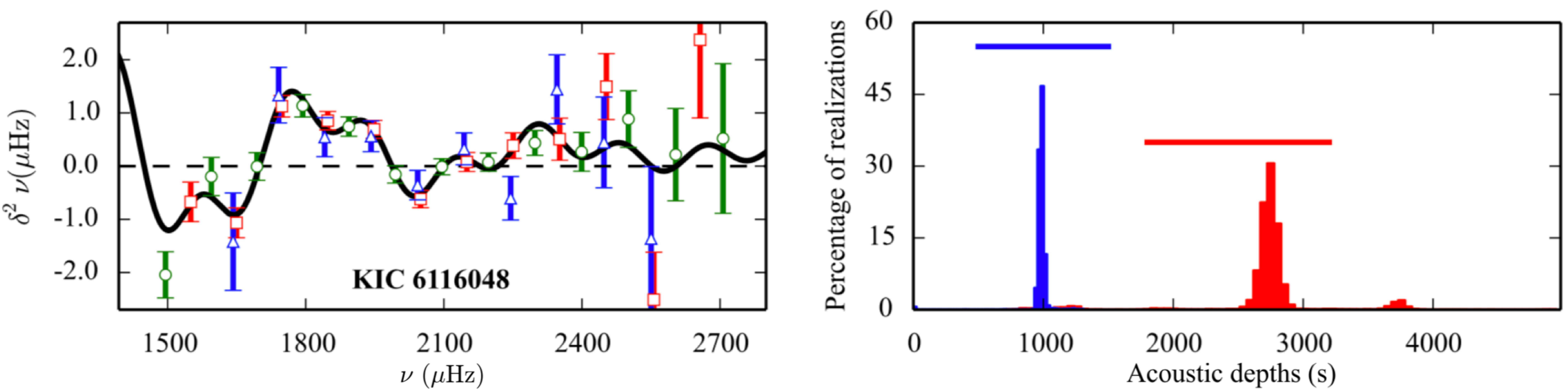}
\end{center}
\caption{\label{fig:glitch}Left panel shows second differences for $\ell=0,1,2$ modes (red, blue, green) observed for KIC 6116048. Solid line show the best fit Eq.~\ref{eq:d2nugl}. Right panel show the distribution of the acoustic depths for the BCE (red) and He{\,\sc II} (blue)  \citep[][]{2017ApJ...837...47V}.}
\end{figure}

Measurement of glitches requires high precision on individual mode frequencies. In CoRoT observations of HD~52265, a BCE glitch was visible \citep{2011A&A...530A..97B} and has been analysed by \citet{2012A&A...544L..13L}. Their analysis suggests that the penetrative distance of the convective envelope in the radiative zone reaches 0.95 $H_{\rm{p}}$ (6\% of the total radius), which is significantly larger than what is found for the Sun. Measuring glitches in F-type stars is challenging because uncertainties of frequencies are high due to mode broadening (Eq.~\ref{eq:err_lib}), reinforced by the overlap of $\ell=0$ and 2 modes. However, \citet{2017MNRAS.466.2123B} succeeded in measuring the position of helium ionisation zone in HD~49933 using CoRoT data.
Glitches have also been intensively studied in \emph{Kepler} solar-type stars. A first work on 19 stars with one-year observations has been carried out by \citet{2014ApJ...782...18M}, then extended to the LEGACY sample (66 stars observed for up to 4 years) by \citet{2017ApJ...837...47V}. Figure \ref{fig:glitch} shows an example of second differences fitted with oscillatory components due to BCE and He{\,\sc{II}} ionisation zone. Helium glitches can be robustly measured in many stars, whereas BCE is more difficult to measure for super-solar mass stars. These measurements can now be used to constrain properties of the helium ionisation zone and helium abundance in the envelope.

The question of the extent of convective cores in solar-type stars is still
debated. It depends on the mixing processes in the core and also on the details of the nuclear reaction rates. A recent study of $\alpha$ Cen A combining observations in astrometry, spectroscopy, interferometry and ground-based seismology shows that the nature of its core is still uncertain \citep{2016MNRAS.460.1254B}.
Signatures of convective cores can be found in seismic variables. Small separations $\delta\nu_{01}$ and $\delta\nu_{10}$ and the ratio of small to large separation $r_{01}$ and $r_{10}$ are very sensitive to the core structure \citep[e.g.][]{2005A&A...432..225P,2005AcA....55..177P,2010A&A...514A..31D,2011A&A...529A..63S,2011A&A...529A..10C}. Convective cores create glitches with very long periods in $\delta\nu_{01},\delta\nu_{10}$, mainly visible as a slope. As opposed to the BCE position directly deduced from observations, interpreting the slope as a core extent is not fully model independent. Indeed, several effects contribute to the slope: the amount of hydrogen in the core (no matter whether a convective core is present or not), the size of the convective core and the sharpness of the discontinuities. Using this idea \citet{2016A&A...589A..93D} have been able to measure the extents of convective cores for eight main-sequence stars. From them, they were able to calibrate the quantity of extra mixing needed in the core to reproduce the seismic observations.

In the case of helioseismology, fine constraints on the internal structure were achieved by inversion techniques. This was possible thanks to the observations of many intermediate- and high-degree modes. Structure inversions for other stars, based only on low-degree modes without strong external constraints on mass and radius are challenging. In this context, \citet{2012A&A...539A..63R} proposed an inverse method to accurately (0.5\% accuracy) estimate stellar mean densities. Inversion efforts have been pursued for several \emph{Kepler} targets, especially 16~Cyg A and B by \citet{2016A&A...585A.109B,2016A&A...596A..73B}. Doing so, they reduced the uncertainties to 2\% on mass, 1\% on radius, and 3\% on the age for 16~Cyg~A. New inversion methods to constrain convective regions have been proposed by \citet{2018A&A...609A..95B} and should be tested soon on \emph{Kepler} observations. Finally, a new method called ``Inversion for agreement'' has been proposed by \cite{2017ApJ...851...80B}. This method takes into account imprecise estimates of stellar mass and radius as well as the relatively small amount of modes available. Because the result is independent of models, it can be used to test their inferences. Thus, the results obtained on 16~Cyg~A and B showed that the core sound speed in both stars exceeds that of the models.

\newpage

\section{Stellar rotation}
One of the main results obtained with helioseismology is the accurate determination of the solar differential rotation in the convective zone \citep[e.g.][]{ThoToo1996,ThoJCD2003,lrsp-2009-1}, as well as the nearly constant rotation rate in the radiative interior down to $\approx$ 0.2 ${\rm{R}}_\odot$, where the measured low-degree p modes do not provide enough sensitivity below this radius \citep[e.g.][]{1999MNRAS.308..405C,2004SoPh..220..269G,2008SoPh..251..119G}.  Gravity modes are needed to properly constrain the rotation within the core \citep[e.g.][]{2008A&A...484..517M}. However, although the detection of individual g modes in the Sun is still controversial \citep[e.g.][]{2010A&ARv..18..197A,2018SoPh..293...95S} the latest results obtained considering these modes or their period spacing suggest a faster rotation rate in the core \citep{2007Sci...316.1591G,2008AN....329..476G,2011JPhCS.271a2046G,2017A&A...604A..40F}. 

When extending the analysis of internal rotation to other stars, the precision and the extension of the region to be explored depends on our ability to probe their inner regions. In MS stars, only pure acoustic modes have been unambiguously characterized so far and, therefore, only the outer convective zone and the outer part of the inner radiative zone can be probed by asteroseismology. Only when stars enter the sub-giant region, can mixed modes be measured \citep[e.g.][]{2010A&A...515A..87D,2012ApJ...745L..33B} and rotation of the inner zones and the core can be obtained \citep[e.g.][]{2012ApJ...756...19D,2014A&A...564A..27D}.

\begin{figure}[!htp]
\begin{center}
\includegraphics[width=0.65\textwidth]{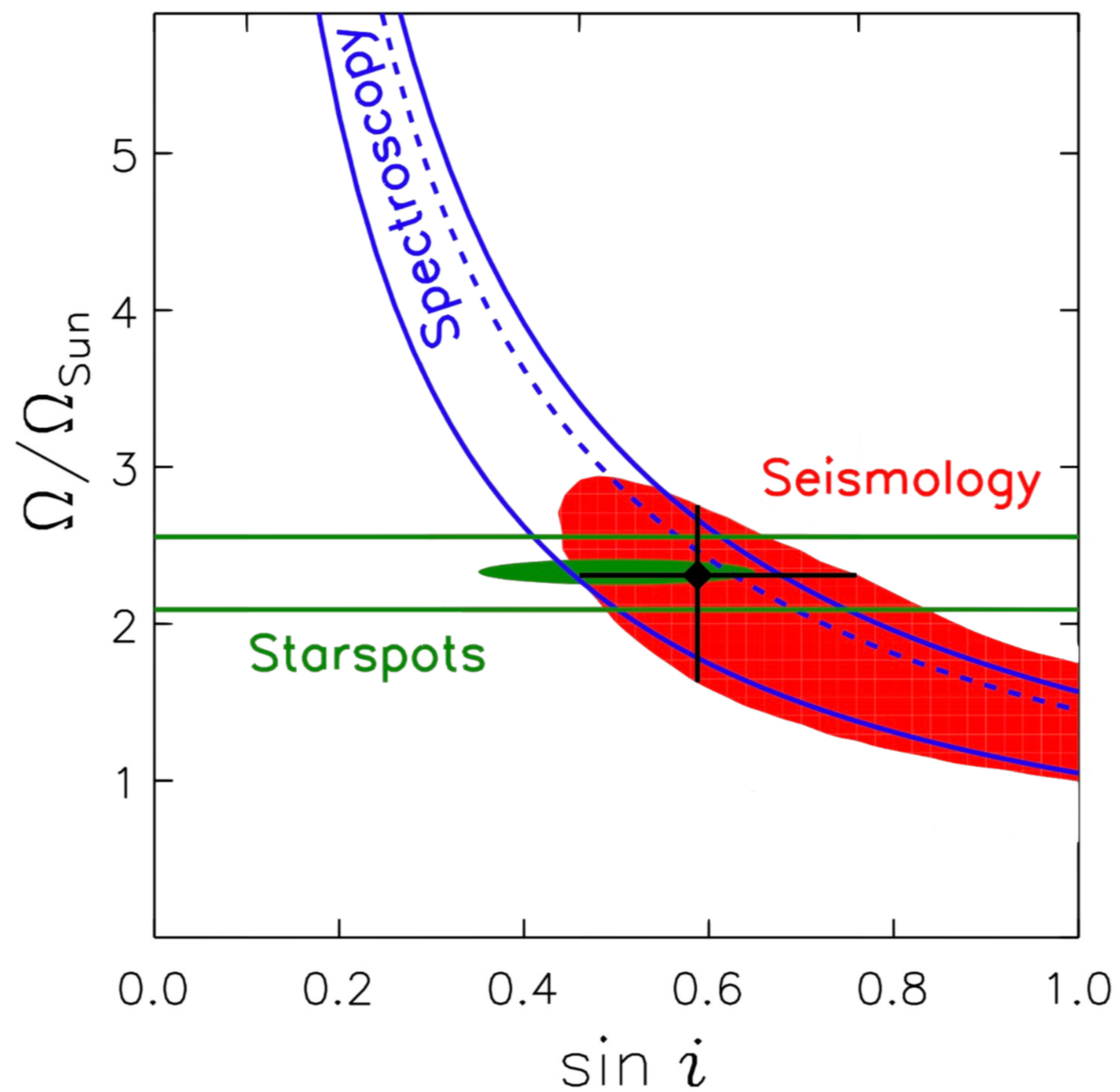}
\end{center}
\caption[Inference of the rotation rate of HD~52265]{\label{FIG_rot_Gizon} Angular velocity of HD~52265, ${\rm{\Omega}}$, normalized by the angular velocity of the Sun (${\rm{\Omega}}_{\rm{Sun}}/2\pi$=0.424 $\mu$Hz) as a function of the sine of the inclination of the stellar rotation axis to the line of sight, $\sin i$. 
The black diamond with error bars is the best-fit seismic result. \modif{The red shaded region represent the 1-$\sigma$ seismic constraint in both parameters}. The two horizontal lines represent the angular velocity obtained form the analysis of the low-frequency range of the power spectrum. The green ellipse is the 1-$\sigma$ result obtained from the starspot modeling of the light curve. Finally,  the dashed (observations) and
the solid (1-$\sigma$ errors) blue curves are the constraints obtained from ground-based spectroscopy. Figure adapted from \cite{2013PNAS..11013267G}.}
\end{figure}

The first unambiguous determination of the average internal rotation and the rotation inclination axis of a MS solar-like star using asteroseismic techniques was performed by \cite{2013PNAS..11013267G} on the CoRoT target HD~52265. To prove the validity of their measurements, \cite{2013PNAS..11013267G}  compared the asteroseismic rotation with the spectroscopic value, the surface rotation period, and starspot modeling obtained from the analysis of this CoRoT light curve. The result is shown in Fig.~\ref{FIG_rot_Gizon}.

This example illustrates all of the information provided by the study of continuous high-precision photometry. On the one hand, a direct determination of the rotation period can be obtained by the analysis of the light curve, either in the time domain, or by studying the low-frequency part of the temporal power spectrum. In addition, starspot modeling can also provide additional information such as the rotation inclination angle \cite[e.g.][]{2009A&A...506..245M,2014A&A...564A..50L}. On the other hand, seismology yields a weighted average of the internal rotation heavily biased towards the surface during the MS, as well as the rotation inclination angle. In the next two sections, we will describe in more detail these two types of analyses. 

\label{Sec:rotation}

\subsection{Photospheric rotation from the study of the photometric light curves}

The crossing of cool spots over the visible disk of a star produces a modulation in the photometric signal which is proportional to the rotation period of the star, $P_{\rm{rot}}$, at the latitudes where spots and active regions exist. Several works have been dedicated to the study of the extraction of these rotation periods for CoRoT, \emph{Kepler} and the K2 missions. Thus, in the last decade, it has been possible to retrieve the rotation periods of thousands of stars \cite[e.g.][]{2013MNRAS.432.1203M,2013A&A...557L..10N,2013A&A...560A...4R,2013MNRAS.436.1883W,2014ApJS..211...24M,2015A&A...582A..85L}. 


There are several different -- but complementary -- ways to extract the information on the rotation period from light curves. In this section, we do not pretend to provide an exhaustive review of all of the techniques and results on this topic; we focus mostly on what has been done related to asteroseismic studies of MS solar-like stars. 

\begin{figure}[!htp]
\begin{center}
\includegraphics[width=1.0\textwidth]{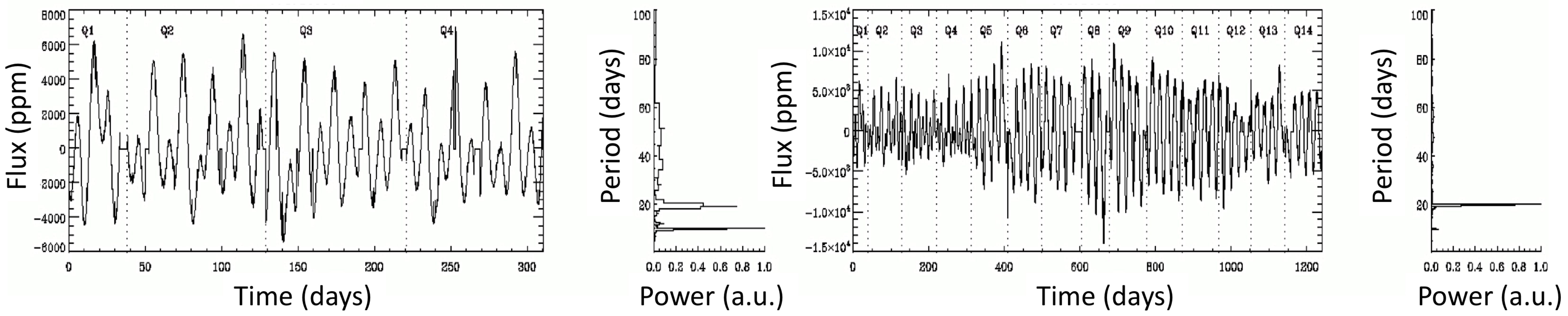}
\end{center}
\caption[Example of a star where the second harmonic is the tallest peak in the PSD]{\label{FIG_rota_halfperiod} Light curves and associated normalized period-power spectrum of KIC~4918333. In the left panel, only the first 310 days are analyzed. On the right, the full light curve is presented.}
\end{figure}

The careful study of the low-frequency part of the power spectrum can be used to extract the rotation period by selecting the highest peak in this frequency region \citep[e.g.][]{2009A&A...506...51B,2011A&A...534A...6C,2013A&A...557L..10N}. However, sometimes the highest peak can be the second or even the third harmonic of the rotation period instead of the first corresponding to the true rotation period. An example is given in Fig.~\ref{FIG_rota_halfperiod}, where the light curve and the longer periods of the period--power spectrum of KIC~4918333 is shown. On the left-hand panel, the first 310 days of the light curve are analyzed and the tallest peak corresponding to a rotation period of 9.8 $\pm$ 0.8 days is shown. A second peak is also visible at a period of 19.5 $\pm$ 1.2 days but with a smaller amplitude. When this analysis is extended to 1250 days, the period of 19.5 $\pm$ 1.2 days is the most prominent one without ambiguity. Hence, when analyzing the rotation period from the study of the power spectrum, it is important to check any possible signal at twice the period corresponding to the highest peak.

This situation occurs when two active regions develop in stars with approximately $180^{\rm{\circ}}$ separation in longitude. This is clearly visible in the light curve shown in the left panel of Fig.~\ref{FIG_rota_halfperiod}. Two modulations of different amplitudes (one large and one small) are separated by approximately 10 days. As a consequence, the largest peak in the spectrum is not the $\approx$ 20-day periodicity but the $\approx$ 10 days. 

\begin{figure}[!htp]
\begin{center}
\includegraphics[width=1.0\textwidth]{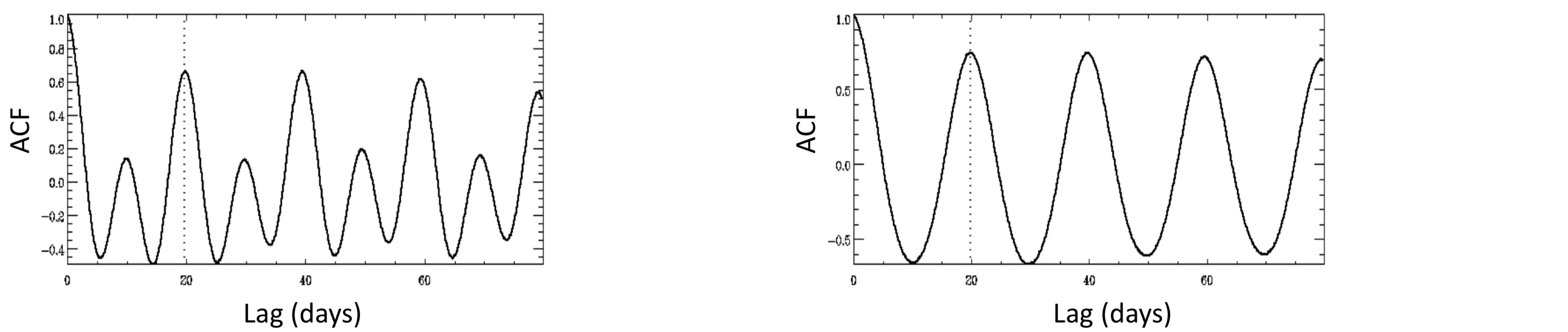}
\end{center}
\caption[Wavelets analysis of KIC~4918333]{\label{FIG_ACF_half_rot} ACF analysis of the light curves shown in Fig.~\ref{FIG_rota_halfperiod}.}
\end{figure}

To avoid this possible ambiguity, \cite{2013MNRAS.432.1203M} used the autocorrelation function, ACF, as a robust method to infer the rotation period directly from the \emph{Kepler} light curves in the time domain. They showed that in stars such as KIC~4918333, the highest peak in the ACF was the first harmonic and not the second and thus the retrieved rotation periods were robust. In Fig.~\ref{FIG_ACF_half_rot} the ACFs of the analysis of the first 310 and 1250 days of KIC~4918333 are shown. With this methodology -- contrary to the direct analysis of the power spectrum -- the highest peak is the one at $\approx$ 20 days for both length of the time series.  

\begin{figure}[!htp]
\begin{center}
\includegraphics[width=1.0\textwidth]{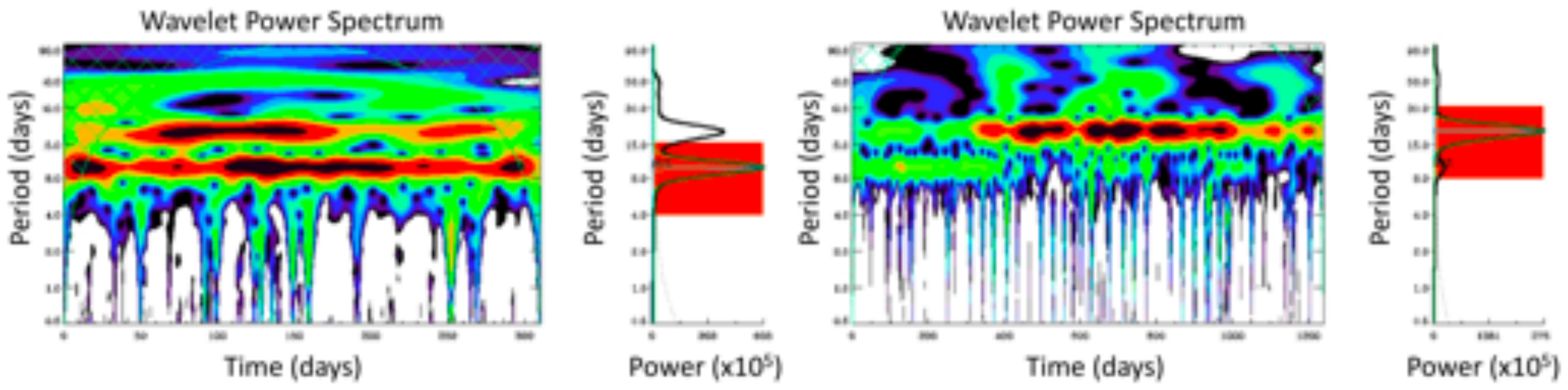}
\end{center}
\caption[time--period analysis of KIC~4918333]{\label{FIG_wav_period_rot} time--period analysis of the light curves showed in Fig.~\ref{FIG_rota_halfperiod} computed using a Morlet wavelet between 0.5 and 100 days on a logarithmic scale following \cite{1998BAMS...79...61T}. The green-crossed area is the cone of influence corresponding to the region where it is not possible to obtain reliable results. As in Fig.~\ref{FIG_rota_halfperiod}, the global wavelet power spectrum is shown in the right of each of the time--period panels. The red-shaded regions depict the tallest peak in this diagram. The green line represents the best Lorentzian fit to this peak. The dotted line represent the 95$\%$ confidence level.}
\end{figure}

Because the photometric light curves are not free from instrumental effects, time--period diagrams have been used since the CoRoT mission to look for the rotation periods in stars. \modifeng{In the mean time, we also checked} any possible ``glitches'' in the signal that could produce a spike in the low-frequency part of the spectrum that could perturb the determination of the rotation \citep[e.g.][]{2009A&A...506...41G,2010A&A...511A..46M}.  This way, segments where instabilities appear in the light curve could be easily inspected and removed from the analysis. To extract the rotation period from the time--period diagrams, a projection on the period axis is done, and a fitting of the highest peak is then performed as if it was a normal power spectrum. The application of this time--period methodology to KIC~4918333 is shown in Fig.~\ref{FIG_wav_period_rot}.

Since the different methods used to calibrate the data can filter the stellar signal in different ways, while leaving other instrumental signals in the final light curves, it is recommended to use several types of data calibrations when studying the rotation of a star \modif{\citep[e.g.][]{2013ASPC..479..129G,2014A&A...572A..34G,2016MNRAS.456..119C,2016JSWSC...6A..38B}}. It has also been shown that each method to extract the rotation (e.g. ACF, time--period diagrams, the direct analysis of the low-frequency part of the spectrum) works better in some circumstances or for some type of stars. Hence, the most reliable procedures to retrieve the rotation periods are those that combine different calibration procedures and analysis techniques \citep{2015MNRAS.450.3211A}.

\modifeng{The average rotation is not the only quantity that can be inferred} from the analysis of the light curves. Differential rotation in latitude ${\rm{\Delta\Omega}}$ can also be obtained in some cases \citep[e.g.][]{2012A&A...543A.146F,2013A&A...557A..11R,2014A&A...564A..50L,2015A&A...583A..65R}, by directly studying the number and structure of the peaks in the low-frequency part of the spectrum or by performing starspot modeling \citep[e.g.][]{2006PASP..118.1351C,2007AN....328.1037F,2013A&A...557L..10N,2014ascl.soft12002W,2016A&A...592A.140L}. Thanks to these analyses, general trends have been found. For example, the dependence of ${\rm{\Delta\Omega}}$ with the rotation period is weak and it slightly increases with $T_{\rm{eff}}$ in the range 3500 to 6000 K \citep{2013A&A...560A...4R}, while the relative differential rotation, ${\rm \Delta\Omega/\Omega}$, increases with the rotation period \citep[][]{2015A&A...583A..65R}. In addition, \cite{2015A&A...576A..15R} were able to discriminate solar and antisolar differential rotation (i.e., to identify the sign of the differential rotation at the stellar surface) using peak-height ratios of the first harmonics of the differential rotation peaks. However, a theoretical and numerical study by \cite{2017A&A...599A...1S} showed that the peak-height ratios are essentially a function of  the fraction of time the spots are visible. This time is related to how strongly the spot modulation follows a sinusoidal form. Hence, depending on the rotation inclination angle with respect to the line of sight and on the location of the spots, the inferred sign of the differential rotation can be wrong. 

Signatures of the change in latitude of the spots have also been measured for example in KIC~3733735 \citep{2014A&A...562A.124M}. During the periods with low activity, the mean rotation period is $\sim$ 3 days, while during the periods of maximum activity the rotation period decreases to 2.54 days (see Fig.~\ref{mag_3733735}). Similarly as for the Sun, this behavior can be explained by the existence of surface differential rotation and a change in the latitude where the spots emerge as the magnetic cycle progresses.

\subsection{Internal rotation through asteroseismic measurements}

\begin{figure}[!htp]
\begin{center}
\includegraphics[width=1.0\textwidth]{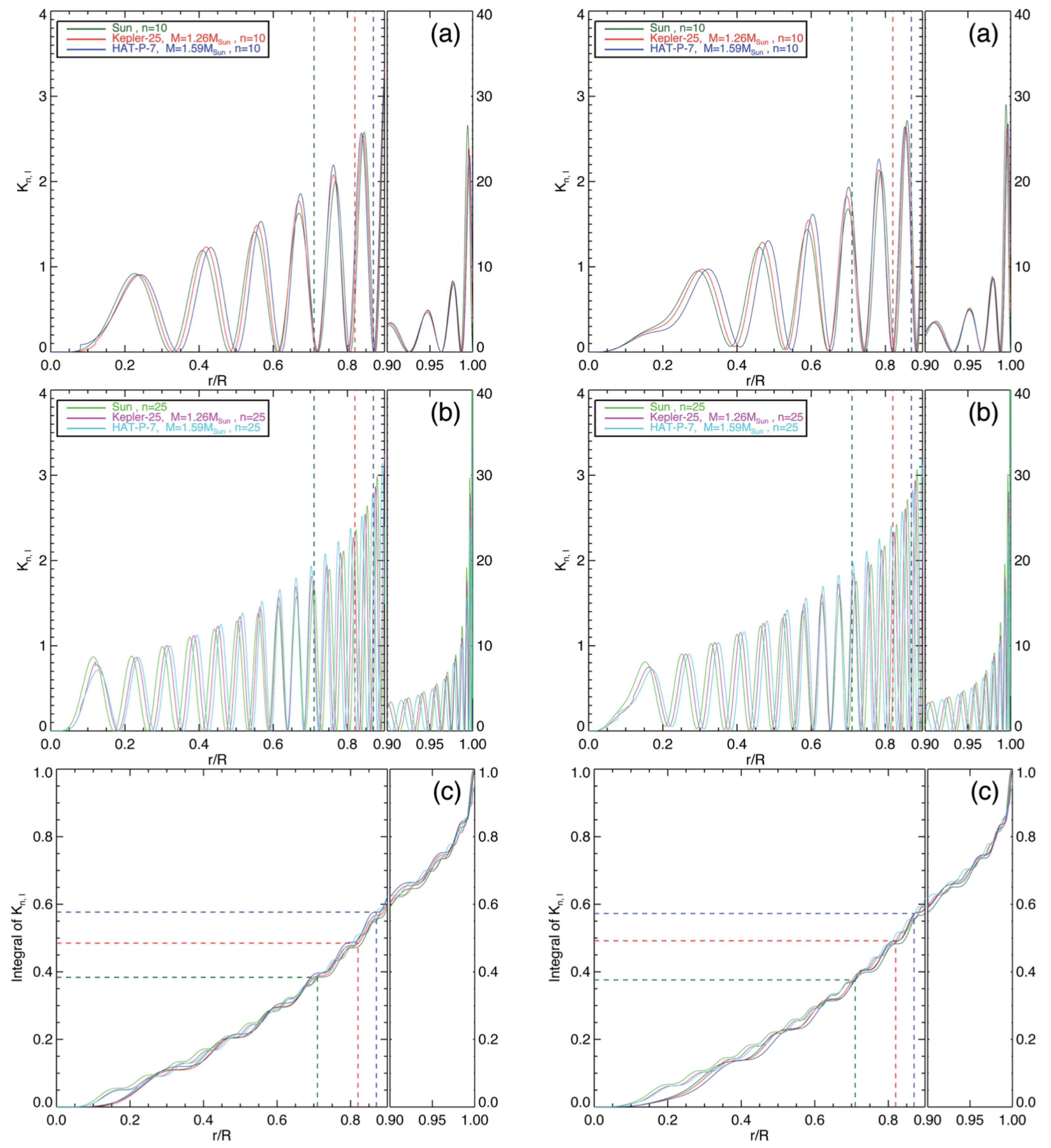}
\end{center}
\caption[Rotational Kernels of three main-sequence stars]{\label{FIG_rot_kernels}  Rotational kernels of a dipolar (left panels a,b) and quadrupolar modes (right panels  a,b)  with $n$=10 and $n$=25 and their integral over the normalized radius (c) for three host stars: Kepler-25 (red lines), HAT-P7 (blue lines) and the Sun (green lines). The horizontal scale has an expansion between 0.9 and 1 $r/$R$_\odot$ to magnify the external layers. The dotted lines indicate the location of the base of the convective zone of each star. Figure from \cite{2015MNRAS.452.2654B}.}
\end{figure}

As we have already said in this review, only acoustic modes have been characterized in main-sequence solar-like dwarfs so far. Therefore, the information that one can obtain from seismology comes from the careful analysis of acoustic modes. Unfortunately, although low-degree p modes penetrate deep in the stellar interior, they spend only a small fraction of their time in the deep interior and the amount of information that they can provide from these regions is small. This is illustrated in Fig.~\ref{FIG_rot_kernels} where we show the rotational kernels, $K_{n,\ell} $, of the dipolar  and quadripolar modes of radial orders $n=10$ and $n=25$ for three planet-hosting stars: Kepler-25 \citep[KIC~4349452, M=1.26 $\pm$ 0.03 M$_\odot$,][]{2014PASJ...66...94B}, HAT-P7 \citep[KIC~10666592, M$\sim$1.59 $\pm$ 0.03 M$_\odot$,][]{2014PASJ...66...94B} and the Sun. The two radial orders $n=10$ and 25 cover the typical observational range for theses stars. In spite of the different masses of the three stars and the different position of the base of the convective zone, the kernels of these modes are nearly identical \citep[see also][]{2014ApJ...790..121L}. However, \modifeng{the position of the base of the convective zone depends on the mass of the star.} Therefore, more massive stars have shallower convective zones and thus, they have a larger contribution from the radiative zone to the average internal rotation rate. It is, however, possible to extract some general properties. These kernels have larger amplitudes near the surface. They are also denser in these outer regions implying that the waves spend more time there than in the inner layers. As a consequence, the sensitivity to the rotation is larger towards the surface of the star. Moreover, the kernels in the radiative zone are almost linear functions of the radius above $\sim$~0.15~R$_\odot$ up to the base of the convective zone. Therefore, each observed mode probes the radiative zone uniformly.

The first consequence is that the rotation rate extracted from seismology and the rotation period extracted from the surface (either from the study of the photometric variability or from spectroscopy once the inclination angle is known) should be similar as already explained when described the first asteroseismic detection of the rotation in HD~52265 (see Fig~\ref{FIG_rot_Gizon}). Interestingly, any significant difference between the surface and the asteroseismic rotation rate should indicate the existence of differential rotation, either from the surface or from the external convective zone inside the stars. \citet{2013PNAS..11013267G} found that both determinations of the rotation agreed within one sigma in all cases, suggesting that the differential rotation should be weak in these stars. 

A comparison of the asteroseismic rotation rate \citep[from][]{2014A&A...568L..12N} and their surface rotation period was performed by \cite{2015A&A...582A..10N} for five \emph{Kepler} targets with masses in the range 1.02 to 1.2 M$_\odot$. Because of the large errors in any inference of the rotation period, the asteroseismic rotation rate can be considered a good proxy of the surface rotation only under the assumption that there is no large differential rotation neither at the surface nor in the convective envelope. Hence, asteroseismic rotation rates can be used for example in  gyrochronology studies when no other precise rotation rates are available \citep[e.g.][]{2015MNRAS.446.2959D,2015A&A...582A..10N}.

\begin{figure}[!htp]
\begin{center}
\includegraphics[width=0.6\textwidth]{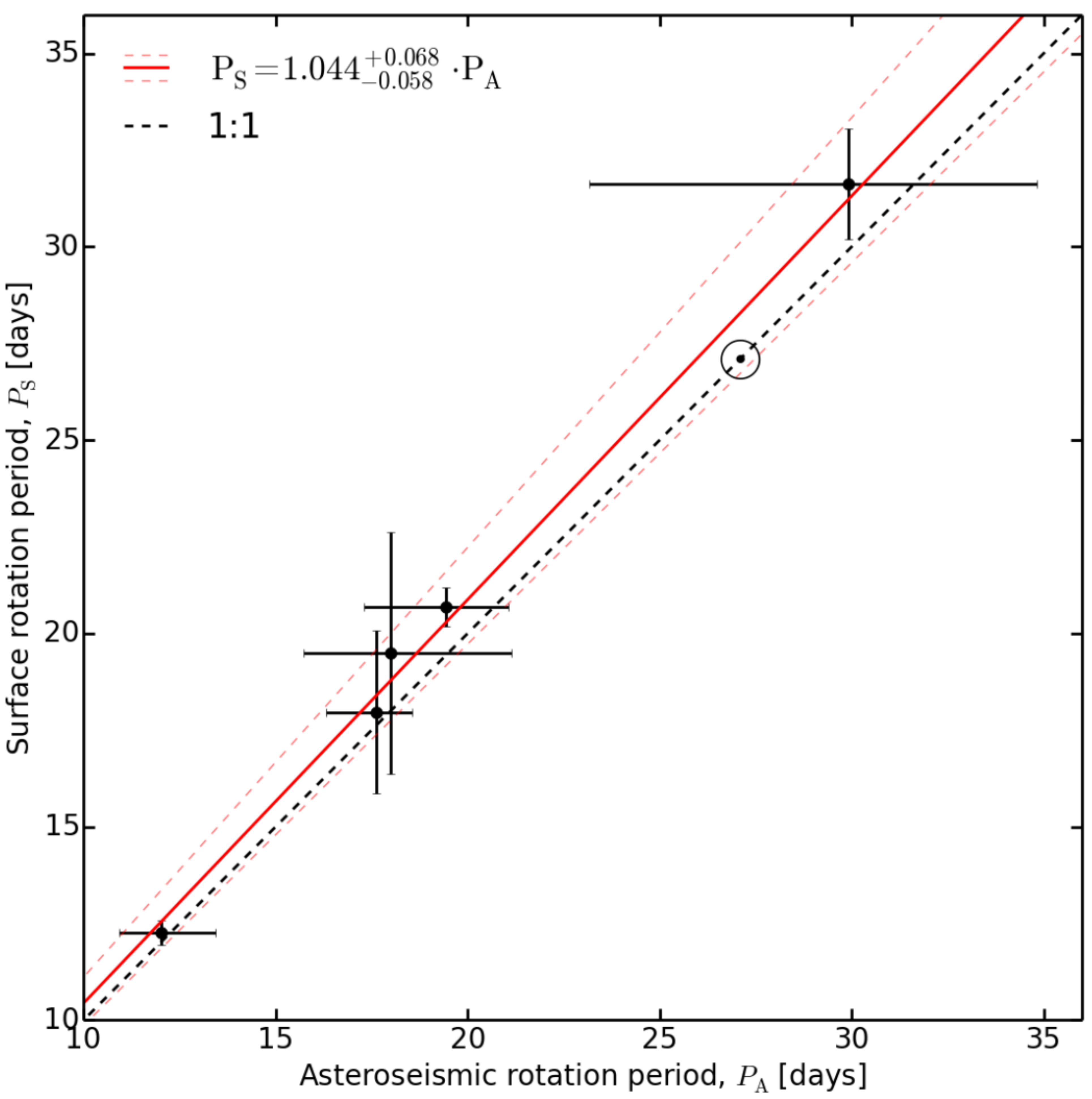}
\end{center}
\caption[Rotation period extracted rom asteroseismology and from the photometric variability]{\label{FIG_rota_Nielsen}Rotation period extracted from asteroseismology and from the study of the photospheric variability. The dashed-black line represents the 1-to-1 line. The red continuous line represents the best fit to the data (excluding the Sun represented by $\odot$) including the errors of the slope (gray dashed lines). Figure from \cite{2015A&A...582A..10N}.}
\end{figure}

Based on similar studies of the rotational kernels but using inversion methods, \cite{2016A&A...586A..24S} suggested that it could be interesting to use many stars in order to reduce the observational errors and be able to, for example, constrain the sign of the radial differential rotation. Moreover, it has been demonstrated using an ``ensemble-fit'' of 15 stars across the main sequence, that it would be possible to distinguish between solid rotation and radial differential rotation of around 200 nHz using observable splittings of angular degrees 1 and 2 \citep{2016A&A...586A..79S}. 

Another simplified, complementary approach to rotational inversion was proposed by \cite{2015MNRAS.452.2654B} taking advantage of the relative simplicity of the stellar interiors of MS solar-like dwarfs. In principle, it is possible to consider that the region probed by the measurable acoustic modes is composed of two different zones: the outer convective and the inner radiative zone. Therefore, the rotational splitting $\delta \nu_{n,l}$ can be expressed as a weighted average of the rotation rate, $f_{\rm{rad}}$, $f_{\rm{conv}}$, in each zone as follows:

\begin{equation}
\label{eq_rotation_ot}
\delta \nu_{n,\ell} \sim I_{\rm{rad}} f_{\rm{rad}} + I_{\rm{conv}} f_{\rm{conv}} \; ,
\end{equation}
where $I_{\rm{rad}}$ and $I_{\rm{conv}}$ are the integrals of the rotational kernels in the radiative and convective zones defined as:
\begin{equation}
I_{\rm{rad}} =  \int_{0}^{r_{\rm{bcz}}} K_{n,\ell} (r) {\rm{dr}} \; , 
\end{equation}
\begin{equation}
I_{\rm{conv}} =  \int_{r_{\rm{bcz}}}^{R} K_{n,\ell} (r) {\rm{dr}} \; ,
\end{equation}
with $I_{\rm{rad}} + I_{\rm{conv}} = 1$. The position of the base of the convective zone is denoted by $r_{\rm{bcz}}$ and $R$ is the radius of the star.

In this simplified model, assuming that $f_{\rm{rad}}$ is a good approximation of the rotation rate in the radiative zone, it is necessary to suppose that: a) the two zones rotate uniformly with different rates (about the same axis), b) $f_{\rm{conv}}$ is approximately equal to the surface rotation (as in the solar case), c) rotational splittings remain nearly constant over the observed range of modes. Hence, the average rotational splitting $<\delta \nu_{n,\ell}>$, can be used as a representative value of the seismically measured internal rotation rate, $f_{\rm{seis}}$. This last assumption is validated because the integrals of the observed mode kernels shown in Fig.~\ref{FIG_rot_kernels} are nearly identical.
Therefore, it is possible to express Eq.~\ref{eq_rotation_ot} as follows:
\begin{equation}
\label{eq_rot_ot_final}
<f_{\rm{rad}}> = f_{\rm{surf}} + \frac{f_{\rm{seis}}-f_{\rm{surf}}}{<I_{\rm{rad}} >}
\end{equation}
where $< >$ denotes that the corresponding parameter is averaged over the observed range of modes.  

Knowing the kernel integrals from a model and the average internal rotation rate from the seismic splittings, it is necessary to have a good determination of the surface rotation rate in order to infer the rotation in the internal radiative region with Eq.~\ref{eq_rot_ot_final}. Two possibilities are available: on the one hand, the surface rotation obtained from photometric variability can be different form the average surface rotation because it is weighted towards the active latitudes ($30^{\rm{\circ}}$ in the solar case). On the other hand, the spectroscopic value is averaging the full visible disk and it can be considered a better approximation. However, in this latter case, what is measured is the projected velocity into the line of sight velocity:
\begin{equation}
v \sin i = 2 \pi R f_{\rm{surf}} \sin i	\; .
\end{equation}

Thus, to extract $f_{\rm{surf}}$ it is necessary to know the rotation inclination angle, $i$, from seismology and the radius of the star from the same models used to compute the rotational kernels.

\begin{figure}[!htp]
\begin{center}
\includegraphics[width=0.6\textwidth]{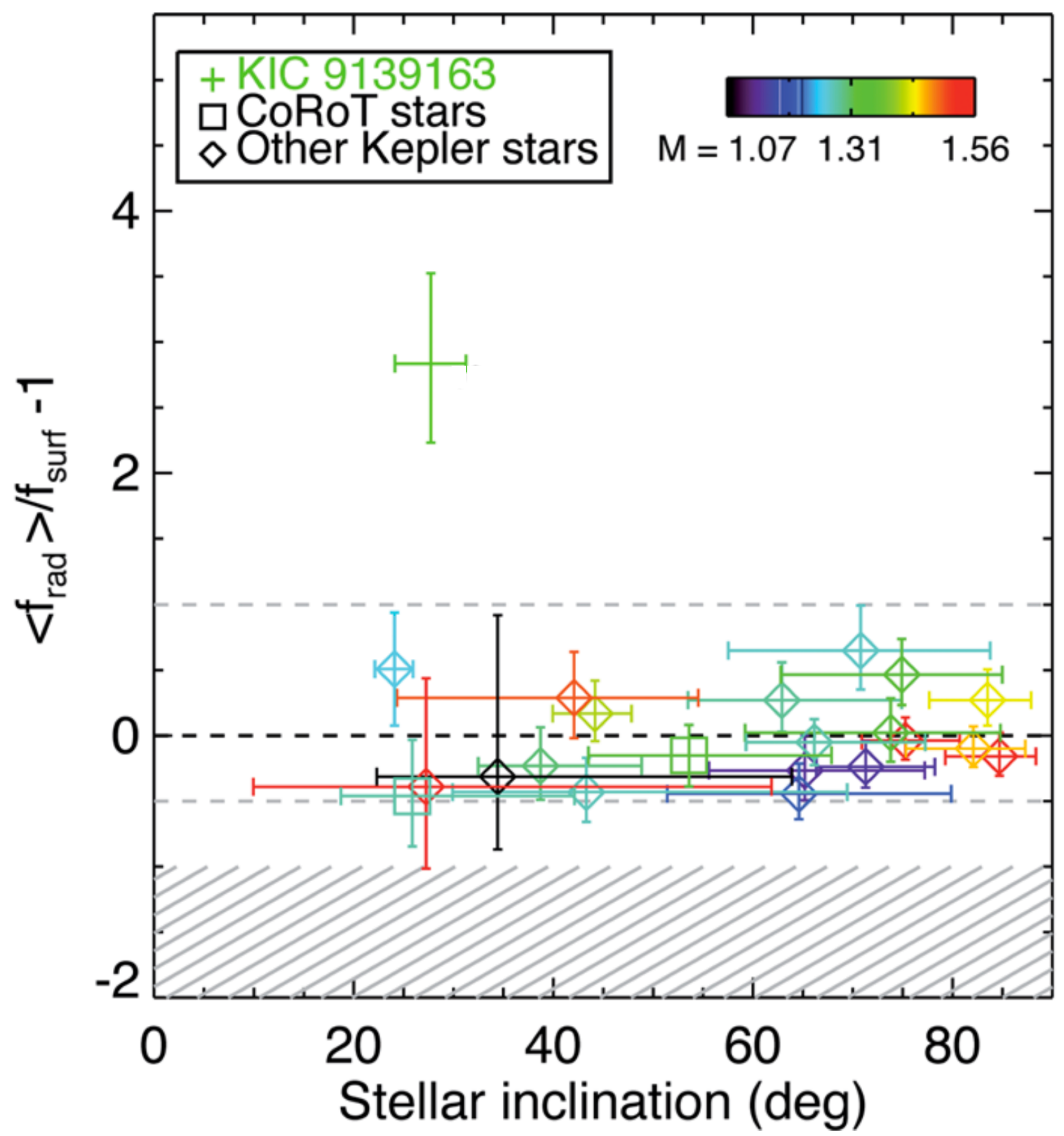}
\end{center}
\caption[Ratio of the radiative rotation to the surface rotation]{\label{FIG_inte_rot_Oth} Ratio of the average radial differential rotation in the radiative zone obtained from Eq.~\ref{eq_rot_ot_final} and the surface rotation (obtained from spectroscopy) as a function of the stellar inclination angle (obtained from seismology).  Colors represent the stellar mass. Grey lines indicate a differential rotation of a factor of 2. The dashed area indicates a negative rotation rate in the interior. Figure adapted from \cite{2015MNRAS.452.2654B}.}
\end{figure}

Using 22 stars observed by CoRoT and \emph{Kepler} -- of masses in a range  $\sim$1.07 to $\sim$ 1.56 M$_\odot$ for which $f_{\rm{surf}}$ is available from spectroscopy and photometric variability -- \cite{2015MNRAS.452.2654B} deduced that nearly uniform internal rotation (between the radiative zone and the surface) is common in other solar-like stars from masses up to $\sim$ 1.5 M$_\odot$. Thus, the Sun is not an isolated case in stellar evolution and efficient angular momentum transport mechanisms are required in the main sequence to achieve such rotation profile \citep[e.g.][]{2013LNP...865...23M,2013A&A...549A..74M}. The ratio $< f_{\rm{rad}}>/f_{\rm{surf}} -1$ is shown in Fig.~\ref{FIG_inte_rot_Oth}.

It is important to note that only one star, KIC~9139163, significantly departed from the one-to-one relation shown in Fig.~\ref{FIG_inte_rot_Oth}, offering a scenario where the interior could be spinning much faster than the surface. Interestingly, this star is among the youngest and more massive of the sample suggesting a scenario where the angular momentum transport processes that are responsible for the quasi-uniform internal rotation might not have had enough time to complete their work.


\section{Stellar magnetic activity and  magnetic cycles}
\label{Sec:magnetism}
In distant stars, as was the case for rotation, asteroseismic observations provide two different, but complementary, ways to study magnetic activity in general and magnetic activity cycles in particular. On the one hand,  magnetic variability can be measured on different time scales by directly analyzing the average luminosity flux modulation in the light curve or its fluctuation as a function of time \citep[e.g.][]{2010ApJ...713L.155B,2011AJ....141...20B,2014A&A...572A..34G}. It is out of the scope of this review to describe in details the methodologies and results associated to these studies. However, in the next paragraph we will provide a brief overview of them. On the other hand, magnetic variability can be studied through the characterization of the temporal evolution of the oscillation modes, i.e., measuring the variations of the frequencies, amplitudes, and line widths of the acoustic modes. 

As discussed in the section about internal rotation, to reach the core, mixed or gravity modes are required. Unfortunately, those modes have not been unambiguously detected in main-sequence solar-type dwarfs. Therefore, nothing can be said about the magnetic field in the core of these stars. However, studying red giant stars, \cite{2016Natur.529..364S} evoked the possibility of having magnetic fields of dynamo origin in the convective core of stars with masses greater than $\sim$ 1.2 M$_\odot$. 
The origin of the depressed dipolar modes seen in \emph{Kepler} observations \citep[e.g.][]{2012A&A...537A..30M,2014A&A...563A..84G} has been suggested to be magnetic, although the exact nature of phenomena is still debated \citep[e.g.][]{2015Sci...350..423F,2017A&A...598A..62M,2017MNRAS.467.3212L,2018MNRAS.477.5338L}.

\subsection{From the direct analysis of the light curves}
\begin{figure}[!htp]
	\centering
		\includegraphics[width=0.8\textwidth] {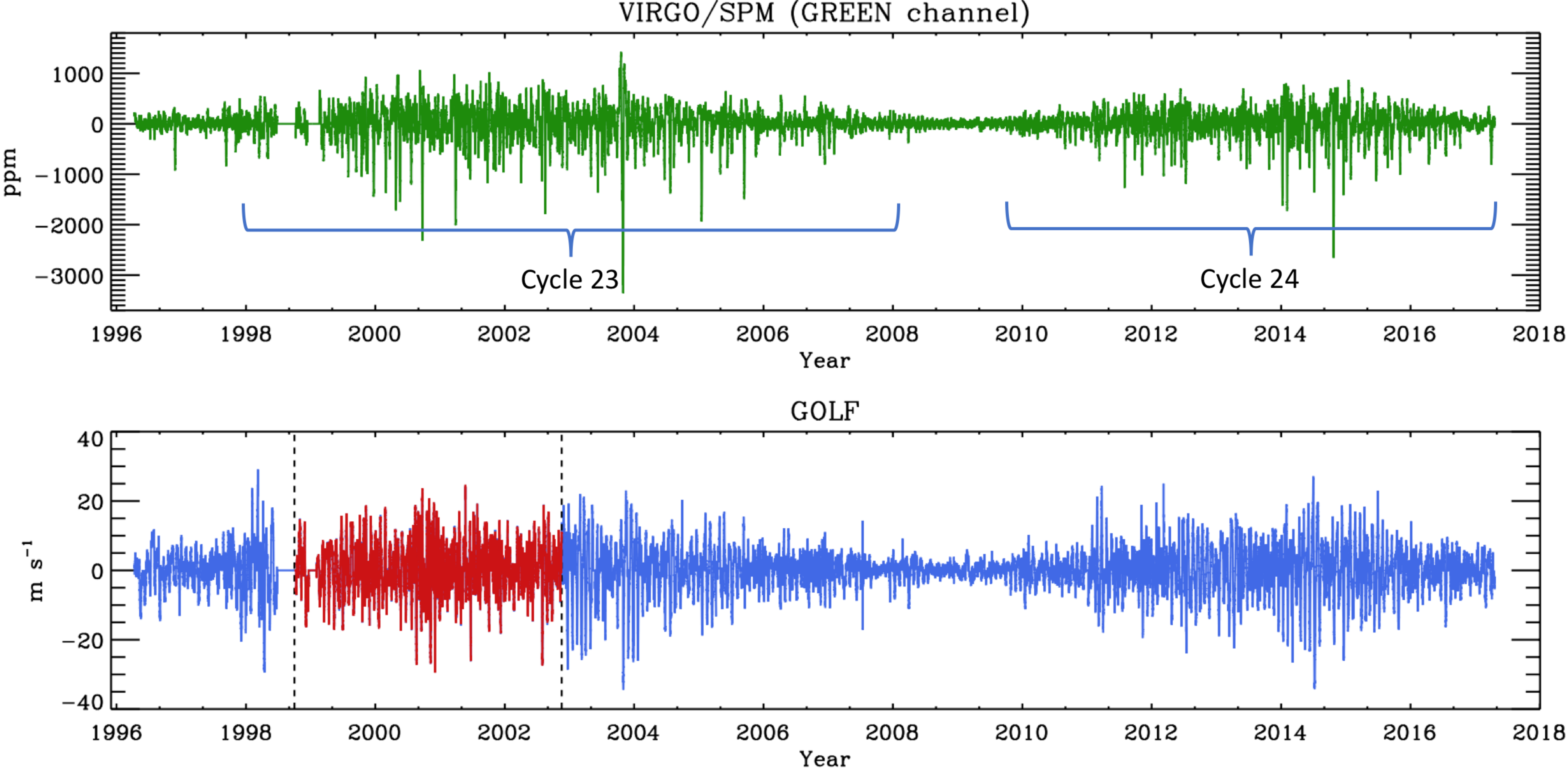}
	\caption[Photometric and Doppler velocity time series obtained from SoHO]{Photometric and Doppler velocity time series respectively from the green channel of the VIRGO SPM (top panel) and GOLF instrument on board SoHO (bottom panel). The GOLF red-wing period denoted by the two vertical dashed lines where the instrument is sensitive to higher layers in the atmosphere is represented in red (see for details  \citet{2004ESASP.559..432G}, \citet{GarSTC2005}, and \citet{JimCha2007}). Figure adapted from \cite{2017A&A...608A..87S}. } 
	\label{sph_svel}
\end{figure}

High-quality, photometric time series of several months to years duration are now available from CoRoT, \emph{Kepler}, K2 and TESS missions. They allow a detailed study of the photospheric stellar variability at different time scales and to compare them with the Sun \citep[e.g.][and references therein]{2011ApJS..197....6G,2013ApJ...769...37B}. However, not all this variability can be directly associated with magnetism.  Different physical phenomena (pulsations, convection or rotation) can indeed coexist on the same time scales. To overcome this problem, it has been proposed to study the time scales associated with the rotation period of the star as an indicator of a magnetic origin \citep[e.g.][]{2014JSWSC...4A..15M,2014A&A...562A.124M,2014A&A...572A..34G}. For magnetically active stars developing starspots at their surface, the global brightness is modulated as a function of the position and the size of the spots over the visible stellar disk with a period that is related to the rotation period of the star at the active latitude where the spot appears \citep[e.g.][]{2010IAUS..264..120L,2012A&A...543A.146F,2014A&A...564A..50L,2014ApJS..211...24M}. As the magnetic cycle evolves, the number and size of the spots change. An example of the solar photospheric and Doppler-velocity time series is shown in Fig.~\ref{sph_svel}. Hence, a photospheric magnetic activity proxy, $S_{\rm{ph}}$, can then be constructed by measuring the dispersion of the light curve on time scales of five times the rotation period \citep{2014JSWSC...4A..15M}. The same procedure can be applied to Doppler velocity observations and another proxy, called $S_{\rm{vel}}$, has been defined. 

The validity of this proxy as a function of magnetic activity has been verified using the Sun \citep{2017A&A...608A..87S} and compared to the chromospheric activity of other \emph{Kepler} targets including 18 solar analogs \citep{2016A&A...596A..31S,2017EPJWC.16001010G}.  This proxy (or other similar metrics) is currently being used to search for magnetic activity cycles or magnetically-related trends that could be part of magnetic cycles longer than the duration of the observations for stars of many different spectral types \citep[e.g.][]{2009A&A...501..703O,2013A&A...550A..32M,2014JSWSC...4A..15M,2014MNRAS.441.2744V,2015A&A...583A.134F}. Finally, by combining magnetic proxies and seismology it has been possible to revisit the age-rotation-activity relations \citep{2013MNRAS.433.3227K,2014A&A...572A..34G} and to establish that the photospheric magnetic activity of the Sun is comparable to other main-sequence solar-like pulsating stars \citep{2014A&A...572A..34G}, in particular, compared to other solar analogs of similar age \citep{2014ApJ...790L..23D,2016A&A...596A..31S,2017A&A...602A..63B}. In this sense, we can conclude that the Sun is a normal star in terms of its surface magnetism when compared to its siblings. 

\begin{figure}[!htp]
	\centering
		\includegraphics[width=0.5\textwidth] {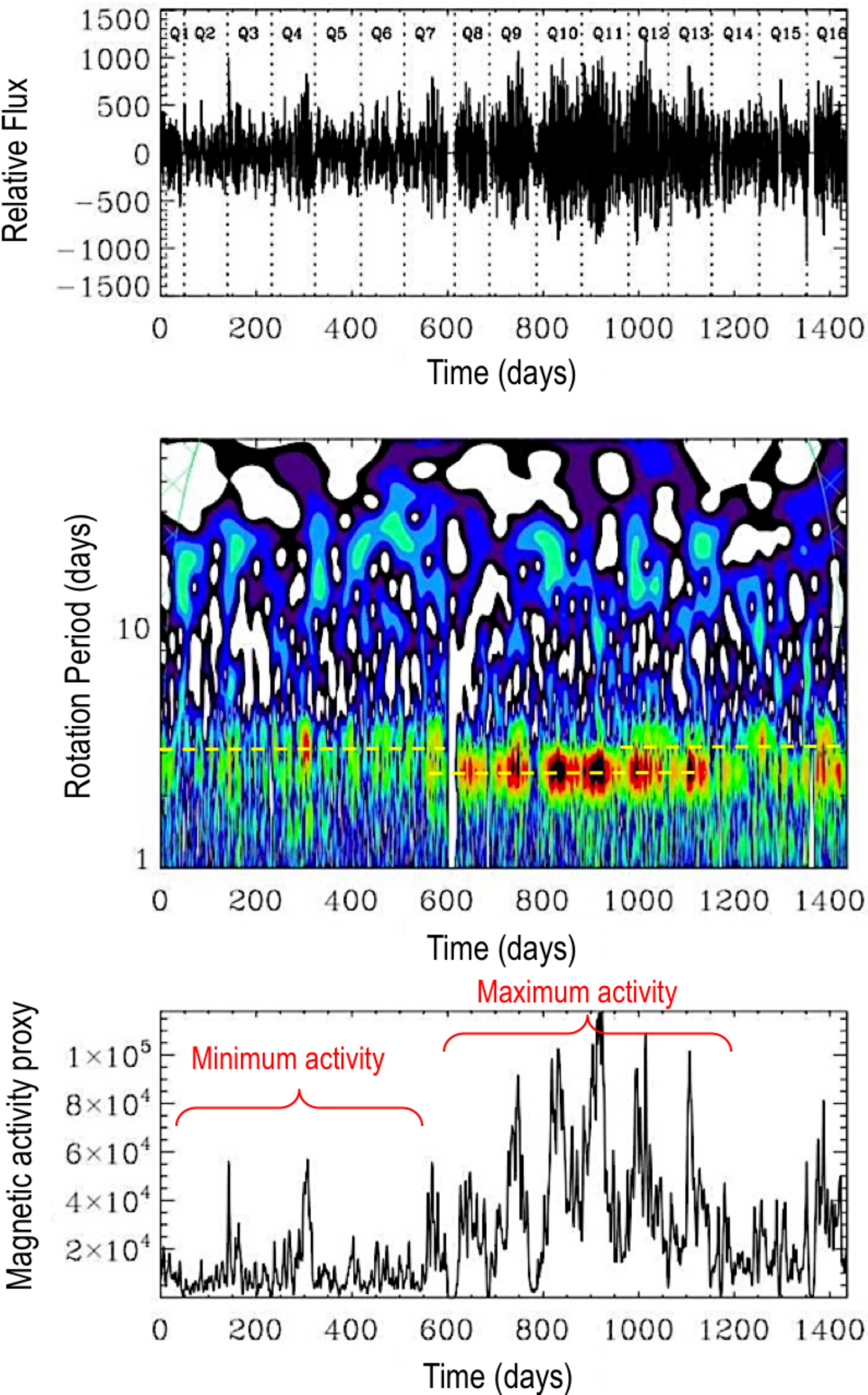}
	\caption[Study of the surface dynamics of the \emph{Kepler} target KIC~3733735]{Relative flux of the \emph{Kepler} target KIC~3733735 (top panel). A time--period analysis is shown in the middle panel. The two horizontal dashed lines indicate the main two rotation periods for this star (3 and 2.54 days).The bottom panel represents a magnetic activity proxy similar to the $S_{\rm{ph}}$ computed by projecting the time--period diagram onto the temporal axis (between 2 and 6 days).} 
	\label{mag_3733735}
\end{figure}

An example of the study of the surface dynamics (rotation and magnetism) using  \emph{Kepler} photometry is given in Fig.~\ref{mag_3733735}. In the top panel the relative measured flux is given. The envelope of the surface brightness shows an increase between the day $\sim$ 550 and $\sim$ 1200 of the mission that can be interpreted as an increase of the magnetic activity of the star. In the middle panel of Fig.~\ref{mag_3733735} a time--period diagram is computed. It shows two main bands of power (depicted by the yellow horizontal dashed lines) at 3 and 2.54 days corresponding to the main average rotation rate of the star during these two seasons.  Projecting the time--period diagram onto the time domain around the periods given before (from to 2 to 6 days), a magnetic proxy can be built (see bottom panel of Fig.~\ref{mag_3733735}), confirming the existence of an on-going stellar activity cycle, with a season of maximum activity where the starspots are located at longitudes of faster rotation (2.54 days), while the band at a slower rotation rate is dominating the location of the spots during the minimum of magnetic activity. This behavior is similar to that observed in the Sun which produces the so-called butterfly diagram \citep[e.g.][]{2015LRSP...12....4H}. 

\subsection{From asteroseismology}

To study magnetic activity cycles with asteroseismic techniques, it is important to use the Sun as a reference because we can characterize in detail its surface magnetism and look for correlations with different observed features that are today impossible to obtain for distant stars. In this way, we can then apply this knowledge to other stars. At the very beginning of helioseismology, a correlation between the acoustic-mode frequencies and  solar magnetic activity was found \citep{1984LIACo..25..215V,1985Natur.318..449W,1987A&A...177L..47F,1989A&A...224..253P,1990Natur.345..322E}, i.e., the frequencies shifted towards higher values as the 11-year magnetic cycle progressed. Later, it was discovered that the frequency shifts increased with frequency for intermediate- and low-degree modes \citep[][]{1990Natur.345..779L,1992A&A...255..363A}, i.e. modes at higher frequency had a larger frequency shift than modes at low frequency. This frequency dependence -- high-frequency modes have outer turning points compared to low-frequency modes -- is the main change of the mode parameters when the effect of the mode inertia is removed. Considering also that there was no significant dependence of the frequency shifts with the degree of the modes led to the conclusion that the perturbations related with the magnetic activity cycle were confined to a thin layer very close to the photosphere \citep[e.g.][]{1990Natur.345..779L,1991ApJ...370..752G,1995ASPC...76..280N,2012ApJ...758...43B}. Moreover, the Sun has not only a 11-year periodicity. Shorter --quasi-biennial-- modulations have been measured in the Sun \citep[e.g.][]{1995SoPh..161....1B} in several magnetic activity proxies and also confirmed by seismology \citep{2010ApJ...718L..19F,2011JPhCS.271a2025B,2012A&A...539A.135S}. The existence of several time scales in the modulation of the magnetic proxies is not peculiar to the Sun and many other stars show several long and short periods that could be interpreted as magnetic cycles with an ``active'' and ``inactive'' phase \citep[e.g.][]{1995ApJ...450..896B,1998ApJ...498L..51B,1999ApJ...524..295S,2007ApJ...657..486B,2017ApJ...845...79B}. Last but not least, important differences were found between the surface magnetic proxies and the frequency shifts during the last extended minima between Solar Cycles 23 and 24 \citep{2009ApJ...700L.162B,2009A&A...504L...1S}. While no activity was measured in standard magnetic proxies, the frequency shifts showed a quasi normal behavior leading to the conclusion that the magnetic perturbations in the subsurface layers were still strong \citep{2015A&A...578A.137S}. 

\begin{figure}[!htp]
	\centering
		\includegraphics[width=0.6\textwidth] {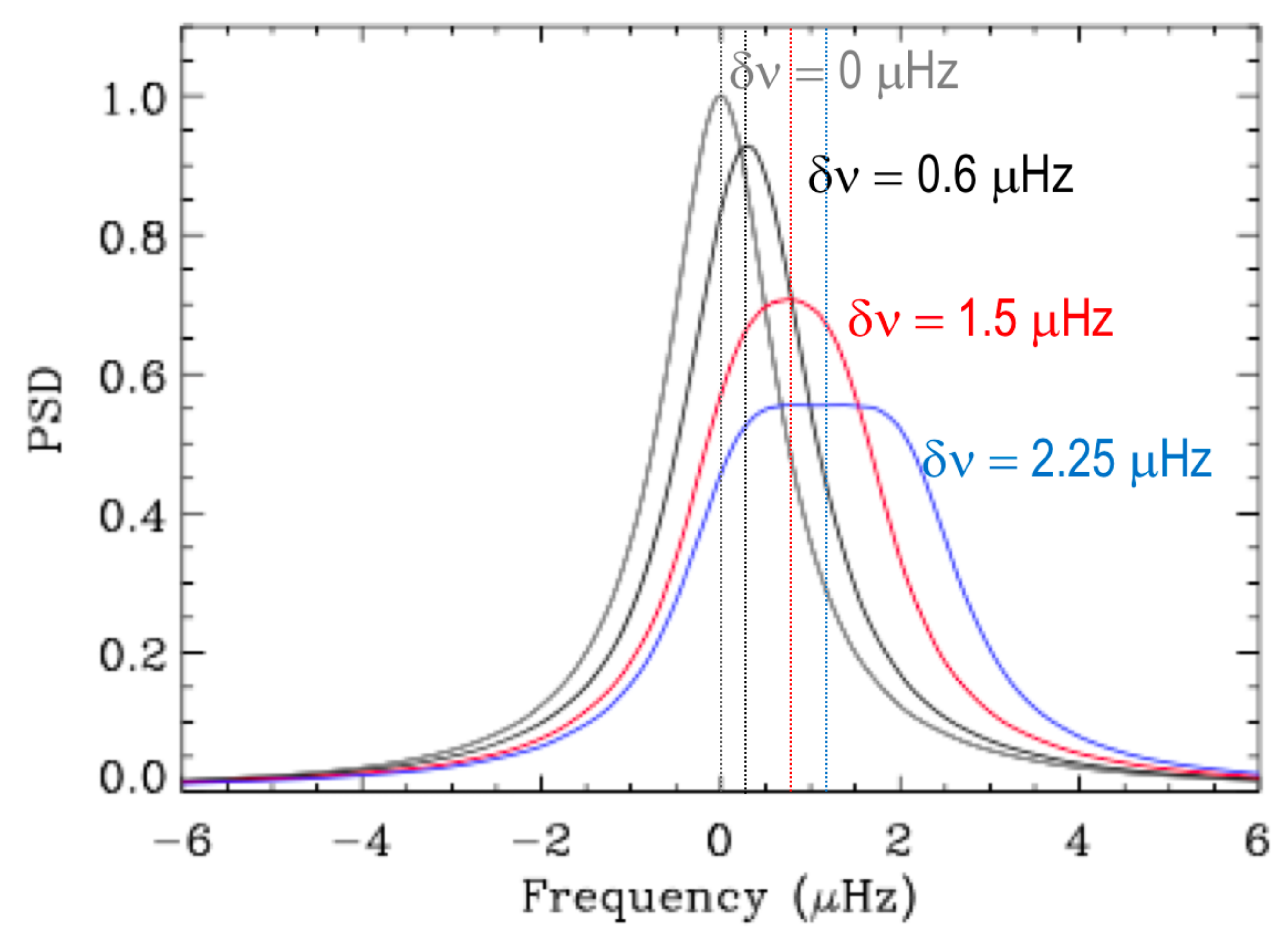}
	\caption{Lorentzian profile of a mode with a linewidth $\Gamma$=1.5 $\mu$Hz (grey), and three other pseudo-Lorentzian profiles resulting from the analysis of a long time series with three different maximum-to-minimum frequency shifts of 0.6, 1.5, and 2.25 $\mu$Hz, black, red and blue lines respectively. Figure adapted from \citep{2014SSRv..tmp...49C}.	}
	\label{FIG_dist_prof}
\end{figure}

All of these early findings on the modulation of the frequency shifts with the magnetic cycle were soon followed by the study of the evolution of all of the other oscillation-mode parameters. Hence, the measurement of a reduction of the mode heights, an increase of the mode linewidths and a reduction of the velocity power with the magnetic activity cycle were found. Finally, the energy that is supplied to the modes remained constant along the magnetic cycle within the uncertainties \citep[e.g.][]{1990LNP...367..129P,2000MNRAS.313...32C,2002ApJ...572..663K,2003ApJ...595..446J,2013JPhCS.440a2020G,2015AdSpR..56.2706B}. It is therefore important to take these changes into account when performing the fitting of the acoustic modes in active stars. All of these variations modify the simple Lorentzian profiles of the modes introducing a bias of the retrieved parameters when simple Lorentzian profiles are used  to fit spectra obtained from time series that are long compared to the timescale of the variations introduced by the activity \citep{2008MNRAS.384.1668C,2008SoPh..251...53C}. In particular, frequency shifts modify the central frequencies of the Lorentzians. Thus, when the resultant profile is fitted by a single Lorentzian function, the inferred seismic parameters are biased such as the large frequency separations as demonstrated by \cite{2011MNRAS.413.2978B}. An example of this distorted Lorentzian profile is given in Fig.~\ref{FIG_dist_prof} for a mode with a linewidth of $\Gamma$=1.5 $\mu$Hz and four different maximum-to-minimum frequency shifts: 0, 0.6, 1.5, and 2.25 $\mu$Hz. 

One exception to this is the characterization of low-degree low-order p modes for which the variations induced by the magnetic activity cycle are too small, around $\Delta_\nu/\nu \sim 10^{-5}$~nHz for the solar case \citep[e.g.][]{GarReg2001}. 

\begin{figure}[!htp]
	\centering
		\includegraphics[width=0.6\textwidth] {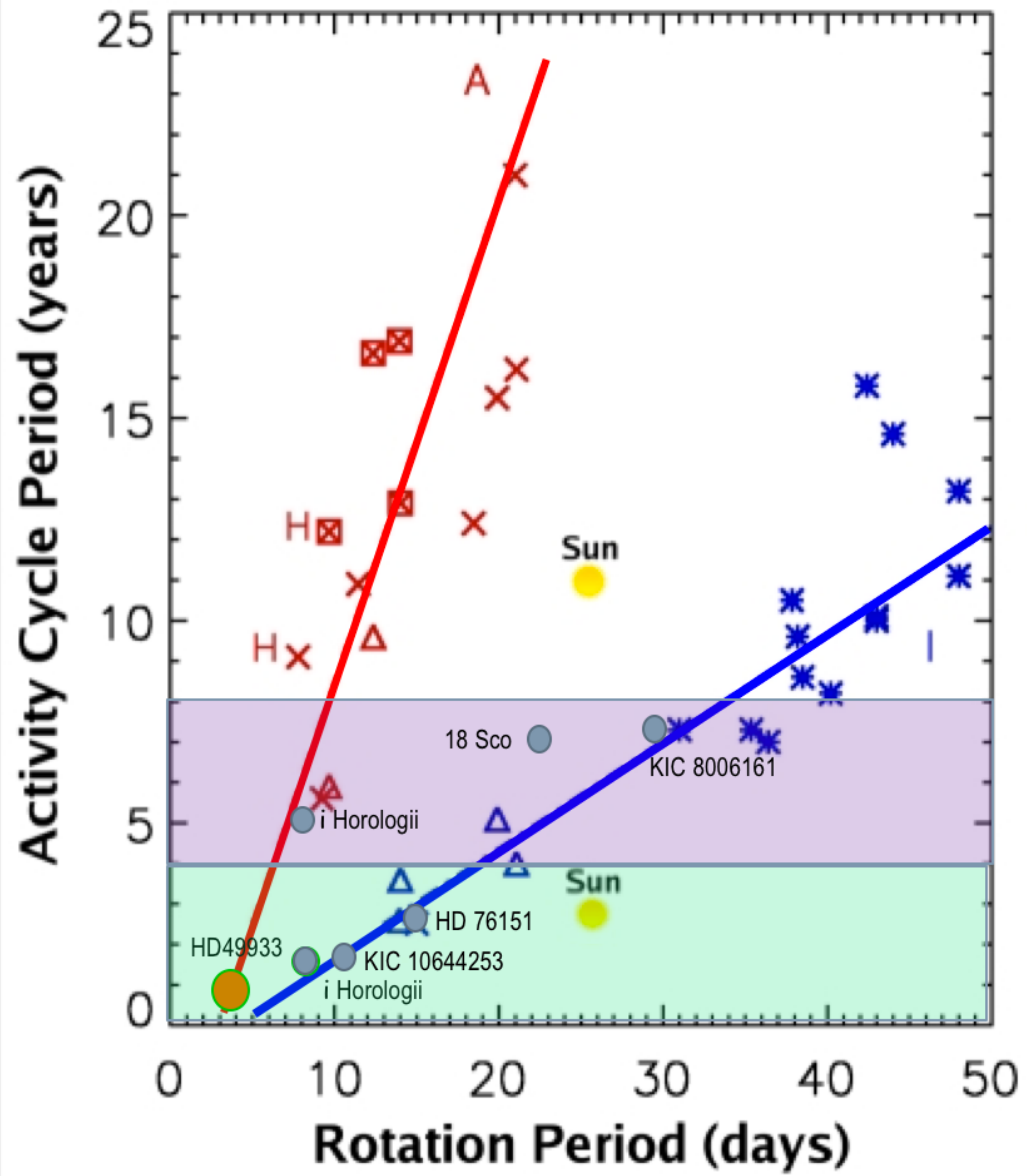}
	\caption[Cycle period versus rotation period]{Magnetic activity cycle period versus the rotation period of stars showing the existence of two branches an ``active'' (red line denoted by A and red symbols) and an ``inactive'' sequence (blue line denoted by B and blue symbols). Several stars have been added to the original diagram from \cite{2007ApJ...657..486B} such as $\iota$ Horlogii \citep{2013ApJ...763L..26M}, HD~76151 \citep{2017PhDT.........3E}, and the seismic targets 18 Sco \citep{2007AJ....133.2206H}, HD~49933 \citep{2010Sci...329.1032G} and the \emph{Kepler} targets KIC~8006161 and KIC~10644253 \citep[][]{2016A&A...589A.118S,2017A&A...598A..77K,2018ApJ...852...46K}. The green shaded region represents the region where full stellar cycles can be studied with the \emph{Kepler} data. The violet shaded region represents the region where longer activity cycles could be uncovered by \emph{Kepler} observations if they are in a regime where the magnetic activity changes significantly. The letter H indicates stars in the Hyades, crosses indicate stars on the ``active'' A sequence, and asterisks indicate stars on the ``inactive'' B branch. Squares around the crosses show stars with a color B-V $<$ 0.62. Triangles indicate secondary periods for some stars on the active sequence.}
	\label{FIG_Bomm}
\end{figure}

To study stellar magnetic activity cycles it is necessary to have long time series covering all -- or a significant part -- of the magnetic cycle. The longest time series available today for asteroseismology were obtained by the \emph{Kepler} mission covering slightly more than four continuous years. Due to the rather well-known relation between the rotation period of stars and the length of the magnetic cycle \citep[e.g.][]{1984ApJ...287..769N,1984ApJ...279..763N,1998ApJ...498L..51B,2007ApJ...657..486B,2017ApJ...845...79B}, the study of full magnetic cycles is limited to fast rotating stars ($P_{rot} \leq$ 7 days) for the active, ``A'', branch and for stars with intermediate rotation rates ($P_{rot} \leq$ 18 days) for the less-active, ``I'', sequence (see Fig.~\ref{FIG_Bomm}). 

The relation between magnetic activity cycle periods and stellar evolution is also well known \citep[e.g.][]{1998ApJ...498L..51B}. Recently, \cite{2017ApJ...845...79B} re-visited this relation in light of the most recent and longest spectroscopic observations from \cite{2017PhDT.........3E} who combined datasets from several instruments. Unlike previous interpretations where young stars would evolve along the active A branch, they now believe that all stars younger than 2.3 Gyr are capable of exhibiting longer and shorter cycle periods. If their calculations are correct, for the G pulsating dwarfs HD~76151 and KIC~10644253 shown in Fig.~\ref{FIG_Bomm}, longer periods in the 12 to 16 year range are expected, and may have already been found in HD 76151 \citep{2017PhDT.........3E} and in $\iota$ Horologii, HD~17051, where \cite{2017MNRAS.464.4299F} measured a long-term activity cycle of about five years fitting the ``active'' branch in the B\"ohm-Vitense diagram. For the solar analogue 18 Sco, it would be interesting to discover either a second shorter modulation as in the Sun or a longer cycle period.

An interesting result was obtained by \cite{2014A&A...562A.124M} from the analysis of the temporal evolution of $S_{\rm{ph}}$. They selected 22 solar-like stars with detected solar-like oscillations and with rotation periods shorter than 12 days, i.e. the limit to observe at least half of a cycle from Fig.~\ref{FIG_Bomm}. Only two of the stars show a cyclic-like variation, while two others showed a decreasing or increasing trend in the $S_{\rm{ph}}$ temporal variation. Five stars show modulations (or beating) due to long-lived spots at two different active longitudes with different rotation rates. The rest of the stars show no cycle-like behavior although they showed surface magnetic activity. Therefore, although it is premature to infer firm conclusions from a so small sample of stars, it seems that fast rotating stars can exhibit magnetic activity but without any magnetic cycle (or at least much longer than it could be expected from Fig.~\ref{FIG_Bomm}. The only correlation found was between the $S_{\rm{ph}}$ and the rotation period for stars showing a beating between long lived spots at different rotation rates, in the direction of higher $S_{\rm{ph}}$ for longer $P_{\rm{rot}}$. 

To perform asteroseismic studies of magnetic activity, it is then necessary to first measure the oscillation properties and then study their evolution with time. As it has been said, solar magnetism reduces the amplitude of the solar modes. Therefore, a very active star would probably have oscillation modes with very small or even not detectable amplitudes. This effect was first observed by CoRoT in two main-sequence targets: HD~175726 \citep{2009A&A...506...33M} and in HD~49933 \citep{2010Sci...329.1032G}. Later, using a \emph{Kepler} sample of main-sequence stars, \cite{2011ApJ...732...54C} showed that there was a clear correlation between stellar magnetic activity and the amplitude of the stellar oscillations. The most active stars in the sample did not show measurable oscillation modes.

To look for the evolution with time of the seismic properties, the light curve is divided into small segments for which the characterization of the oscillation modes is carried out. The length of the sub-series is found as the best trade-off between frequency resolution and the number of sub-series to be analyzed. The longer the series the better the precision on the extracted parameters \citep[e.g.][]{2016A&A...589A.103R}. However, it can be very challenging to obtain individual p-mode frequency shifts that can be as small as half a $\mu$Hz for short time series. This is why a global method was developed in the early days of helioseismology by \cite{1989A&A...224..253P} to obtain averaged frequency shifts by computing the cross correlation of the p-mode hump computed from the PSD of each sub-series in comparison to a reference one. This reference can be either the PSD of one of the sub-series or the average spectrum of all of them \citep[e.g.][]{1989A&A...224..253P,2010Sci...329.1032G,2016A&A...589A.103R,2016A&A...589A.118S,2017A&A...598A..77K,2018ApJS..237...17S}.

The first attempts to detect differences in the seismic parameters associated with magnetic activity were reported by \cite{2006MNRAS.371..935F}. They analyzed observations made with the star-tracker on the WIRE satellite of  $\alpha$ Cen A and compared the obtained frequencies with previously obtained frequencies measured by \cite{2002A&A...390..205B} and \cite{2004ApJ...614..380B}. \cite{2006MNRAS.371..935F} conjectured that the average difference of about $0.6 \pm 0.3$ $\mu$Hz ($\sim 2\sigma$) in the oscillation frequencies could be due to an ongoing activity cycle in $\alpha$ Cen as the WIRE observations were taken 19 months before the others.

\begin{figure}[!htp]
	\centering
		\includegraphics[width=0.6\textwidth] {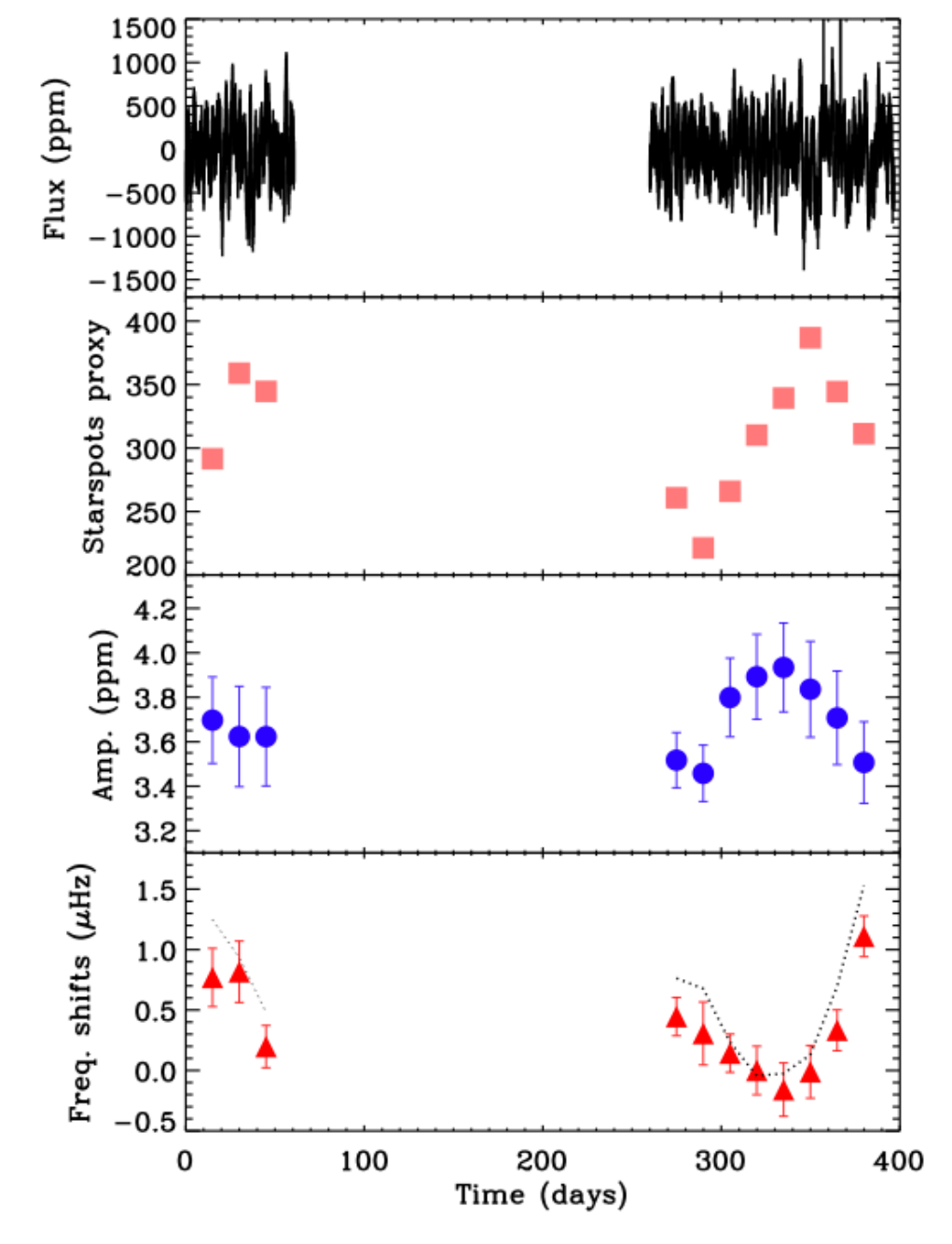}
	\caption[Temporal evolution of some magnetic activity proxies of the CoRoT target HD~49933 ]{From top to bottom: lightcurve of the CoRoT target HD~49933 including the first two runs on this star; second panel: temporal evolution of the dispersion of the light curve; third panel: temporal evolution of the frequency shifts using cross-correlation techniques (red triangles) and the individual mode fitting (dotted line); bottom: maximum amplitude per radial mode versus time (extracted using the A2Z pipeline \citep{2010A&A...511A..46M}). Modified version of the figure presented in \cite{2010Sci...329.1032G}.}
	\label{activ_hd49933}
\end{figure}

Successful asteroseismic measurements of magnetic activity started in 2010 when \cite{2010Sci...329.1032G} unveiled the presence of a magnetic activity cycle in a star other than the Sun: the CoRoT target HD~49933, which was observed during the first two CoRoT runs with a six month interval \citep{2008A&A...488..705A,2009A&A...507L..13B}. They found a modulation of more than 120 days in the three indicators considered: frequency shifts -- measured globally through cross-correlation techniques as well as by extracting the individual p-mode frequencies --, amplitude modulation of the p-mode bump and in the dispersion of the light curve. Moreover, as also observed in the Sun, the modulation detected in the amplitude and the frequency shifts of the p modes was anticorrelated (see Fig.~\ref{activ_hd49933}).

\begin{figure}[!htp]
	\centering
		\includegraphics[width=0.6\textwidth] {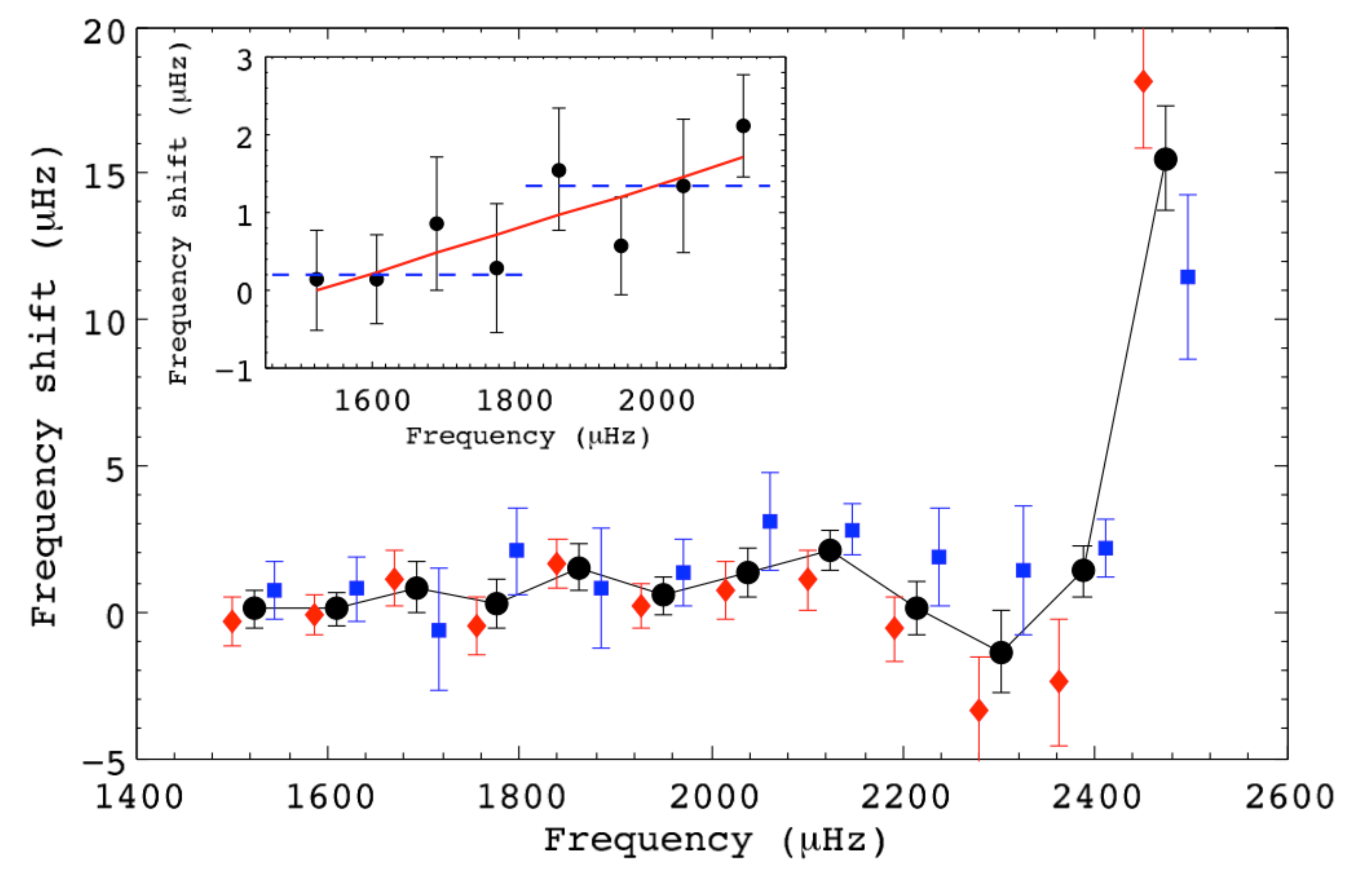}
	\caption[p-mode frequency dependence of the frequency shifts in HD~49933]{Blue squares, red diamonds and black circles are respectively the frequency shifts of the modes $\ell=0$, $\ell=1$, and their weighted means. Mean frequency shifts are computed in the inset  up to 2100 $\mu$Hz.The horizontal blue dashed lines correspond to the frequency shifts obtained with the cross-correlation method in two frequency bands: from 1500 to 1800 and from 1800 to 2100 $\mu$Hz. The horizontal blue dashed lines correspond to the frequency shifts obtained with the cross-correlation method in two distinct frequency regions. The solid red line corresponds to a weighted linear fit. Figure from \cite{2011A&A...530A.127S}.}
	\label{Salabert_hd49933}
\end{figure}

The analogies between the magnetic cycles of both stars continue. As in our Sun, the frequency shifts measured in HD~49933 present a frequency dependence with a clear increase with frequency. However, the maximum frequency shift is about 2 $\mu$Hz around 2100 $\mu$Hz, 4 times bigger than in the solar case \citep{2011A&A...530A.127S}. Similar variations are obtained between the $\ell= 0$ and $\ell = 1$ modes computed independently (see Fig.~\ref{Salabert_hd49933}). At higher frequencies, the frequency shifts show indications of a downturn followed by an upturn for both low-degree modes $\ell$ = 1 and 2. Therefore, the frequency variation of the p-mode frequency shifts of this star has a comparable shape to tthat observed in the Sun, which is understood to arise from changes just beneath the photosphere \citep[e.g.][]{2016LRSP...13....2B}.

\begin{figure}[!htp]
	\centering
		\includegraphics[width=0.55\textwidth] {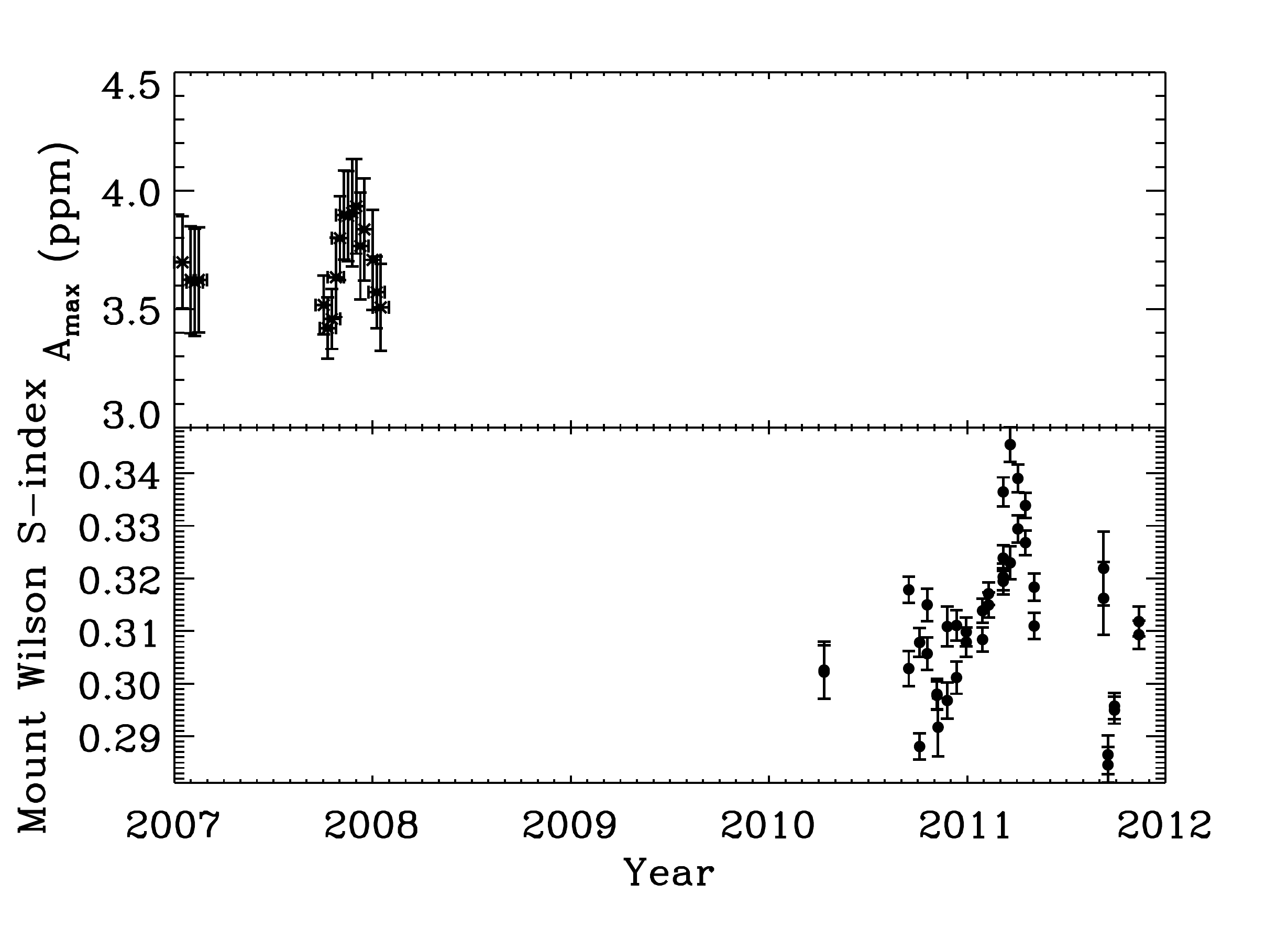}
	\caption[Temporal evolution of the Mount Wilson S-index of HD~49933]{Temporal evolution of the Mount Wilson S-index (T.~S. Metcalfe private communication) and $A_{\rm{max}}$ obtained from the first two CoRoT runs. Figure courtesy of S. Mathur.}
	\label{CaHK_hd49933}
\end{figure}

Complementary observations of HD~49933 were taken in the Calcium H and K lines since 13 April, 2010, showing that this is an active star with a Mount Wilson S-index of 0.3 confirming previous conclusions \citep{2005A&A...431L..13M}. The temporal evolution of the Mount Wilson S-index showed that a magnetic activity cycle is ongoing in this star.  However, longer observations are needed to clearly establish the length of the cycle and whether or not the magnetic cycle is regular. In Fig.~\ref{CaHK_hd49933}, the Mount Wilson-S index of HD~49933 is shown. For comparison, the variations of $A_{\rm{max}}$ obtained from the two first runs of CoRoT are depicted in the top panel of the figure.

The same methodology was applied to three other CoRoT targets complemented by ground-based analysis of the chromospheric magnetic activity done with the NARVAL spectropolarimeter located at the Bernard Lyot 2 m telescope at the Pic du Midi Observatory \citep{2003EAS.....9..105A} in order to study any possible hint of magnetic activity cycles \citep{2013A&A...550A..32M}. Interestingly one star, HD~181420 \citep{2009A&A...506...51B}, seems to be in a stationary regime without any visible change of the activity during the observations. This is an unexpected result because this star rotates rapidly (2.6 days) and we were expecting to see some indications of a magnetic activity cycle. For the other two stars, HD~49385 \citep{2010A&A...515A..87D} shows a small increase of activity at a 1-$\sigma$ level but not confirmed by our spectroscopic measurements, while HD~55265 \citep{2007A&A...471..885S,2011A&A...530A..97B} presents a small variation of the seismic parameters, also at a 1-$\sigma$ level, and in the spectroscopic observations performed by NARVAL. That could indicate that this star was observed during the rising phase of a long magnetic activity cycle.

\subsubsection{Influence of metallicity on magnetic activity}

\begin{figure}[!htp]
	\centering
		\includegraphics[width=0.75\textwidth] {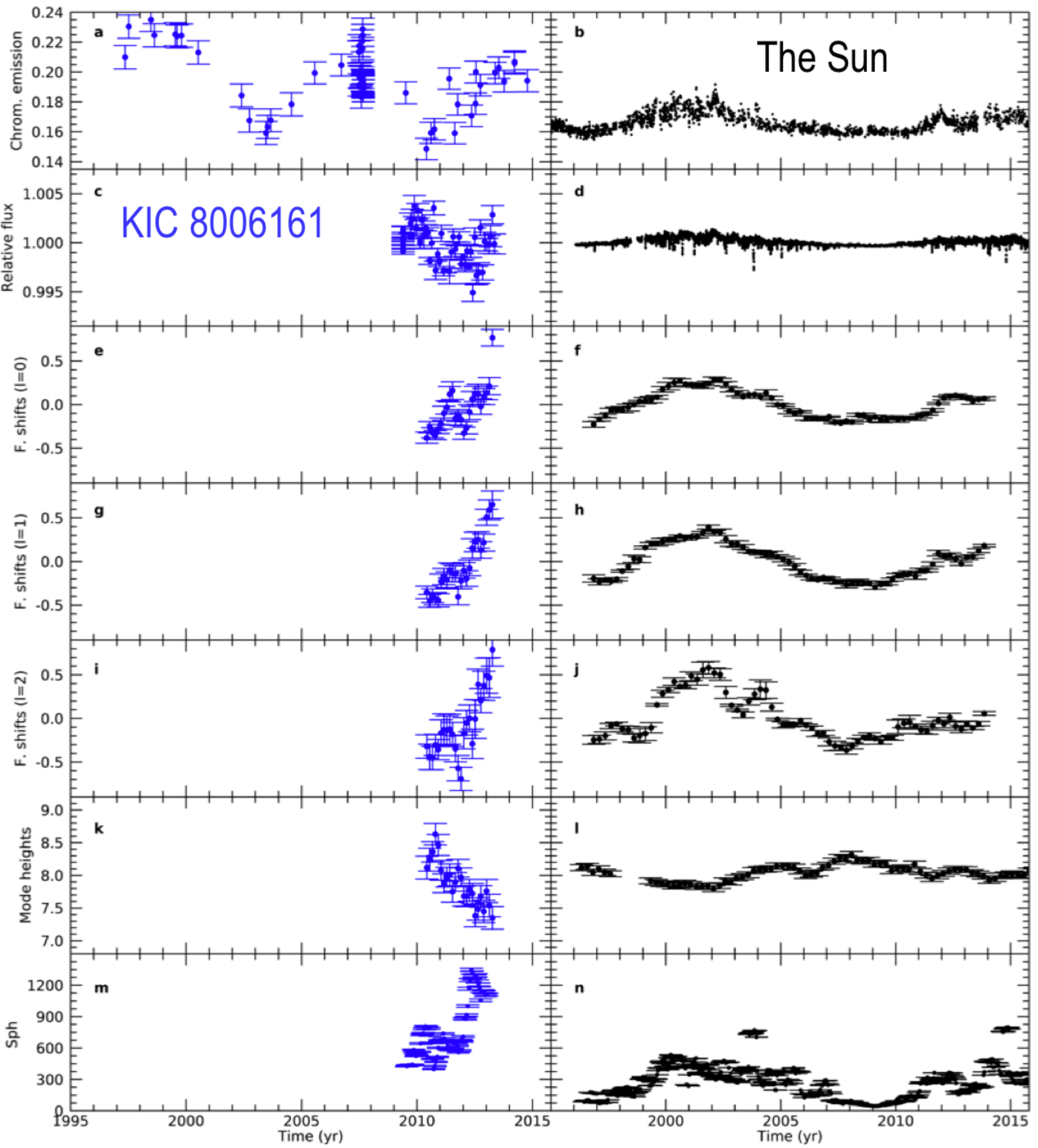}
	\caption{Activity proxies as a function of time for KIC~8006161 (left panels) and the Sun (right panels). The panels show from top to bottom: the chromospheric emission, the relative flux, radial frequency shifts, frequency shifts of dipolar modes,  frequency shifts of quadrupolar modes, logarithmic mode heights of the modes and $S_{\rm{ph}}$. Figure from \cite{2018ApJ...852...46K}.}
	\label{FIG_Doris_activity_karoff}
\end{figure}

The longer observation period of the \emph{Kepler} main mission (up to four continuous years) allows one to better track temporal changes in the seismic parameters for main-sequence solar-like stars. This is illustrated in Fig~\ref{FIG_Doris_activity_karoff} where the temporal evolution of the seismic parameters and the photospheric and chromospheric activity proxies are shown for KIC~8006161 (HD~173701) and for the Sun. All of the figures are in the same scale for both stars. KIC~8006161 is a very interesting target because it is a solar analogue \citep{2018ApJ...852...46K} with $M=1.00 \pm 0.03$~M$_\odot$, $R=0.93 \pm 0.009$~R$_\odot$, $T_{\rm{eff}}$=5488 $\pm77$~K, $P_{\rm{rot}} =21 \pm 2$ days, age = 4.57 $\pm$ 0.36 Gyr, and a metallicity that is twice the solar value (0.3 $\pm$ 0.1 dex). 

KIC~8006161 has a comparable magnetic activity cycle period ($\sim$ 7.4 yr) deduced from more than thirty years of chromospheric observations (top plot in Fig.~\ref{FIG_Doris_activity_karoff}), being in the rising phase of its cycle during the four years of \emph{Kepler} observations. Interestingly, the amplitude of all the temporal variations recorded are much larger than the corresponding solar values.  \cite{2018ApJ...852...46K} conjectured that these differences could be a consequence of the higher metallicity of the star. An increase of the stellar metallicity produces larger opacities and thus a larger internal temperature gradient. Therefore, the Schwarzschild criterion for convection \citep{1906WisGo.195...41S} is satisfied deeper inside the star leading to a deeper convective zone \citep{2012ApJ...746...16V}. Theoretical studies and numerical simulations have shown that larger convective zones induce larger differential rotation \citep{2017ApJ...836..192B} and thus a stronger magnetic dynamo \citep{2011ApJ...728..115B}. Although it is not possible with the present observations to infer how strong the dynamo is, the surface differential rotation is larger than that in the Sun, reinforcing the conclusions. Unfortunately, firm conclusions cannot be derived with only one observations and the study of a larger sample of stars with different metallicities is necessary to completely understand the influence of metallicity on the length and strength of the magnetic activity cycles. 

\subsubsection{Relation between the frequency shift strength with effective temperature and age}
\begin{figure}[!htp]
	\centering
		\includegraphics[width=0.75\textwidth] {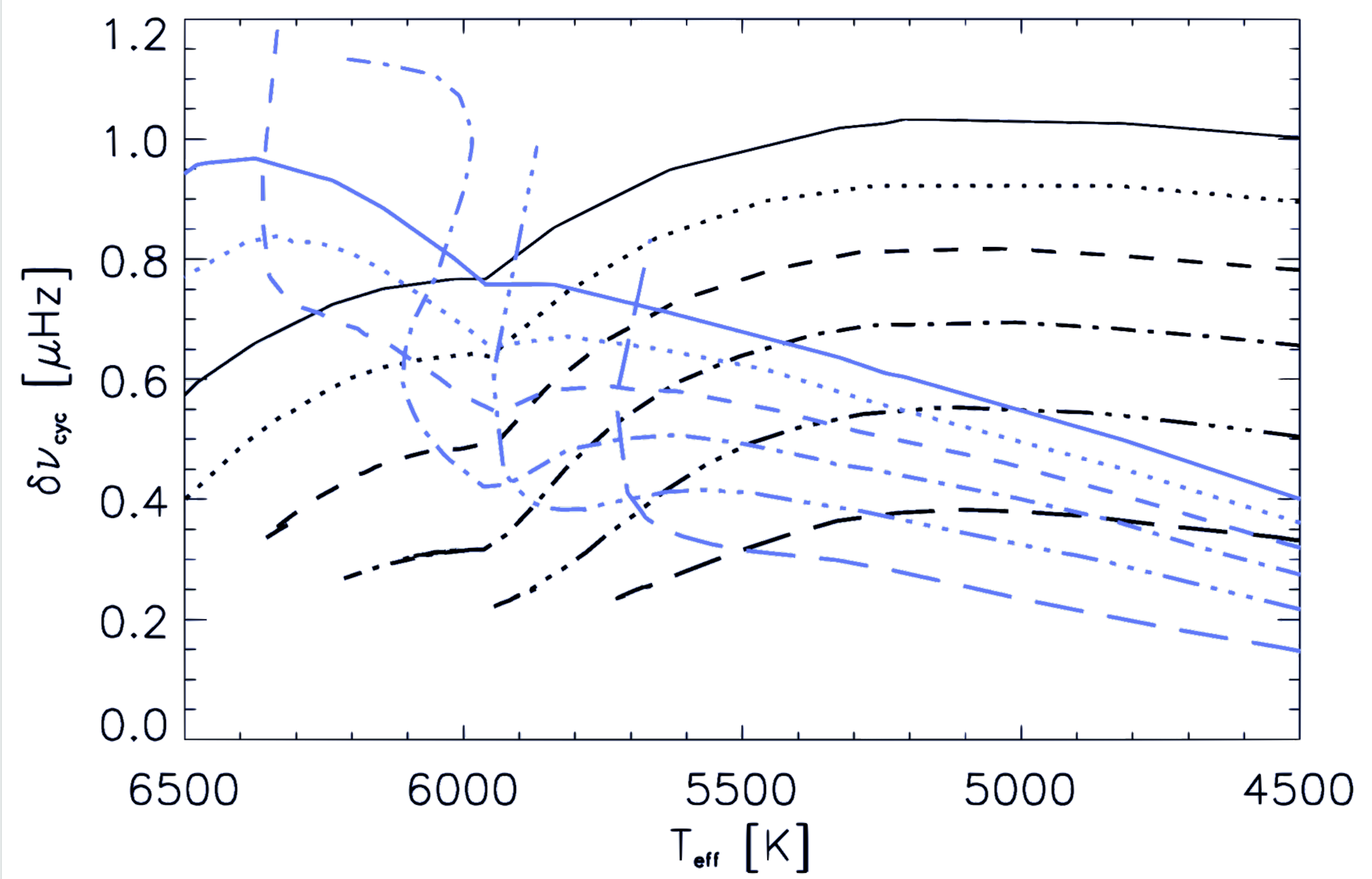}
	\caption{Frequency shifts, $\delta \nu_{\rm{cyc}}$, as a function of effective temperature computed for different Padova isochrones \citep[different line styles,][]{2002A&A...391..195G,2004A&A...422..205G}. Black lines follow the scaling relation of \cite{2007MNRAS.377...17C} while blue lines follow \cite{2007MNRAS.379L..16M}. Figure adapted from \cite{2009MNRAS.399..914K}.}
	\label{FIG_variation_time_freq_shifts}
\end{figure}

We have just presented an explanation for a possible relation between metallicity and the magnetic cycle strength. Using the ensemble analysis of frequency shifts from asteroseismology, it is possible to depict other correlations between different parameters. The relation between the effective temperature and the frequency shifts is particularly interesting as two different scaling relations have been proposed by \cite{2007MNRAS.377...17C} and \cite{2007MNRAS.379L..16M}. On the one hand, \cite{2007MNRAS.377...17C} proposed that the frequency shifts linearly scale with the amplitude of the activity cycles $\Delta R_{\rm{HK'}}$ \citep[defined by][]{2002AN....323..357S}. This is validated in the Sun because the frequency shifts change linearly with the temporal evolution in the Mg~{\sc{II}}  lines \citep{2007ApJ...659.1749C}. With this scaling, the frequency shifts generally increase towards lower temperature and decrease with age as seen in Fig.~\ref{FIG_variation_time_freq_shifts} \citep{2009MNRAS.399..914K}.

\begin{figure}[!htp]
	\centering
		\includegraphics[width=0.75\textwidth] {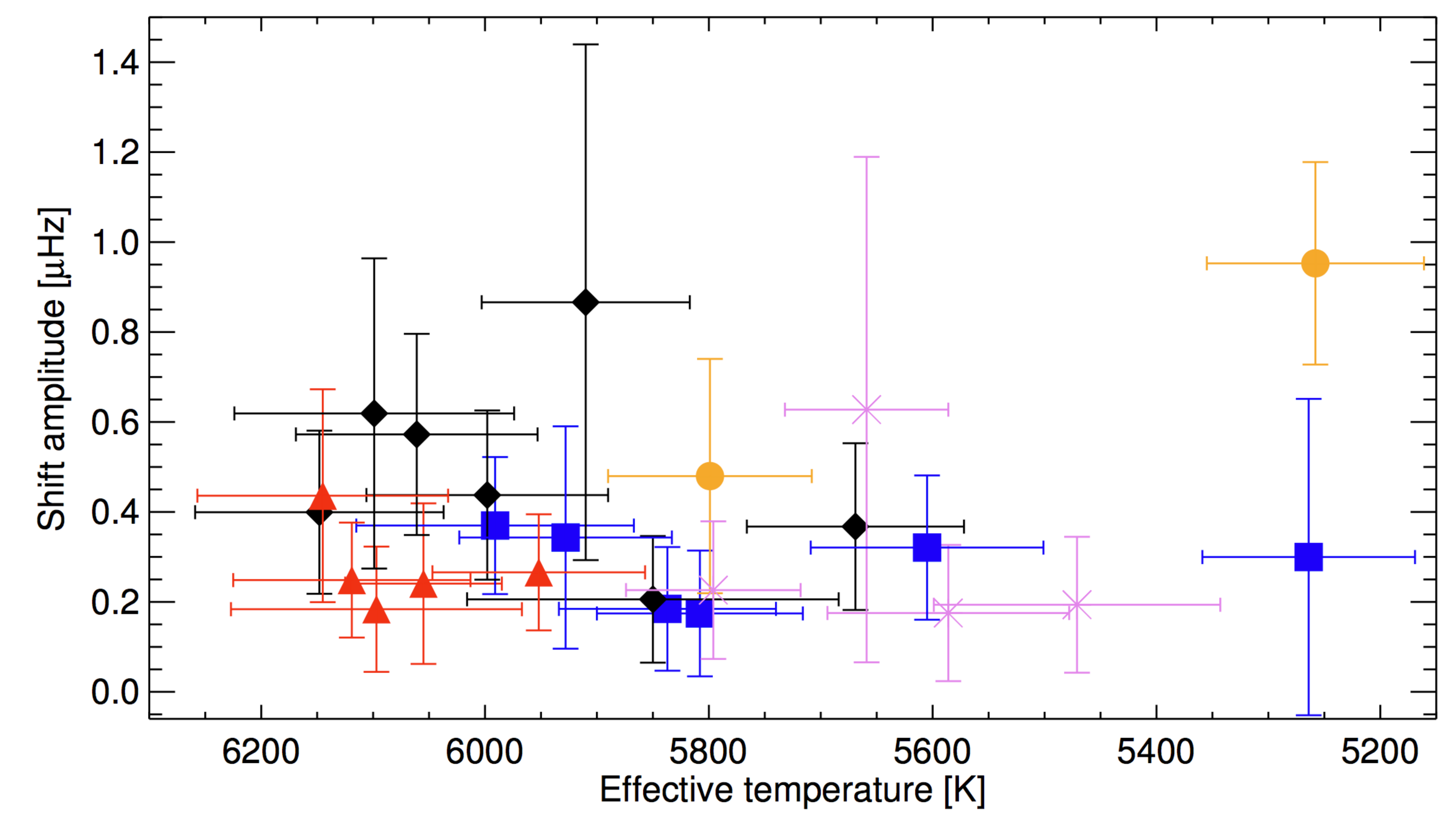}
	\caption{Average frequency shifts as a function of effective temperature for 24 \emph{Kepler} targets. Black diamonds, red triangles, orange circles and purple asterisks represent stars younger than 4 Gyr, between 4 to 5 Gyr, between 5 to 6 Gyr, and between 6 to 7 Gyr respectively. Figure from \cite{2017A&A...598A..77K}.}
	\label{FIG_freq_shift_teff}
\end{figure}

On the other hand, \cite{2007MNRAS.379L..16M}, proposed that the frequency shifts are also proportional to $\Delta R'_{\rm{HK}}$ but scaled with a factor that depends on the depth of the perturbation normalized by the mode inertia. In this case, the frequency shifts generally increase towards higher temperature and decrease with age (see  Fig.~\ref{FIG_variation_time_freq_shifts}). While both scalings yield a decrease of the frequency shifts with age, they have different predictions with effective temperature. The result of the hot F star HD~49933 and the Sun validates the second methodology. Unfortunately, the ensemble analysis of 24 \emph{Kepler} targets by \cite{2017A&A...598A..77K} (see Fig.~\ref{FIG_freq_shift_teff}) was unable to distinguish between both scaling relations due to the large uncertainties in the frequency shifts. However the authors made a comment about a weak correlation with $T_{\rm{eff}}$ supporting the scaling relation by \cite{2007MNRAS.379L..16M} but with one clear exception, KIC~8006161 the high metallic solar analogue already discussed in detail in previous paragraphs. Another small trend was found between the frequency shifts and age (excepting again KIC~8006161) in the direction predicted by both scaling relations, i.e., reduction of the frequency shifts (activity) with age as found by \cite{1972ApJ...171..565S}.
 
\subsubsection{Relation between frequency shifts and amplitude shifts for a large sample of stars}

The analysis of the temporal variations of the seismic parameters and $S_{\rm{ph}}$ of a larger sample of 87 solar-type \emph{Kepler} stars \citep{2018arXiv180600136S}, showed similar results to those of \cite{2017A&A...598A..77K}. They found that about 60$\%$ of the stars in the sample show ``(quasi-)periodic variations'' in the frequency shifts. Moreover, 20$\%$ of the stars show frequency and amplitude shifts correlated instead of anticorrelated. Although these results seem to be puzzling, they could be explained in a simple way. First, \cite{2018A&A...611A..84S} showed that small variations in frequency shifts -- that could be interpreted as due to magnetic origin -- can be explained by different realizations of stochastic noise. Second, the presence of hysteresis effects between different magnetic proxies implies that at several stages of the cycle, two indexes could be in phase while they would be most of the time in anti-phase. Longer time series covering several stellar magnetic cycles would be required to properly understand all of these observations. 

\subsection{On the variation of the frequency shifts with frequency}

The linear dependence of the frequency shifts with frequency found in the Sun and in HD~49933 was also found in the young solar analogue KIC~10644253 \citep{2016A&A...589A.118S}. Therefore, in these three stars the perturbation inducing the variation of the mode parameters needs to be located outside of the resonant mode cavity of the modes, i.e., in a thin layer very close to the photosphere. Otherwise, an oscillatory signal would be expected as was first discussed by \cite{1991ApJ...370..752G}. They explained the apparent oscillatory signal superimposed on the linear dependence of the frequency shifts with frequency in the Sun as the consequence of a perturbation located near the He {\sc{I}} ionization zone. This depth was found by analyzing the periodicity of this oscillatory signal. Later, \cite{Gough1994} proposed that the perturbation was much deeper and that it was due to changes in the acoustic glitch of the He {\sc{II}} ionization layer. Indeed several authors have found variations in the amplitude of the depression in the adiabatic index, $\Gamma_1$, at this ionization layer that could be the consequence of the changing activity on the equation of state of the gas in that layer \citep{2004ApJ...617L.155B, 2006ApJ...640L..95V}.

\begin{figure}[!htp]
	\centering
		\includegraphics[width=1.01\textwidth] {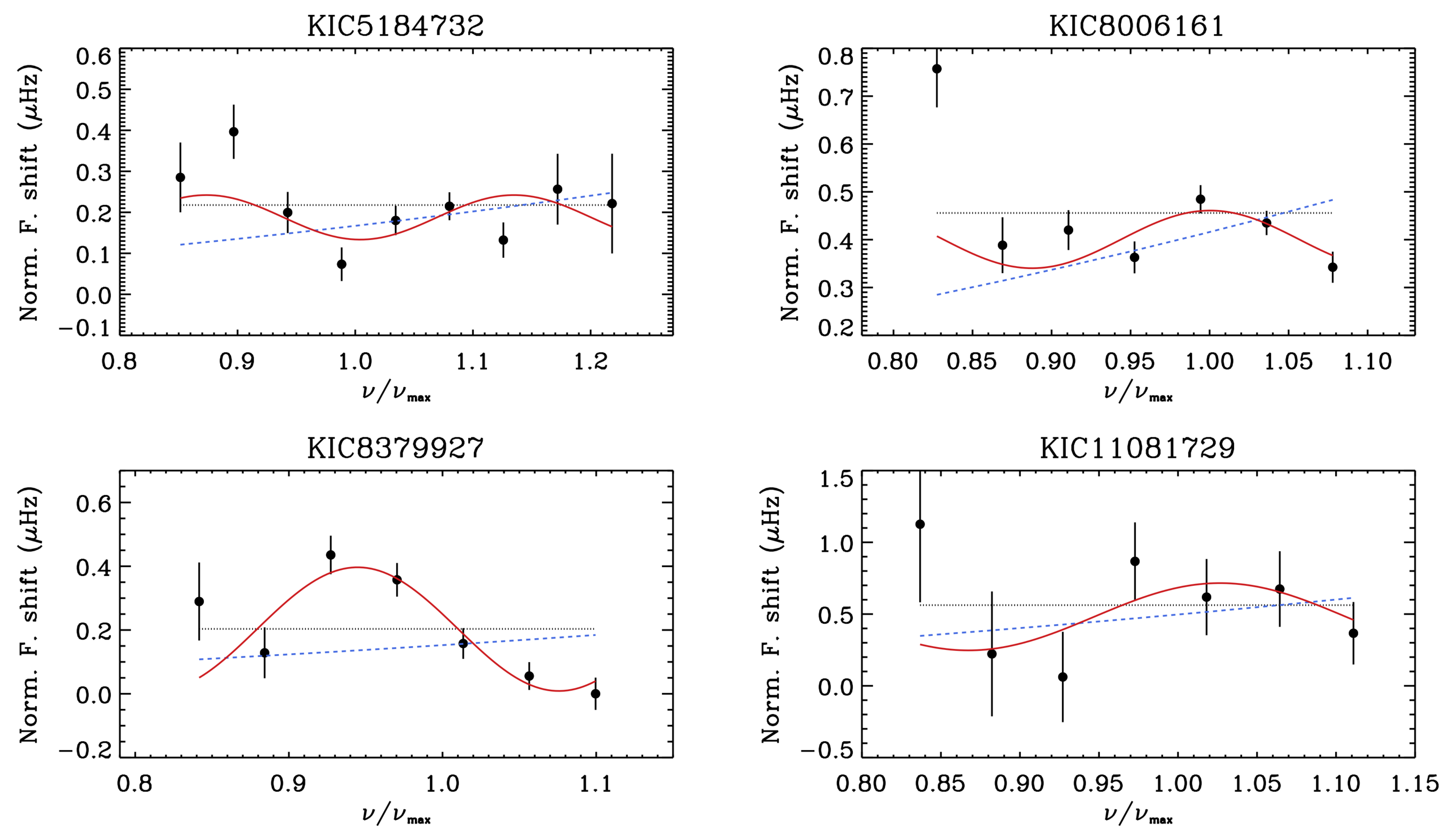}
	\caption{Frequency shifts normalized by the mode inertia as a function of frequency normalized by $\nu_{\rm{max}}$ for four \emph{Kepler} targets. The solid red line is the result of a sine-wave fitting. The black dotted line is the average of the frequency shifts while the blue dashed line is the results of doing a linear fit. Figure taken from \cite{2018A&A...611A..84S}.}
	\label{sinusoidal_freq_shift}
\end{figure}

\cite{2018A&A...611A..84S} found that in four stars of the \emph{Kepler} sample (KIC~5184732, KIC~8006161, KIC~8379927, and KIC~11081729), the frequency shifts normalized by the mode inertia show an oscillatory behavior instead of a linear one (see Fig.~\ref{sinusoidal_freq_shift}). Before deriving firm conclusions about the positions and mechanisms responsible for the frequency shifts in stars, it is necessary to better determine these frequency shifts as a function of frequency (with smaller uncertainties) as well as in a larger number of stars.  However, it seems that the picture is much more complicated than was outlined from the analysis of the Sun. 

\section{Conclusions and Perspectives}

In this work we have reviewed the theory behind the asteroseismic techniques applied to study main-sequence solar-like dwarfs, as well as the latest results obtained from ground-based and space-borne missions such as CoRoT, \emph{Kepler}, K2 and TESS. This is a young research topic that has reached its constant pace during the last decade. 

It has been shown that asteroseismology of MS solar-like stars is providing strong constraints about the structure and dynamics from the surface to the core of stars. It is possible today to infer masses, radii and ages with a precision and accuracy never reached before. This is impacting many fields in stellar astronomy. For example, precise and accurate ages of field stars at all stages of evolution in the main sequence have shown that stars seem to stop braking when they reach a given Rossby number and from there, they seem to continue with a quite constant surface rotation \citep{2016Natur.529..181V}. This could be the consequence of a change in the properties of external magnetism, which could also impact stellar magnetic cycles \citep{2017SoPh..292..126M} and the possibility of giving a correct age to middle aged stars through gyrochronology. An extended sample of stars and complementary ground-based observations of the magnetic activity will be necessary to progress in this area. 

Asteroseismology is also helping to improve exoplanet research \citep[for a full review on the synergies between asteroseismology and exoplanetary science see][]{2018ASSP...49..119H}. For example, precise stellar ages are a key parameter to date the full planetary systems and thus, better understand the theory of formation and evolution of planet-hosting stars and the extrasolar planet systems as a whole. The asteroseismic improvement in the determination of stellar radius is directly impacting the precision of planet radius extracted with the transit method. Thanks to this new increased precision, it is now possible to properly characterize the so-called ``radius valley'' or ``photoevaporation desert'' at around 2 R$_\oplus$ \citep{2016NatCo...711201L,2018MNRAS.tmp.1712V}. Finally we mention that the combination of transit photometry with asteroseismology allows a systematic measurement of orbital eccentricities of transiting planets \citep[e.g.][]{2015ApJ...808..126V}, which was only possible before in relatively large gas-giant planets, or for multiplanet systems where the effects of eccentricities and masses could be successfully distinguished, 

New internal rotation profiles have encouraged stellar astrophysicists to study angular momentum transport and efficient mixing processes and develop new mechanisms explaining the quasi-uniform rotation found in the outer part of the radiative zone, the external convective zone and the surface. 

In conclusion, asteroseismology of solar-type stars is in very good shape. At the time of writing these conclusions, the community is actively analyzing K2 data, where dozens of new pulsators are foreseen, as well as the first two sectors of  TESS data that are already available. New missions will contribute to enhance our known sample of pulsating star. The future is already here because of the work engaged to prepare the ESA's M3 PLATO mission, which will be able to characterize tens of thousand of these MS cool dwarfs after 2026 for which many of them will be planet hosts. Asteroseismology of MS solar-like dwarfs is just at the dawn of its potential.


\newpage



\section{Acknowledgements}
\label{sec:acknowledgements}
The authors wish to thank the entire SoHO, CoRoT and \emph{Kepler} teams, without whom many of the results presented in this review would not be possible. The authors received funding from the European Community seventh programme ([FP7/2007-2013])  under grant agreement no. 312844 (SPACEINN) and under grant agreement no. 269194 (IRSES/ASK).  The authors also acknowledges funding from the CNES. RAG  acknowledges the ANR (Agence Nationale de la Recherche, France) program IDEE (n. ANR-12-BS05-0008) ``Interaction Des \'Etoiles et des Exoplan\`etes''. The authors also want to thank Dr. J. Leibacher who did a thorough lecture of the manuscript helping us to improve the manuscript.

\section{Appendix A}
\label{app:A}
\modif{In table~\ref{tab1}, the values of the large frequency spacing and the effective temperature used to plot Fig.~\ref{Seis_HR_simple} are listed, as well as the references where these values are from. A complete version of this table is available on line.}

\begin{table}
\caption{\modif{Large frequency spacing, effective temperature and references of the stars shown in Fig.\ref{Seis_HR_simple}. A complete version of this table is available on line.}}
\label{tab1}       
\begin{tabular}{llll}
\hline\noalign{\smallskip}
Identifier & $\Delta\nu$ & $T_{\rm{eff}}$ & Reference   \\
\noalign{\smallskip}\hline\noalign{\smallskip}
	 Sun  &   5777  & 135.0    & 		\\
     1430163  &   6796  &   84.60  &   Chaplin et al. 2014\\
     1435467  &   6433  &   70.80  &   Chaplin et al. 2014\\
     1725815  &   6550  &   55.40  &   Chaplin et al. 2014\\
     2010607  &   6361  &   42.50  &   Chaplin et al. 2014\\
     2309595  &   5238  &   39.30  &   Chaplin et al. 2014\\
     2450729  &   6029  &   61.90  &   Chaplin et al. 2014\\
     2837475  &   6688  &   75.10  &   Chaplin et al. 2014\\
     2849125  &   6158  &   41.40  &   Chaplin et al. 2014\\
     2852862  &   6417  &   53.80  &   Chaplin et al. 2014\\
     ...	&	&	&\\
\noalign{\smallskip}\hline
\end{tabular}
\end{table}

\newpage



\bibliography{jbbib,BIBLIO}

\end{document}